\documentclass[utf8]{frontiersFPHY} % for Physics and Applied Mathematics and Statistics articles
\setcitestyle{round} % for Physics and Applied Mathematics and Statistics articles
\usepackage[onehalfspacing]{setspace}
\usepackage{url,hyperref,lineno,microtype,multirow,graphicx}
\usepackage{amssymb}
\usepackage[export]{adjustbox}
\usepackage{tablefootnote}

\def\keyFont{\fontsize{8}{11}\helveticabold }
\def\firstAuthorLast{Yap {et~al.}} 
\def\Authors{Jacinta Yap\,$^{1,2*}$, Adam Steinberg\,$^{2,3,4}$, Hannah Norman\,$^{2,3,4}$, Konrad Nesteruk\,$^{5}$, and Suzie Sheehy\,$^{2,6}$}
\def\Address{
$^{1}$Department of Radiation Oncology, Peter MacCallum Cancer Centre, Melbourne, VIC~3000, Australia \\
$^{2}$School of Physics, University of Melbourne, Melbourne, VIC~3010, Australia \\
$^{3}$Cockcroft Institute, Daresbury, Warrington WA4~4AD, United Kingdom \\
$^{4}$Department of Physics and Astronomy, University of Manchester, Manchester M13~9PL, United Kingdom \\
$^{5}$Department of Radiation Oncology, Massachusetts General Hospital and Harvard Medical School, Boston, Massachusetts~02114, United States \\ 
$^{6}$Australian Nuclear Science and Technology Organisation (ANSTO), Lucas Heights, NSW~2234, Australia
}

\def\corrAuthor{Jacinta Yap}
\def\corrEmail{jacinta.yap@petermac.org}

\begin{document}
\onecolumn
\firstpage{1}

\title[Toward Ultra-fast PBT Treatments]{Toward Ultra-fast Treatments: Large Energy Acceptance Beam Delivery Systems and Opportunities for Proton Beam Therapy}

\author[\firstAuthorLast ]{\Authors} %This field will be automatically populated
\address{} %This field will be automatically populated
\correspondance{} %This field will be automatically populated

\maketitle

\begin{abstract} 
The availability of proton beam therapy (PBT) continues to grow exponentially worldwide, driven by technological advancements to reduce the facility size and costs, and toward more efficient and higher quality treatments. The characteristic physical and biological advantages of protons can provide superior clinical outcomes for patients, as modern techniques enable a highly configurable and conformal dose delivery. Although active scanning methods allow precise beam control, PBT beams are highly sensitive to range and motion errors which impact treatment quality. Treatment delivery is largely determined by capabilities of the beam delivery system (BDS), where faster delivery can have many potential benefits including improved dosimetric quality, utility, cost effectiveness, patient throughput and comfort. Despite significant developments in accelerators, delivery methodologies, dose optimisation and more, the energy layer switching time (ELST) is still a persisting limitation in existing BDS. The ELST can contribute significantly to beam delivery time (BDT) and extend treatment times, requiring compensation by optimisation planning approaches, motion mitigation strategies, or active beam modification. This fundamental constraint can be addressed by increasing the narrow energy acceptance range of conventional beamlines to minimise the ELST, enabling ultra-fast delivery. A large energy acceptance (LEA) BDS has the potential to revolutionise PBT through immediate improvements to current treatment delivery and emerging delivery modalities: the complete exploitation of PBT -- and unlocking its full potential – can only be made possible with advances in beam delivery technologies. We review the abundant opportunities offered by an ultra-fast BDS: shorter treatment times, reduced motion induced dose degradation, improved effectiveness of motion management techniques, possibilities for volumetric rescanning, bidirectional delivery, further planning optimisation, and novel delivery strategies. We overview the design concepts of several LEA proposals, technology requirements, and also discuss the remaining challenges and considerations with realising a LEA BDS in practice. There are multiple avenues requiring further development and study, however the clinical potential and benefits of this enabling technology are clear: ultra-fast delivery offers both immediate and future improvements to PBT treatments.

\tiny
 \keyFont{\section{Keywords:} proton beam therapy, charged particle therapy, beam delivery, large energy acceptance, rapid delivery, novel delivery modalities, compact facilities} 
\end{abstract}

\section{Introduction} 
Proton beam therapy is a well established modality of radiotherapy cancer treatment and one of the most advanced and precise techniques available. Using a beam of protons can be clinically advantageous compared to conventional X-ray radiotherapy (XRT) as the characteristic `Bragg Peak' (BP) enables radiation to be delivered to targeted sites with greater accuracy and radiobiological impact\footnote{Heavier ions such as carbon or helium also exhibit these characteristic benefits and are also used for treatment in charged particle therapy (CPT), but to a lesser extent.}. Modern PBT delivery -- pencil beam scanning (PBS) -- utilises the high energy deposition at the BP to achieve an extremely precise and conformal dose distribution throughout the entire tumour volume, sparing healthy surrounding tissue. This makes PBT highly suited for deep-seated cancers and tumours situated near critical organs. The dose sparing of normal tissue lowers the risk of damaging side effects, improving tolerance to treatment and a better quality of life for patients. New possibilities with multi-modality or combined therapies (i.e. immunotherapy) could also lead to better clinical outcomes \cite{Durante2021}. 

\begin{figure}[htb!]
\centering
\includegraphics*[width=\textwidth]{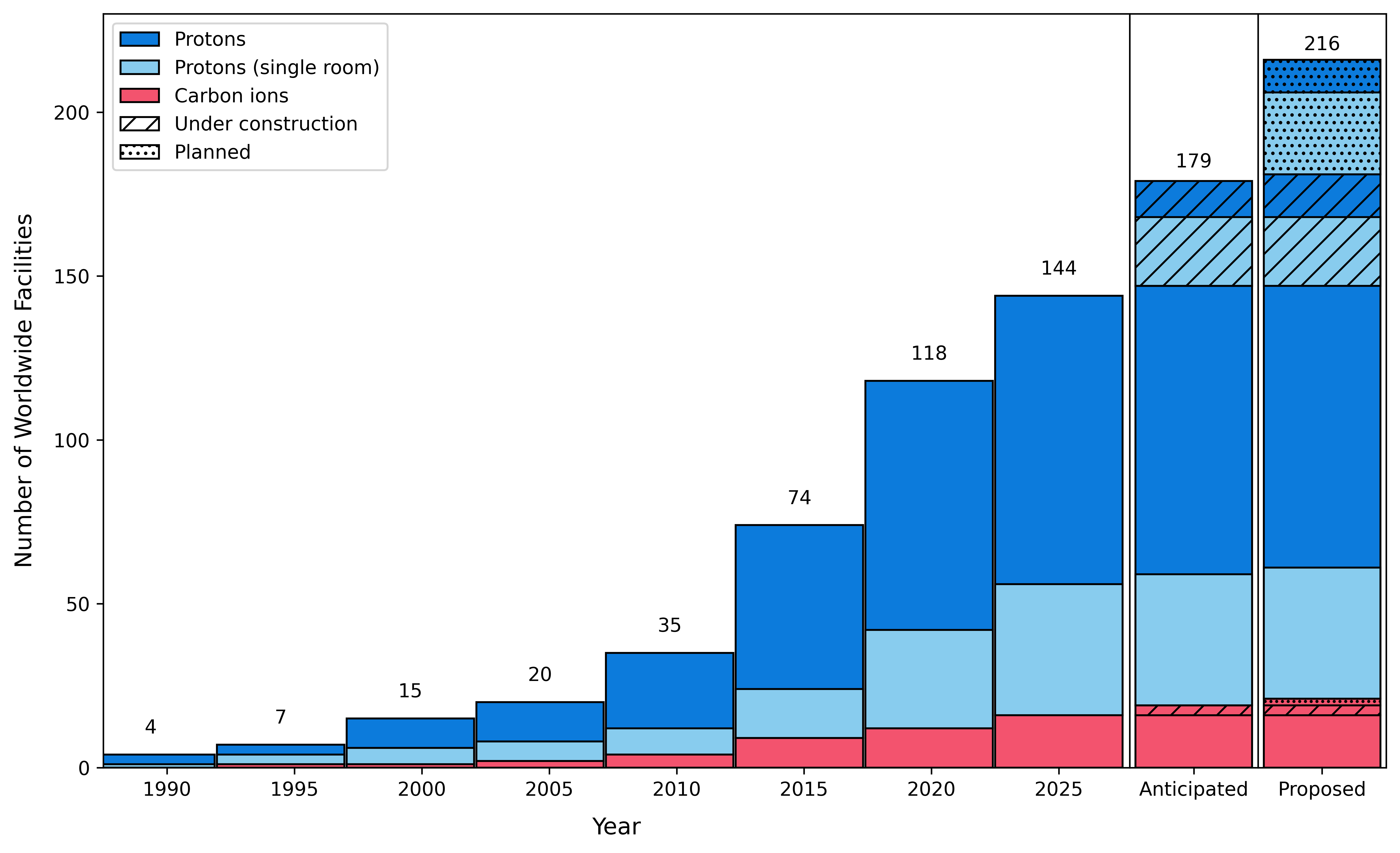}
\caption{Worldwide PBT and carbon ion beam therapy (CIBT) facilities: operational, anticipated (under construction) and proposed (planned), total number of CPT facilities listed above bars. PTCOG data updated Dec 2025 \cite{PTCOGfacilities}. Single room PBT facilities are those reported with one beam or gantry -- including dedicated ocular and upright centers.}
\label{F_CPTstats}
\end{figure}

Radiotherapy is involved in the treatment of approximately 40\% of all cancer patients \cite{Delaney2005,Barton2014}: an estimated 50\% of these patients could benefit from PBT, yet only $<$1\% of radiotherapy patients currently receive it \cite{Yan2023}. Presently there are $>$140 operating charged particle therapy facilities (128 PBT and 16 carbon ion centres, Dec 2025 \cite{PTCOGfacilities}) and although PBT availability is increasing worldwide (Figure \ref{F_CPTstats}), the global burden of cancer is growing: 50-70\% of cancer patients need radiotherapy \cite{Abdel-Wahab2024} yet many barriers still prohibit widespread adoption. Although considerable advances in accelerator and delivery technologies -- particularly associated with the size and complexity of machines -- have made it easier and more affordable to provide PBT, a significant and prevailing hurdle to access is the high cost \cite{Bortfeld2017}. 

\begin{figure}[htb!]
\centering
\includegraphics*[width=\textwidth]{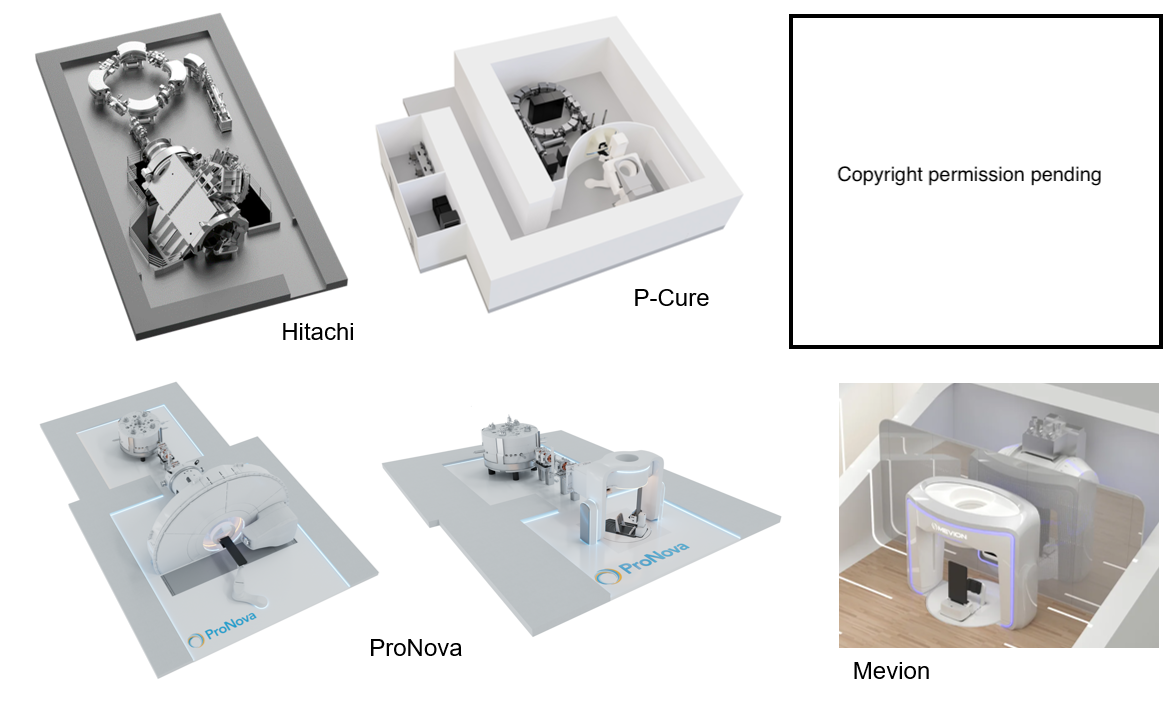}
\caption{Compact, single-room PBT systems with gantries (left) and upright systems (middle, right). Hitachi \cite{Rossi2022}, P-Cure and MedAustron synchrotron solutions (top). ProNova and Mevion upright solutions with the Leo Cancer Care Marie chair (bottom, right). Images reproduced with permission, P-cure, ProNova Solutions and Mevion Medical Systems, pending permission from MedAustron.} 
\label{F_CompactFacilities}
\end{figure}

Recent growing interest in reduced footprints and costs have led to significant investment in smaller accelerators and therefore `single-room' systems\footnote{$\sim$65\% of CPT facilities either under construction or planned are listed as having only one treatment room \cite{PTCOGfacilities}.}. The majority of upcoming PBT facilities are single-room systems with superconducting (SC) synchrocyclotrons such as the IBA ProteusOne \cite{Pidikiti2018}, gantry-less Mevion S250-FIT or ProNova SC360. Compact synchrotrons are now available from a range of vendors and have been installed at several sites by Hitachi (PROBEAT) \cite{Rossi2022}, P-cure \cite{Feldman2024} and Protom (Radiance 330) \cite{Balakin2021}. An alternate design has also been proposed by MedAustron (Compact 200+) with upright, couch or gantry configurations (Figure \ref{F_CompactFacilities}). A facility which can accommodate both the accelerator and treatment room within a single room (or similarly sized XRT linac bunkers) can be $\sim$USD~\$30-50M \cite{Bortfeld2017,Kerstiens2018a} compared to $\sim$USD~\$100-200M \cite{CADTH_145,Xia2022} for a multi-room facility -- the standard route in previous decades. There is developing ambition for compact technologies, given that the BDS\footnote{Here we refer to BDS as comprising all the components after the accelerator complex in the bunker, downstream to the nozzle in the treatment room.} is a major contributor to the capital costs and physical space requirements. Most BDS include a gantry, a massive mechanical structure supporting a series of magnets to steer the beam to the patient. A gantry-less solution enables the possibility of retrofitting into existing facilities \cite{Bortfeld2020,Clasie2022}, also encouraging greater worldwide adoption \cite{Yan2023}. 

The future technological requirements to fully exploit PBT are well recognised \cite{Yap2021}: there is a clear evolution of the field striving for better treatment efficiency (cost reduction, improved patient throughput and caseload utility) and efficacy (treatment quality). Much of the R\&D progress made in recent years has resulted in improvements to accelerators, beam transport, delivery methodologies, dose optimisation methods and many more \cite{Paganetti2021,Schreuder2020,Myers2019,Vidal2021,Flanz2017,Mohan2017,Farr2018,Collings2022}. This has led to cost reduction in providing PBT and continues to be the trend today \cite{Graeff2023,Mohan2022,Durante2021}. Despite progress, even if cost-effective treatments and universal patient access to PBT are realised, patients will be unable to completely benefit from the dosimetric and therapeutic advantages offered by PBT unless persistent limitations in current technologies are overcome. 

A key constraint in the beam delivery process is the time delay incurred in varying the magnetic fields and beamline components: the ELST. One approach which could address this is to increase the momentum acceptance range by design (i.e. a large momentum acceptance or large energy acceptance BDS) to allow a wider range of beam energies to be transported and delivered to the patient. In principle, accepting a larger beam momentum range bypasses restrictions imposed by existing BDS designs and eliminates the existing ELST bottleneck, speeding up energy changes for longitudinal layer coverage. The potential improvement to PBT treatments enabled with a LEA system could be revolutionary: the immediate clinical benefits with ultra-fast delivery and overall shorter treatment times, and also the emerging techniques and future therapies it makes possible. 

One crucial aspect which forms the focus of this review article is the delivery speed, including the role of the BDS and restrictions with treatment efficiency. Faster beam delivery brings potential clinical and economic benefits, in addition to an improved patient experience. We explore opportunities to improve the beam delivery process by increasing beamline momentum acceptance and minimising the ELST, enhancing scanning capabilities. We examine the clinical advantages of a LEA BDS enabling ultra-fast delivery, given the direct benefits to treatment times, motion management, optimised delivery, and new possibilities. Increasing the momentum acceptance range has been a design objective explored in many conceptual proposals, but a LEA BDS for PBT has yet to reach clinical implementation. An overview of existing LEA proposals are presented, comparing their design features, optics, magnet technology and feasibility. Finally, we identify key considerations for beam quality and translation of this technology into clinical practice, examining strategies for realistic performance and operation.  

\section{The Case for Faster Beam Delivery: Opportunities and Challenges}
\label{Section_ClinicalMotivation}
Most modern PBT facilities have adopted technologies to deliver treatments using active PBS: a highly conformal and optimised dose distribution is achieved by changing the beam intensity, energy, and transverse position i.e. intensity modulated proton therapy (IMPT). The beam can also be delivered from multiple angles, typically with gantries which rotate to direct the beam to the patient.

Many interconnected factors determine how quickly treatment can be delivered. The main methods of spatially controlling the beam is by spot, line, or raster scanning, however the entire delivery process itself is multifaceted: there are many different technological components and systems (accelerator, beam transport, and control systems etc.) responsible for beam production, transport, and delivery. The BDT  -- which includes the duration of active irradiation and beam preparation -- can vary significantly depending on the target volume, characteristics of the BDS, accelerator output, transmission capabilities and various other technical parameters as described in \cite{Yap2021}. 

\begin{figure}[htb!]
    \centering
    \includegraphics*[width=\textwidth]{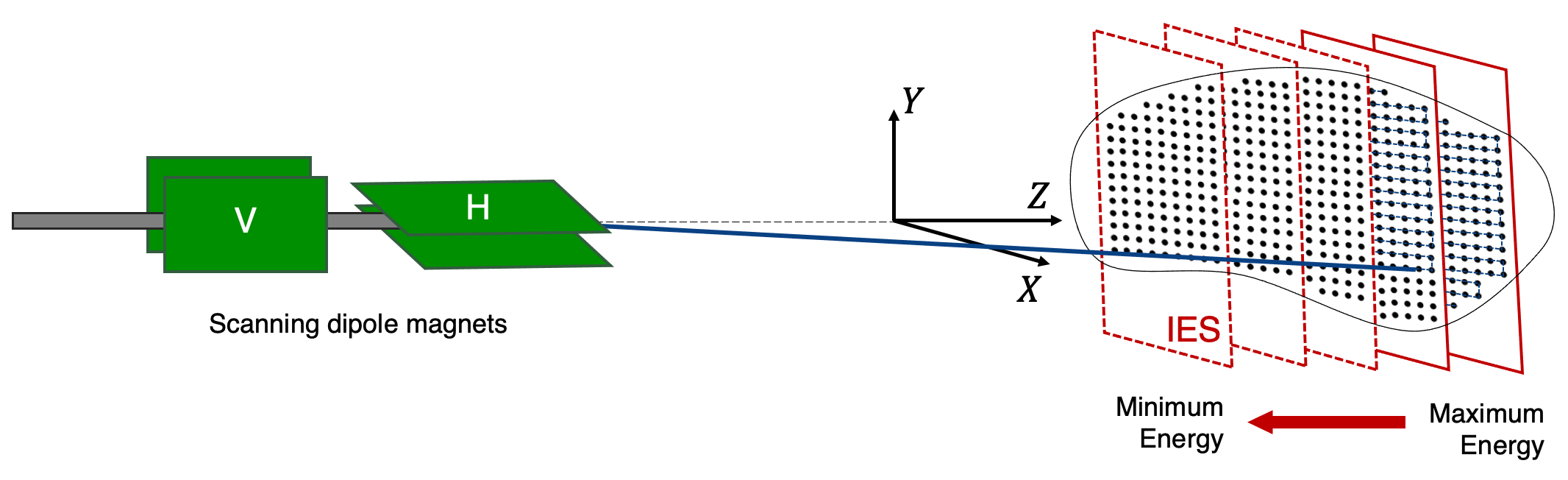}
    \caption{Schematic showing typical PBS delivery. 3D coverage of the treatment volume is achieved by scanning the beam across a layer (in the X--Y plane) in a predetermined pattern before the energy is lowered reaching a shorter depth (in Z), then scanned across and repeated again for each consecutive layer. Adapted from \cite{Yap2021}.} 
    \label{F_PBS}
\end{figure}

In PBS, a narrow beam scans across a transverse layer or `iso-energy slice' (IES) in a pre-programmed pattern before the energy of the particle beam is reduced and the beam is scanned again. This is repeated for successive layers until there is complete coverage of the whole volume (Figure \ref{F_PBS}). This 3-dimensional distribution of dose reduces excess radiation outside the target site, minimising damage to healthy tissue and the risk of side effects. 

\begin{figure}[htb!]
\centering
\includegraphics*[width=\textwidth]{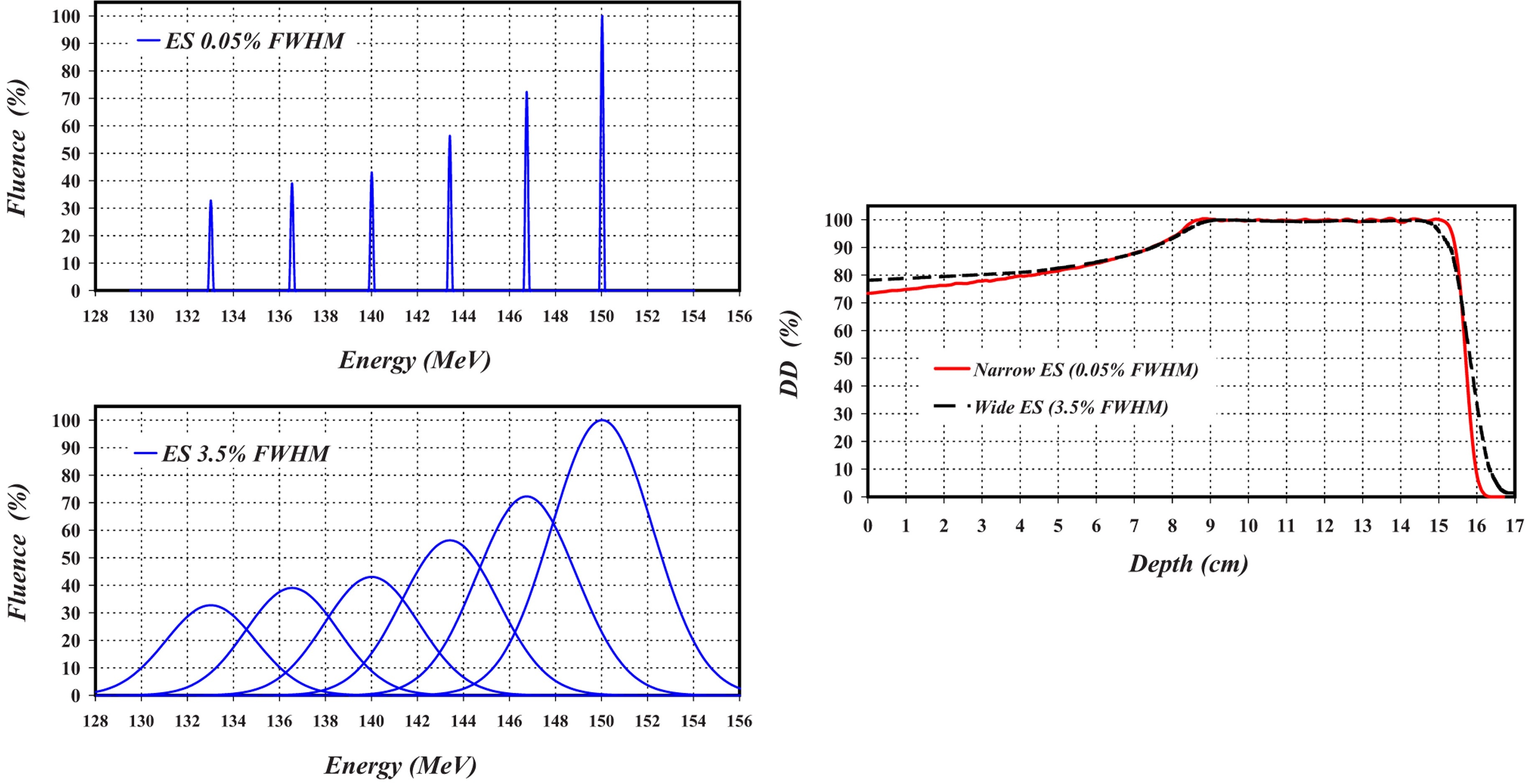}
\caption{Beams with a narrow (0.05\%, top left) and wide (3.5\%, bottom left) incident energy spread (ES) and their resulting spread out BPs (right). A shallower surface dose and sharper distal fall-off can be achieved with a smaller beam energy spread. Reproduced with permission  \cite{Hsi2009}.}
\label{F_SOBP}
\end{figure}

The resulting dose distribution is highly configurable: delivery parameters can in principle be varied according to the treatment objectives and plan, including the scanning path or pattern, number of fields or spots, amount of time the beam dwells at an individual position (or it may be irradiating whilst moving along a path, such as in continuous scanning \cite{Zenklusen2010}), size of the beam spot, longitudinal beam spread, and spacing between the IES and spots. Many of these can be varied with the treatment planning software (TPS), using different optimisation algorithms or strategies (i.e. single field vs. multi-field optimisation) to provide the best quality treatment given a balance of computational demand, robustness and treatment time \cite{Giordanengo2017}. For example, smaller spot sizes can achieve greater conformity \cite{VanDeWater2012,Kraan2018} but larger spots have higher plan robustness \cite{Liu2024}, and larger spot spacing can reduce treatment time but will degrade the homogeneity \cite{Alshaikhi2019}. Many narrow BPs modulated will result in a smaller entrance plateau and steep distal SOBP fall-off but require many closely spaced IES for a uniform distribution \cite{Hsi2009} (Figure \ref{F_SOBP}). In contrast, a wider BP increases the IES spacing, requiring less layers for coverage but may reduce conformity \cite{VanGoethem2009}. Further to this, fundamental issues such as patient and tumour motion will also have an inevitable impact on treatment quality \cite{Pakela2022,Schaub2020}. This requires additional considerations with planning and optimisation, discussed later in Section \ref{Section_ReducingBDT}.

Consequently, as the delivery of hundreds of thousands of narrow beams are required for coverage of the entire tumour volume, PBS is inherently slow. Changing the depth of the beam requires complex adjustments -- with the machine, magnets and control systems -- and the time taken can be particularly lengthy relative to the total BDT: this energy layer switching time is a recognised, underlying bottleneck \cite{Yap2021}. Especially for large or complex cases which require many energy layers, and facilities with machines with longer ELSTs, the accumulation of time spent unnecessarily waiting becomes increasingly significant. Studies performed at clinics to model delivery parameter timings estimate the ELST can contribute up to 70-90\% to the total BDT (independent of tumour characteristics) \cite{Suzuki2016,Shen2017}.  

Efficiency gains in the BDT achieved through better system and software optimisation of the ELST, and facility scheduling utilisation have also demonstrated substantial improvements in uptime, clinical capacity, and workflow. A facility which implemented these upgrades was able to shorten their BDTs by more than a factor of 3, resulting in more fields and fractions being delivered and therefore higher patient throughput even with a mixed case load \cite{Mah2020}. The authors of this study also remark that faster treatments have the added benefit of allowing clinical staff more time to engage patients within the treatment time slot and less time may be spent immobilised in discomfort: important improvements to the overall patient experience. 

\subsection{Minimising Energy Layer Switching Time}
\label{Section_Minimising_ELST}
The energy layer switching time is a limiting factor in delivery efficiency, ranging anywhere around the order of $\sim$100's ms up to a few seconds, depending on the accelerator and method of energy variation. An overview of reported ELSTs by several facilities given the main PBT accelerator types are listed in Table \ref{T_ELSTs}. 

\renewcommand{\arraystretch}{1.5}
\begin{table}[htb!]
\caption{Overview of reported ELSTs by accelerator type (vendor and model, if any) and facility. ELSTs achieved using multiple energy extraction (MEE) are indicated.} 
\resizebox{\textwidth}{!}{
\begin{tabular}{|l|l|l|l|}
\hline
  \multicolumn{1}{|c}{ELSTs} &
  \multicolumn{1}{|c|}{Accelerator} &
  \multicolumn{1}{|c|}{Facility} &
  \multicolumn{1}{c|}{Reference} \\ \hline
  \begin{tabular}[c]{@{}l@{}}80~ms (Gantry 2)\\ 200~ms (Gantry 3)\end{tabular} & Cyclotron & PSI, Switzerland & \cite{Pedroni2011} \\ \hline
   800~ms & Cyclotron (Varian ProBeam) & 	Emory Proton Therapy Center, USA & \cite{Zhu2023} \\ \hline
   900~ms & Cyclotron (IBA C230) & Willis-Knighton Cancer Center, USA & \cite{Pidikiti2018} \\ \hline
   500~ms & Cyclotron (IBA ProteusPLUS® 235) & ProCure Proton Therapy Center, USA & \cite{Mah2020} \\ \hline
  220~ms (MEE) & Synchrotron & HIMAC, Japan & \cite{Mizushima2017} \\ \hline
   2.1~s & Synchrotron (Hitachi PROBEAT) & MD Anderson, USA & \cite{Suzuki2016} \\ \hline
  \begin{tabular}[c]{@{}l@{}} 1.91~s \\ 200~ms (MEE) \end{tabular} & Synchrotron (Hitachi PROBEATV) & Mayo Clinic, USA & \cite{Shen2017,Younkin2018} \\ \hline 
  2~s & Synchrotron & MedAustron, Austria & \cite{Lebbink2022} \\ \hline
  700~ms & Synchrocyclotron (IBA ProteusONE® S2C2) & Beaumont Proton Therapy Center, USA  & \cite{Zhao2022} \\ \hline
  50~ms & Synchrocyclotron (Mevion S250i) & MAASTRO Proton Therapy, The Netherlands & \cite{Vilches-Freixas2020} \\ \hline
\end{tabular}
}
\label{T_ELSTs}
\end{table}

For a synchrotron based facility the ELST is on the order of seconds, as it is dependent on the ramping speeds of the storage ring magnets, extraction mechanism, or other features which are described in \cite{DeFranco2021}. This includes the time it takes to inject, accelerate, extract, decelerate and/or dump the remaining particles in each spill, for a single energy layer. To speed up IES changes, hybrid methods have been developed implementing a downstream range modulating device coupled with enhanced extraction operation (multiple-energy, flattop extraction), allowing different energies to be produced within a single spill \cite{Iwata2010,Noda2017}. The ability to modulate energies with the accelerator itself provides the advantage of not needing a degradation system, which impacts activation, radiation protection and transmission loss considerations. However, for C-12 or heavier ions which have a narrower BP, the accumulation of layers can result in a fluctuation of dose at the top of the SOBP distribution. To produce a smoother SOBP plateau, beam modulation devices (i.e. range modulators \cite{Simeonov2017} or ripple filters \cite{Weber1999}) are used to broaden BPs: these also act to reduce the number of IES and ELST, therefore speeding up the BDT. 

\begin{figure}[htb!]
\centering
\includegraphics*[width=\textwidth]{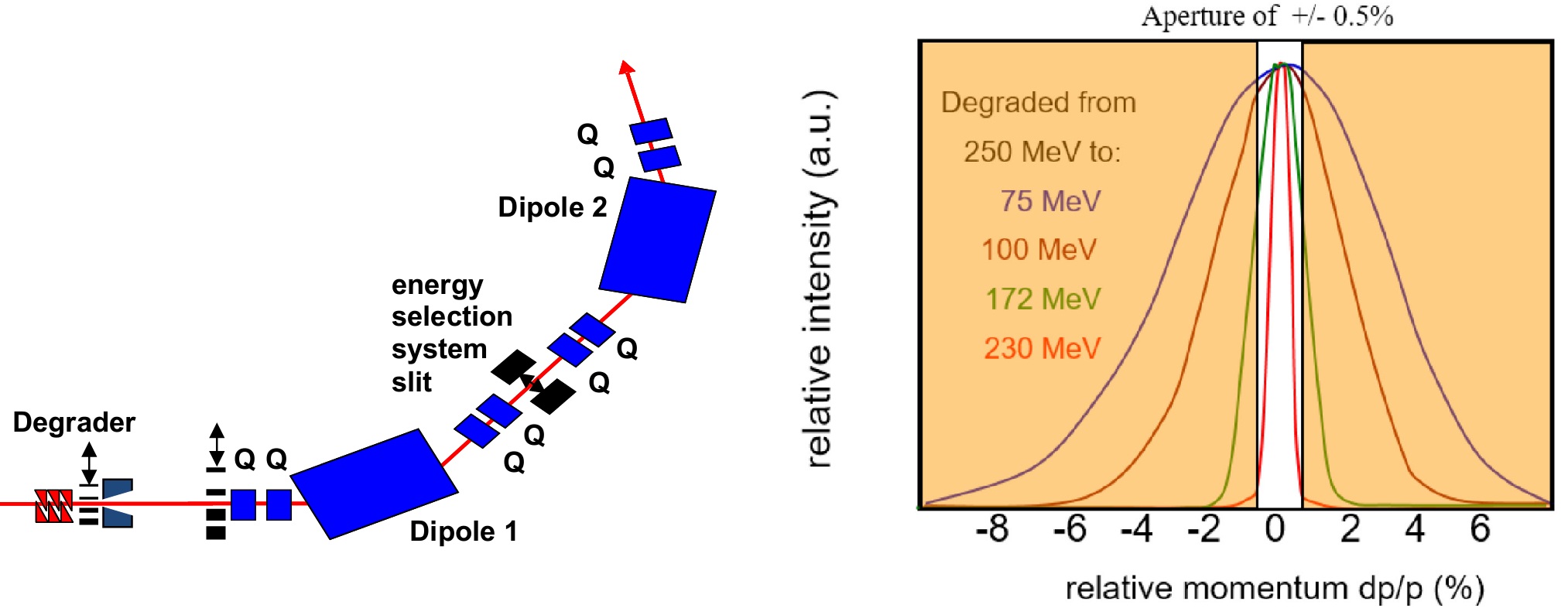}
\caption{Schematic showing components which may typically comprise an ESS (left), adapted from \cite{Gerbershagen2017}. Beam energies are reduced after passing through the degrader, resulting in particles with a spread of momenta (right). The ESS aperture indicates the distribution of particles which would be transported within a 0.5\% momentum acceptance window \cite{Schippers2017}.} 
\label{F_ESS}
\end{figure}

For facilities with isochronous cyclotrons or synchrocyclotrons, the ELST is typically determined by the time it takes for the magnets in the beam transport line or gantry to respond: to ramp, change and settle \cite{Gerbershagen2016,Psoroulas2018}. As the extraction energy is fixed, a degrader and energy selection system (ESS) -- comprising sections of absorbing material, collimators, magnets and diagnostics -- is used to physically attenuate the energy, selecting particles within a small momentum spread (i.e. dp/p 0.5-1\%) to produce the appropriate beam energy (Figure \ref{F_ESS}). Although these degrading devices are mechanically actuated, they can be rapidly positioned as fast as $\sim$10's msecs \cite{Chaudhri2010,Fattori2020,Nesteruk2019} and therefore the longer magnet ramping time for beam transport is the primary contribution to the ELST. 

Synchrocyclotrons such as the gantry-mounted Mevion S250 and S250i (which offers PBS), and the compact S250-FIT report a $\sim$50~ms ELST as there is no transport line, rather the beam is modified in the nozzle. The beam energy is specified by moving 18 polycarbonate plastic plates of varying thickness (the thinner plates are positioned upstream) in the beam path, driven by individual motors. The available combination of multiple plates provides a range resolution of 2~mm where switching between plates is optimised for delivery time and does not need to adhere to conventional high-to-low energy sequencing \cite{Kang2020a}. As there is no refocusing or energy selection, this consequently produces large beam spots with increased penumbra due to scattering \cite{Silvus2024,Grewal2021}. Therefore beam shaping hardware is necessary for adequate clinical performance, where two opposing sets of leaves -- adaptive aperture (AA) -- are used to collimate the low dose tails of each spot either dynamically or in static mode \cite{Vilches-Freixas2020}. The increased spot sizes due to this energy modulation method require AA for penumbra sharpening: this is important particularly for targets at shallow to intermediate depths as scattering in the patient dominates the lateral dose at large radiological depths \cite{Baumer2021,Grewal2021a}. 

\subsubsection{Beamline Energy Acceptance}
\begin{figure}[htb!]
\includegraphics*[width=\textwidth]{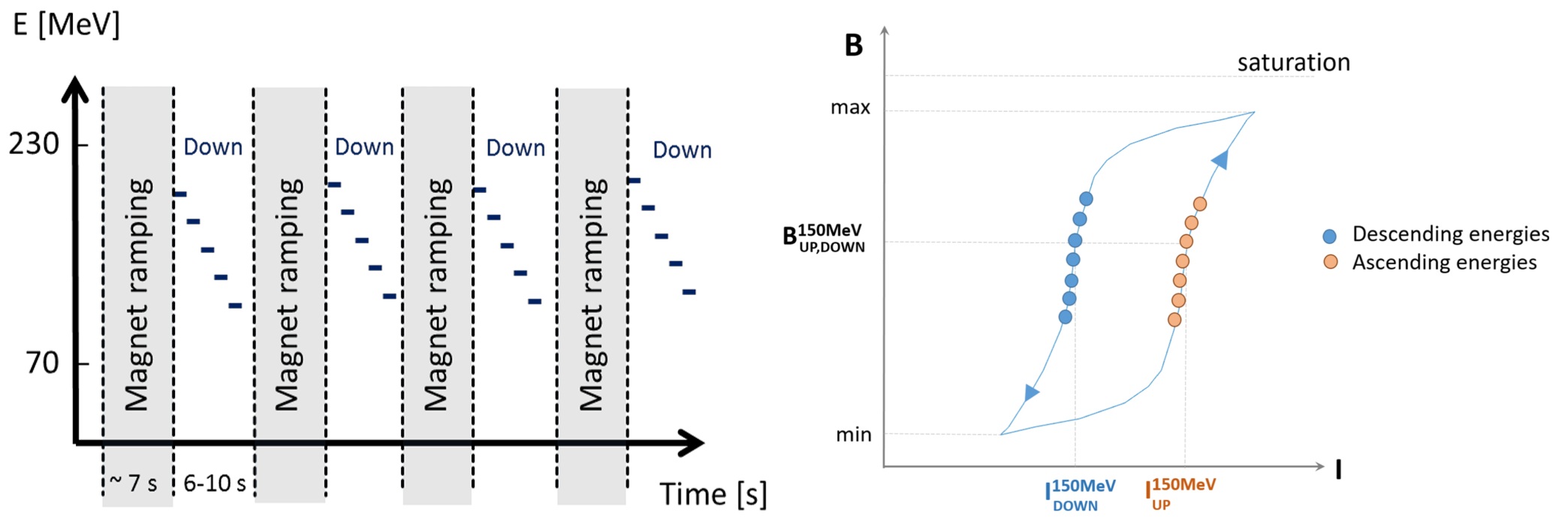} 
\caption{Magnet ramping sequence (shown for rescanning), conventionally from high to low energies, adapted from \cite{Actis2018}. Schematic of a hysteresis loop showing the field current correlation for an electromagnet. Stable delivery follows the cycle from maximum to minimum energy when delivering layers (blue circles). Delivering on the ascending side of the loop (orange circles) is performed for energy meandering, a novel delivery strategy \cite{Actis2023}.} 
\label{F_Hysteresis}
\end{figure}

Beamlines at treatment facilities typically have a momentum acceptance of 0.5-1\%. This means that delivery of each subsequent IES (generally varying 2\% in energy or $\sim$5~mm in depth) requires changing the beamline settings: this is the main bottleneck, as the fields produced by each magnet must be changed in order to accept and transport beams of a different energy \cite{Schippers2015,Gerbershagen2016b}. This is controlled by a set of values corresponding to magnet currents for the nominal fields at each requested beam energy/range \cite{Giovannelli2021}. The magnets are ramped proportionally to the beam momentum and all involved components (i.e. elements in the ESS) must also be adjusted synchronously. Magnet ramping is driven by current variation with the power supplies but this is limited by hysteresis effects, AC losses and eddy currents \cite{Psoroulas2018}. Stabilisation of the field (B to within 0.01\%) must be retained as small errors can lead to range deviations downstream \cite{Farr2018,Myers2019,Gerbershagen2016}. For example, a variation of just 0.5\% in B in the last bending magnet (in a gantry) can cause a 1~mm spot displacement \cite{Psoroulas2018}. Therefore, to preserve the reproducibility and position of the beam at isocentre, this process of beamline energy regulation can only be performed in small steps, typically moving in one direction (highest to lowest energy) along the hysteresis loop (Figure \ref{F_Hysteresis}). This cycle is established by a first full ramping to initialise the magnet and must be followed to sustain this current to field and energy relation: re-irradiating or changing energies between fields requires re-ramping, still following this sequence \cite{Fattori2020,Actis2018}. One way to move faster between energy steps is by enhanced magnets and optimising the beamline tuning process, where PSI have reported achieving their fastest ELST of 80 ms \cite{Pedroni2011,Myers2019}. However for most cases, ELSTs at the current timescales (see Table \ref{T_ELSTs}) are a bottleneck delay for PBS treatments: as mentioned, this is a constraint largely imposed by the optics design and could be overcome by increasing the momentum acceptance.

A BDS with a larger energy acceptance can minimise the ELST, leading to faster overall treatments, higher patient throughput and therefore lower costs \cite{Artz2025}. In addition to better treatment efficiency, a faster BDT does not need to sacrifice or compromise treatment quality \cite{Fu2024,OGrady2023,Muller2016}. In contrast, improvements can be achieved with faster delivery, particularly given dose uncertainties due to moving targets which necessitate the application of motion mitigation techniques \cite{Taasti2025,Mori2018,Graeff2014,Knausl2024,Engelsman2013}. An ideal scenario for ultra-fast delivery would be the possibility to deliver the treatment quicker than the timescale of patient motion i.e. within a single breath cycle, which would negate the need for motion compensation strategies. Nevertheless, even if treatment fields cannot be delivered in such time frames, faster delivery can benefit the implementation of mitigation techniques. 

\subsection{Interplay and Motion Mitigation} 
\label{Interplay_effects}
\begin{figure}[htb!]
\centering
\includegraphics*[width=0.7\textwidth]{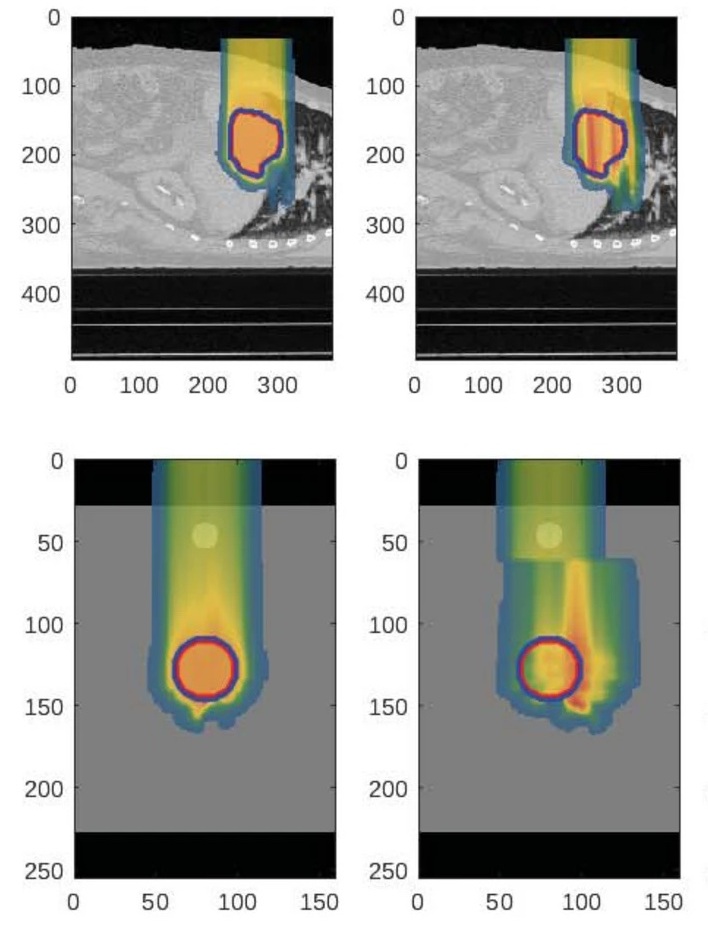} 
\caption{Dose distributions for a stationary target (left) and with motion applied (right), modelled on natural breathing, axes in mm. Resulting inhomogeneities and hot spots are shown when considering a liver case (top) or a 4~cm diameter spherical target (bottom) in a moving phantom, adapted from \cite{Fattori2020}.} 
\label{F_Interplay}
\end{figure}

Motion has a major impact on the accuracy of radiotherapy treatment and is a significant challenge in PBT, particularly with dynamic delivery methodologies. Interplay effects \cite{Pakela2022,Phillips1992,Bert2008} are caused by motion between the beam and patient (individually or simultaneously), resulting in a deviation between the planned and delivered radiation to the tumour. Consequently, this produces localised regions of over- or under-dosage which degrade the expected dose distributions and are detrimental to the quality of treatment (Figure \ref{F_Interplay}). 

Interplay effects can be highly variable as they are dependent on machine capabilities, combined with patient and tumour specific parameters \cite{Shan2020}. The overall magnitude of motion determines both the suitability of PBT and the extent of dose error caused by these effects: treating moving targets, such as tumours within or nearby mobile organs with PBS is problematic as greater motion will generally result in poorer target coverage \cite{Kraus2011,Li2015}. For complex cases or sites where there will be large motion (i.e. lung, liver, pancreas etc.) intensity modulated XRT or other methods (i.e. double scattered proton therapy) may be recommended instead given fewer associated risks and better clinical outcomes \cite{Mohan2022,Kang2017,Liu2011}. In some cases, the benefit of PBT may still compensate for degradation induced by motion \cite{Kothe2022}. 

Intrafractional motion is difficult to control and therefore mitigate, and must be considered on top of interfractional changes (for example temporal and/or anatomical changes such as tumour shrinkage or growth, weight loss etc.); commonly experienced during the course of any standard radiotherapy treatment \cite{Keall2004}. Any differences in geometry can have a significant impact on treatment quality and must be accounted for. Especially for PBT, dose or delivery errors are a greater concern due to the increased sensitivity to inaccuracies and dosimetric precision required. The impact of target motion can be more significant than interplay effects and is shown to increase with time \cite{Hoekstra2021}. Target drifting is evident even when using immobilisation devices and has a greater impact on planning target margins, but can be minimised by reducing total treatment time \cite{Hoogeman2008}.  

Furthermore, motion within the irradiated areas and in the beam path can affect the delivered dose distribution as this changes the radiologic path length \cite{Bert2011}. Treatment quality is correlated to the utilisation of the steep dose gradient and therefore influenced by the ability to predict the position of the BP, and deliver the beam to the correct depth \cite{Kraan2015}. Range uncertainties, limitations with in-vivo imaging and verification, and inaccuracies with CT conversion factors, all pose added challenges with motion \cite{Knopf2013}. There are many active respiratory motion management methods which are used by clinics including breath-hold techniques and beam gating or tracking -- these are not specifically discussed here, more can found in \cite{Zhang2023,Rietzel2010}. The effectiveness of these can depend on applicability to the patient, clinical criteria and facility experience, as well as the capability of the equipment and accelerator. Dose inhomogeneities can also be exacerbated if the patient motion synchronises with the beam delivery timing \cite{Poulsen2018}. For example, the time structure and dynamic control of pulsed beams with synchrotrons may result in worse quality treatments if the spill corresponds with patient respiratory cycles (i.e. 2--6~s) \cite{Younkin2021}. This may also be a factor with using techniques such as MEE, if the spill length is only able to cover a limited number of IES. Implementing one or a combination of these methods can help compensate for motion however often add complexity to the workflow, requires additional resources, expertise and most significantly, prolongs treatment times. Consequently, improving quality again comes at the cost of treatment efficiency. 

\subsection{Optimisation and Novel Delivery Schemes}
\label{Section_ReducingBDT}
There is clear value in having an increased energy acceptance to speed up delivery times, however the ideal range is yet to be defined. To fully encompass the 70--230~MeV therapeutic energy range (i.e. equivalent to 4.8--33~cm water equivalent thickness), a momentum acceptance range of $\pm$30\% is required (or -49\% to +62\% in the energy domain, given a reference energy of 140~MeV). This is dramatically larger than the standard range and designing a system with this capability has proven challenging (discussed in Section \ref{Section_LEA-systems}). Alternatively, some proposals split up the acceptance range into multiple `treatment bands', delivering a smaller subset of energies such that the BDS only needs to vary if treatments require switching between these bands. It is noted that not all energies across the entire clinical range are consistently used for treatment, rather it is selective on the treatment indication and patient plan \cite{Sengbusch2009,Beltran2024}. Also incorporating proton imaging -- provided the accelerator can produce suitable particle types at a high enough energy -- would increase the required acceptance range i.e. up to 330~MeV protons. In general, a LEA may reduce the ELST and BDT such that the entire field could be delivered at much faster timescales negating or minimising motion effects, but this has not yet been investigated in detail. 

\begin{figure}[htb!]
\centering
\includegraphics*[width=\textwidth]{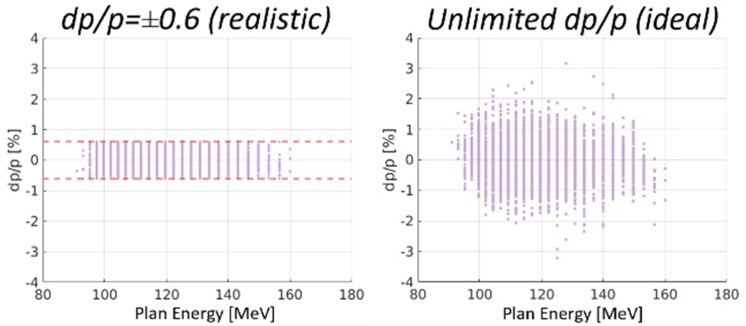}
\caption{Delivery optimised for a simulated 4DCT plan accounting for breathing motion during tumour tracking for a momentum acceptance range within a realistic treatment band (left) and without limitation (right), adapted from \cite{Giovannelli2023}.} 
\label{F_acceptancelimits}
\end{figure}

Implementing optimisation schemes which account for delivery with a LEA is a brand new paradigm: assessing treatment planning and dosimetric impact is only just starting to be explored in the literature but more must be done to understand the potential benefits and effects. A recent study by Wang et al. \cite{Wang2023} for a BDS with a -6\% to +7\% acceptance showed that delivery efficiency could be significantly improved without reducing plan quality and robustness, however there was a sensitivity to spot position uncertainty and concern about beam shape distortion which was not modelled. Another study by Giovannelli et al. \cite{Giovannelli2023} demonstrated that a momentum acceptance range of $\pm$5\% was sufficient to cover the variability and large organ motion for a set of lung patients when using rescanning and tumour tracking strategies (Figure \ref{F_acceptancelimits}). 

The number of energy layers is not typically a variable parameter in the optimisation process by commercial TPS \cite{Jensen2018,Zhu2023}. However, several studies have employed energy layer reduction strategies to optimise plans for efficiency, demonstrating negligible difference in plan quality. It was also noted in \cite{Cao2014} that IMPT optimisation is highly degenerate: different parameters and plans can yield equivalent dose distributions, and additional variables can offer multiple degrees of freedom with plan optimisation. In this study, the authors were able to decrease the number of proton energies and spots used whilst maintaining the dosimetric criteria. Greater benefits were shown for larger tumours and slower ELSTs; other strategies have also been developed to reduce layers whilst maintaining plan quality \cite{Lin2019,Gao2020,VanDeWater2015,Kang2008}. Ultimately, utilising a faster machine will yield greater efficiency gains than implementing planning optimisation methods, as there is a limit to the achievable reduction in BDT before it comes at the expense of treatment quality \cite{Muller2016}: the ELST will still remain a barrier to the BDT \cite{Lin2024}.

Further gains in efficiency can be realised by combining fast delivery with improvements to existing magnet and accelerator technologies, by increasing both scanning speeds and beam current. Particularly with the emergence of methodologies such as FLASH and arc therapy, a LEA BDS may prove essential as an enabling technology given that ELSTs need to be fast enough for rapid energy modulation. For example for BP (or conformal) FLASH, the whole field needs to be delivered within $\sim$1~s \cite{Kang2022}, provided that the high dose rate can be sustained for the whole treatment volume, as necessary to elicit the FLASH effect. Transmission or `shoot through' beams are also being used however, they are not an optimal solution as these neglect advantages of the BP \cite{Schwarz2025}. Passive devices such as ridge filters \cite{Zhang2022} and 3D-printed modulators \cite{Hagmann2025} are also being explored to produce a conformal depth distribution. However, these must be verified for production accuracy and robustness \cite{Hotoiu2025} and require adjustment to the accelerator, BDS and nozzle to achieve increased transmission \cite{Nesteruk2021b}. The feasibility of a LEA BDS with a range shifter for BP FLASH delivery was also studied in \cite{Zeng2024,Zeng2025}, showing better normal tissue sparing. Nevertheless, there are still many existing challenges and developments needed even with current system capabilities to be able to deliver FLASH dose rates safely and accurately for treatment \cite{Farr2022}. These are not discussed in this review but more can be found in \cite{Mazal2021,Jolly2020,Esplen2020,Fenwick2024}. 

For particle arc therapy, flexibility in the complex delivery sequencing and ensuring robustness with good treatment efficiency necessitates a fast BDT: ELSTs are also a bottleneck \cite{Zhao2023,Engwall2022,Liu2020a,Gu2020}. Arc therapy is proposed to be delivered either statically by stacking discrete (step-and-shoot) fields -- potentially at multiple layers -- at each angle; or dynamically (beam stays on during the gantry rotation) with single energy layers per discrete direction \cite{Janson2024}. As there are still many technical barriers to dynamic delivery, static delivery has been investigated as a feasible pathway toward implementing arc therapy in the clinic. Treatment planning studies  performed in \cite{Fracchiolla2025} showed dosimetric benefits specifically for head and neck (H\&N) treatments and delivery times comparable to  multi-field optimisation plans, leading to world first patient treatments with static proton arc therapy (PAT). In a recent study in \cite{Cong2024}, the feasibility of PAT given both delivery modes and the fundamental limitations with a synchrotron accelerator were assessed. The authors demonstrated the possibility for higher quality plans but at the cost of significantly increased treatment times relative to conventional IMPT delivery -- even up to 131\% longer BDT for a chordoma case -- primarily constrained by their 2~s ELST. Additionally, gantry rotation speed is also a limitation \cite{Ding2016} and upright patient rotation has been suggested as a method of facilitating arc PAT \cite{Schreuder2022,Volz2024}. The renewed interest and development into upright chairs and gantry-less systems also offer another parameter space for improvements to treatment, delivery efficiency and the patient experience \cite{Underwood2025}. These opportunities and considerations are discussed further in \cite{Volz2022,Mazal2021}. There are still several challenges to be solved to be able to treat a wide range of sites efficiently with PAT. Research and development to understand the hardware and software complexities of arc delivery, alongside considerations of safety, workflow, quality assurance, treatment planning optimisation and design have been growing: more is outlined in \cite{Mein2024}. 

\subsection{Rescanning and Bidirectional Delivery}
\label{Section_rescanning}
Rescanning (repainting) is a technique commonly used to mitigate interplay, where the treatment volume is re-irradiated multiple times to ensure sufficient dosage and to average out motion effects which may cause `hot' or `cold' spots. A minimum number of rescans is required to achieve a more homogenous distribution however, rescanning is usually combined with other active motion management techniques to compensate for motion induced dose conformity losses \cite{Zhang2023}. Different rescanning strategies exist; layered rescanning (LR) is most typically used, where an IES is repeatedly irradiated until the spot objectives are fulfilled (spots are weighted and/or divided per number of rescans) before changing to the next energy layer. In volumetric rescanning (VR), the delivery of rescans is instead ordered by energy layers where each IES is irradiated once before moving to the consecutive layer, scanning across the complete target volume repeatedly. Rescanning can reduce the effects of motion but there is often a trade-off with treatment time. The the feasibility and effectiveness is dependent on several factors including key motion parameters (respiration phase, period and amplitude) and technical specifications such as scanning speeds, ELST and beam current \cite{Kang2017,Zhang2016}. Therefore the characteristics and performance of the BDS plays a major role: effective rescanning requires a system capable of precise, fast beam delivery, and quick energy changes \cite{Engelsman2013,Schatti2013a,Lambert2005,Han2019a}. 

Both rescanning methods have been examined in numerous studies demonstrating ranging benefits, dependent on patient requirements and machine characteristics. The time structure of delivery is significant, where deliberate interruptions such as random pauses or variation in the scan path can interfere with correlated motion between the beam and target \cite{Rietzel2010}. VR has further potential advantage, offering a greater number of scan paths \cite{Bert2011} to stochastically distribute the rescans over the breath cycle, rather than a limited subset as with LR \cite{Engwall2018}. Published results have shown similar plan improvements using both LR and VR \cite{Zhang2016}. A comparative study in liver patients found that LR was more sensitive to the starting phase of the breathing period but for an optimal number of rescans, there was little difference in dose homogeneity \cite{Bernatowicz2013}. For scenarios with more rescans, LR was largely preferred as VR became unviable due to slow ELSTs and extended treatment times. Although VR may have statistical advantages \cite{Schatti2013a}, its utility is reliant on BDS capability as ELSTs are prohibitive to both BDT and rescanning efficacy \cite{Zenklusen2010,Schatti2014b}.

\begin{figure}[htb!]
\centering
\includegraphics*[width=\textwidth]{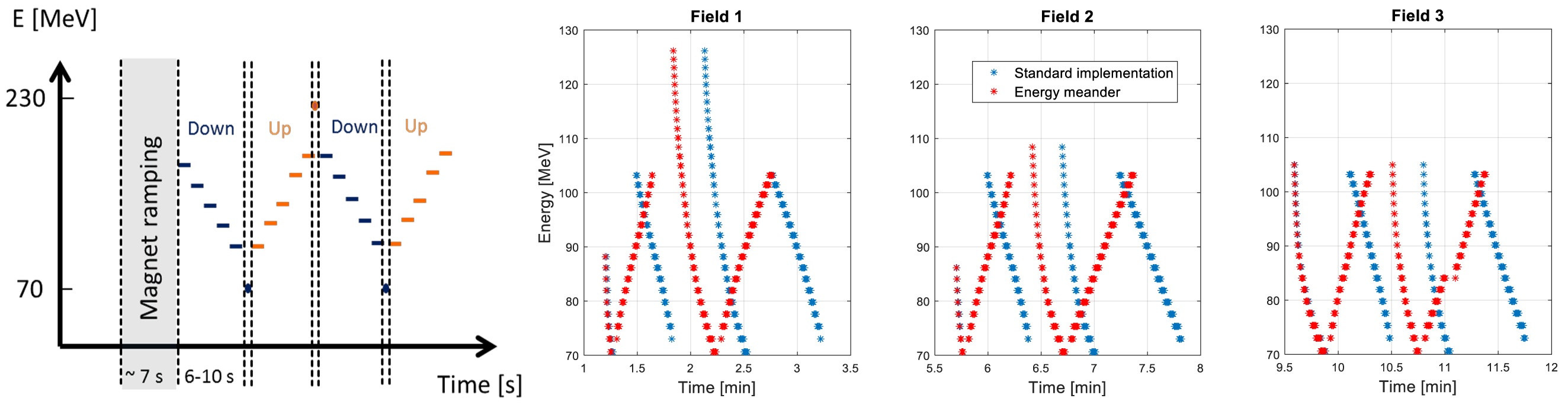}
\caption{Energy meandering scheme showing ramping profile for both `up' and `down' sides of the hysteresis loop (see also: Figure \ref{F_Hysteresis}), adapted from \cite{Actis2018}. Energy and time characteristics of 3 field clinical plan using standard delivery (blue) and energy meandering (red) \cite{Actis2023}.}
\label{F_EMeandering}
\end{figure}

Slow ELSTs are especially challenging for VR, requiring reinitialisation of the magnets in the BDS before each rescan \cite{Actis2023}. To improve treatment efficiency, operational developments at PSI have shown the possibility of `energy meandering' (Figure \ref{F_EMeandering}) to allow delivery on both sides of the hysteresis loop to avoid beamline ramping delays \cite{Actis2018}. This would mean for VR, an additional field could be repainted (in the reverse loop direction) without waiting for another full ramping cycle. Energy meandering has already been implemented for treatments, reporting a reduction of up to 55\% in BDT. Although small differences in beam characteristics were measured for equivalent beam energies (of each ramping direction) in this study, QA checks verified outputs were within tolerance. Gamma pass criteria were met showing no dosimetric differences and deviations did not exceed 1-2\%, comparable to daily variations. Therefore, precise delivery is possible in both ascending or descending energy directions, enhancing standard ramping schemes as restricted by hysteresis effects. 

Similarly, `Field Regulation' is a feature for IBA systems to reduce the ELST in the ascending irradiation direction for VR \cite{Rana2020a}. Hall probes are mounted inside certain groups of magnets to perform real-time measurements providing reference field values (corresponding to the desired range/energy), which are used instead of the magnet current set points. There is some redundancy in these lookup table values as each current can correspond to more than one magnetic field value. The magnets must be ramped again if changing to the opposite direction, thus regulating layer changes by field measurements instead establishes a direct correlation \cite{Rana2020}. There are still considerations with measurement reliability and QA procedures before it becomes clinically adopted however, there is incentive for implementation given ELSTs for the ascending energy direction is an order of magnitude greater (from $\sim$700~ms to 5.5~s, reported for a IBA S2C2 synchrocyclotron) \cite{Zhao2022}.

Energy meandering and field regulation may offer as alternative strategies to speed up delivery in existing systems but there is still a maximum limit to how fast you can regulate or change magnetic fields: these constraints would not be necessary for a LEA system. Ideally, the advantages of a LEA BDS could be fully exploited if combined with an accelerator capable of rapid beam energy variation. A linac and Fixed Field Accelerator (FFA) have repetition rates ranging from 100's Hz to kHz, meaning pulse-by-pulse changes could achieve ELSTs an order of magnitude faster than currently possible (anything faster than $\sim$1~ms would defer to the timing resolution of dose monitoring \cite{Giordanengo2018} and control system interlocks \cite{Schoemers2015}). The Linac for the Image Guided Hadron Therapy (LIGHT) system was previously under development, using novel high frequency RF accelerating structures for a commercial PBT solution. Construction and testing of the initial prototype was carried out, demonstrating fast energy modulation with repetition rates of 100~Hz (i.e. 10~ms ELST) \cite{Degiovanni2024}. Several FFAs have been constructed for various applications: for CPT, a conceptual design study (PAMELA) was proposed with a goal of 1~kHz repetition rates to enable ms ELSTs \cite{Peach2013}. These and other technologies (discussed in Section \ref{Section_FutureDev}) are still under development but may be commercially available in future, as the field evolves to capitalise on the benefits of faster treatment delivery.

The potential of a LEA BDS as an enabling technology has not yet been explored in detail and another possibility is longitudinal, bidirectional scanning. With a LEA, there would be minimal difference to the BDS in changing the direction of depth scanning: delivery would not need to follow the conventional sequential approach of maximum to minimum energy layer. Comparable to using the ascending energy mode in VR, bidirectional delivery could be especially helpful for beam (or tumour) tracking for mobile organs, as well as reduce BDT for future techniques such as PAT \cite{Zhao2022}. The option to go either direction -- whilst at any point during the treatment -- also offers additional parameters to optimise for treatment planning. Given this is a new paradigm, very little currently exists in literature and one factor with LEA delivery which may need to be considered is the division of dose across multiple layers. At faster timescales IES delivery may occur effectively simultaneously, in non-discrete layers -- a related concept is explored as oblique raster scanning in \cite{Amaldi2019}. For some facilities which use continuous PBS (rather than discrete spot scanning) some fraction of dose outside the prescribed dose may be delivered between spots. This occurs transiting between spots or from a time lag between the system controls to turn off the beam (flap dose) \cite{Tsubouchi2023}, however can be considered negligible in practice \cite{Tsubouchi2024}. In any case, the intra-layer dose must be evaluated accurately and rapid dose delivery must be able to be controlled and effectively monitored to ensure patient safety.

Evidently, the capability of the BDS determines how fast a treatment can be delivered: minimising the ELST can reduce treatment times and motion induced dose degradation effects, whilst also improving the effectiveness of motion management techniques, and the range of sites PBT could be used to treat. Ultimately, the optimal treatment is one where adequate coverage, conformity and robustness can be achieved, to meet patient requirements. Future delivery techniques are under development and have the possibility to revolutionise the capabilities of PBT \cite{Kang2023}, but these can only be made possible with advances in beam delivery technologies. The fundamental limitations of existing beam delivery systems can by addressed by increasing the acceptance range, minimising conventional ELST and enabling ultra-fast delivery.

\section{Technology Developments: Increasing Beamline Acceptance to Enhance Beam Delivery}
\label{Section_LEA-systems}

In principle, facility designs aim to fulfil several operational requirements for treatment, where the accelerator and BDS must be able to reliably provide precise (longitudinal and transverse) beam control and sufficient beam intensity with good overall up-time \cite{Schippers2015}. The beam intensity, emittance and momentum spread are important parameters as the machine must be able to efficiently accelerate particles to consistently generate a reproducible and stable beam with a small spread to reduce losses. The BDS design and optics must match and accommodate beams depending on the accelerator and delivery requirements. In the past, the choice of the accelerator was restrictive due to the available magnet and accelerator technology which determined several factors including size, cost, maximum energy, temporal beam structure, delivery characteristics, and achievable treatment quality \cite{Fukumoto1995,Schippers2011a}. Significant technological developments have since enabled modernisation to PBS delivery regardless of accelerator type.

A beam with a large energy spread (or momentum spread) is not ideal for treatment as this determines the modulated dose gradients -- the sharpness of the beam -- which can impact the uniformity and achievable conformity of the resulting dose distribution (Figure \ref{F_SOBP}). When degraders are used to introduce energy losses, it results in particles with a spread of energies. It is useful to design a degrader which reduces scatter and therefore spread, as this increased beam distribution results in lower transmission and beam intensities, particularly at lower energies. A significant proportion of the primary beam is lost at the ESS which includes dipoles, collimators and an adjustable slit which control the energy spread and emittance: only particles at a dispersive point corresponding to the nominal beam energy are transported. A small acceptance in momentum (i.e. 0.5\%) allows the ESS to restrict particles to within a nominal energy window to retain a sharp, steep distal BP fall-off for individual energy layers \cite{Schippers2015}. The sections of beam transport downstream of the ESS may have an equivalent acceptance range for achromaticity: matching this requirement seeks to maximise transmission and also ensures accurate longitudinal positioning \cite{Schippers2009,Anferov2007}. Lower transmission through the degrader and ESS can result in more unwanted radiation being produced upstream, which may require greater shielding and therefore costs \cite{Myers2019}. The output beam intensity can be increased by reducing the occurrence of beam losses at any point within the BDS, as lost particles can also lead to the production of neutrons and excess radiation and damage, or cause quenching of magnetic components \cite{Gerbershagen2017}. 

Historically, passive delivery techniques needed a beam with a fixed maximum energy, producing single fields practically instantaneously through range modulating devices, scatterers and absorbers positioned far downstream. Matching beamline properties was advantageous to maximise transmission and made it easier to use a single library of devices \cite{Anferov2007}. Similarly for synchrotrons, matching conditions supported beam stability and transmission, to produce beams with suitable currents and parameters before modification by passive devices \cite{Coutrakon1994}. Nevertheless, now to deliver PBS with these existing BDS capabilities, the small matched acceptance conditions have become a limiting factor, introducing an ELST each time a new nominal energy is selected and transported for treatment. As long as the beam is defined with the desired characteristics prior to transport through the BDS -- which could be designed to be able to accept all particles in range -- the momentum acceptance range of the system does not need to be prohibitive. The benefit of a BDS with larger acceptance is not to deliver beams with a large energy spread, where the potential detriment to treatment quality is clear: rather it enables the delivery of beams of a large range of energies. Therefore, increasing the acceptance range can address the design constraints of existing beamlines and decrease time spent waiting for system adjustment.

\subsection{Large Energy Acceptance Arcs}

Beam transport and delivery systems for PBT have several boundary conditions. The main requirement for the delivered beam is to have an approximately circular spot shape with minimal distortion, that is well defined in both position and momentum. In cases where the beamline momentum acceptance is small, with magnets or collimators to ensure correct matching, it can be assumed that this spot is energy-independent in the absence of errors. When considering a large energy acceptance beamline, there are additional constraints on both the accelerator design. In order from most to least important, these are:

\begin{enumerate} 
    \item Trajectories for all momenta converge at the start and end of the beamline (geometric achromaticity).
    \item All momenta have the same beam parameters at the end of the beamline (optical achromaticity).
    \item Minimal momentum-dependent displacement in the dispersive plane during transport (small excursion).
\end{enumerate}

In accelerator literature, a beamline or insertion is usually regarded as an `achromat' if the beam trajectory is energy-independent at the start and end of the section i.e. the beam dispersion and its derivative are zero. Here, we extend this to make a clear distinction between `geometric achromaticity' -- matching the definition previously given -- and `optical achromaticity', meaning that the beam shape and size is also energy-independent (i.e. zero chromaticity in the optical functions of the lattice). 

Geometric achromaticity is of paramount importance for an LEA BDS, as any residual dispersion at the nozzle would be directly detrimental to treatment quality. Optical achromaticity is also desirable, as this ensures consistency in beam spot shape and size, though perfect optical achromaticity may be impractical over the full range of treatment energies. Finally, though the size of the beam excursions has no impact on the treatment quality, large excursions necessitate large magnet apertures which would increase the cost of a PBT facility, and may prevent retrofitting in existing facilities.

\subsubsection{Accelerator Considerations for LEA Beamlines}

In most accelerators, optical achromaticity is fulfilled sufficiently well by ramping the strength of focusing magnets to match the beam energy, and any residual chromaticity is countered by the addition of sextupole and other higher order magnets. This is not necessarily the case in LEA arcs, which usually do not include magnet ramping. Here, we define a `large energy acceptance' arc as one comprising a geometric achromat over the full range of energies, as this is the minimum requirement for rapid beam delivery, with optical achromaticity a desirable but not necessary condition.

Many of the design requirements of LEA beamlines are the same as those for a conventional BDS. For example, the horizontal and vertical beam parameters at synchrotron-based facilities must be matched at some stage in the BDS, which can be achieved in conventional systems using scattering foils with collimators or matching quadrupoles \cite{Pavlovic2024}. There is significant additional complexity introduced in a LEA system as particle-matter and particle-magnet interactions are strongly dependent on the beam energy. For example, the focusing strength of a given magnet is inversely proportional to the beam momentum meaning that optical achromaticity is impossible without either rapid ramping magnets or by using energy-dependent trajectories to cancel out the differences in focusing. As such, though the challenges for conventional and LEA beam delivery systems are similar, the solutions are different.

In conventional beam delivery systems, the strength of all magnets must be ramped synchronously with the beam energy to ensure proper beam steering and focusing. This magnet ramping bottleneck is circumvented in LEA arcs by fixing the field of some or all magnets in the beamline, at the expense of energy-dependent particle trajectories and beam dynamics. Though Fixed Field Accelerator (FFA) facilities have been proposed several times for charged particle therapy accelerators \cite{Keil2007, Peach2013, Garland2015, Taylor2018, Meot2019, Aymar2020}, none have yet been realised in the clinic; the same is also true for FFA beam delivery systems.

\subsubsection{LEA Proposals for Charged Particle Therapy}
\label{Section_LEAproposals}

There are several proposed LEA beamline designs for CPT, relying on different techniques to achieve achromaticity. The mechanisms used to achieve the large energy acceptance have varied over time with the discovery of new FFA techniques, ranging from linear magnets with a high packing factor to highly nonlinear arcs. The key details of notable proposals are given in Table \ref{tab:proposal_details}. Here we define the acceptance $\Delta_p$ of a beamline as

\begin{equation}
    \Delta_p = \pm \frac{p_\text{max}-p_\text{min}}{p_\text{max}+p_\text{min}},
\end{equation}

where $p$ is the momentum of a particle beam and the acceptance is given by a single number \footnote{Note that, for accelerators where only one charge state is transported, the momentum and rigidity acceptances are identical.}. The rigidity acceptance is the critical parameter when considering a LEA beamline, as there is direct proportionality between the beam rigidity and the necessary magnetic field strength.

\begin{table}[htbp]
\centering
\caption{Details of proposed large momentum acceptance beamlines for CPT. The final two columns detail whether the arc comprises a geometric and optical achromat.}
\label{tab:proposal_details}
\begin{tabular}{|l|l|l|l|l|l|}
\hline
Source                                & Principle                    & Rig. Range     & KE      & Geom. & Optic. \\
                                      &                              & [Tm] ($\pm$\%) & [MeV]   &       &        \\ \hline
Keil et al. (2007)                    & Linear FODO with matching    & 1.40-2.43      & 90-250  & No    & No     \\
\cite{Keil2007}               &                              & (26.8)         &         &       &        \\ \hline
Fenning (2011)                        & Nearly-scaling FFA           & 1.23-2.43      & 70-250  & Yes   & No     \\
\cite{Fenning2011}             & dispersion suppression       & (32.8)         &         &       &        \\ \hline
Wan et al. (2015)                     & AG-CCT dipole/quadrupole     & 1.4-1.8        & 90-144  & No    & No     \\
 \cite{Wan2015} & (3 settings)                 & (12.5)         &         &       &        \\ \hline
Brouwer et al.                        & Large rectangular bends with & 1.23-2.27      & 70-220  & Yes  & Yes\tablefootnote{Optical achromaticity is achieved using fast ramping quadrupole magnets rather than with a `fixed field' solution.}     \\
(2019) \cite{Brouwer2019} & variable matching quads      & (30)           &         &       &        \\ \hline
Nesteruk (2019)                       & Combined function up to      & 1.70-2.32      & 130-230 & Yes   & No     \\
\cite{Nesteruk2019a}             & sextupole                    & (15.5)         &         &       &        \\ \hline
GaToroid (2020)                       & Superconducting toroidal     & 1.23-2.43      & 70-250  & Yes   & No     \\
\cite{Bottura2020a}          & magnetic field               & (32.8)         &         &       &        \\ \hline
Trbojevic (2021)                      & Edge angles         & 1.19-2.43      & 65-250  & Yes   & No     \\
\cite{Trbojevic2021}                                       &                              & (34.4)         &         &       &        \\ \hline
Dascalu and Sheehy                    & Adiabatic transition between & 1.23-2.32      & 70-230  & Yes   & No     \\
(2021) \cite{Dascalu2021}       & arcs and straights           & (30.7)         &         &       &        \\ \hline
Liao et al. (2024)                           & AG-CCTs and CF quad/sext     & 1.69-1.99      & 129-172 & Yes   & No     \\
\cite{Liao2024}                                      & magnets                      & (8)            &         &       &        \\ \hline
Steinberg et al. (2024)               & Combined function FFA        & 0.1-0.25       & 0.5-3.0 & Yes   & Yes    \\
\cite{Steinberg2024a}                 & demonstrator beamline        & (42)           &         &       &       \\ \hline
\end{tabular}
\end{table}

The first BDS designed to transport all treatment energies without changing magnet excitation was proposed by Keil, Sessler, and Trbojevic \cite{Keil2007} in 2007. The `KST' gantry, designed as part of a wider facility for CPT, comprises many repeating cells of combined function dipole/quadrupole magnets. An insertion of two quadrupoles and drifts is used to connect the beam transfer line to the gantry, matching a circular beam spot to the rotating gantry. The gantry is made up of three quarter-circles with a radius of 5.27~m, giving an overall footprint of approximately 16.5$\times$8.5~m$^2$. The magnet packing factor for the gantry is not given but the drifts appear appear much shorter than the magnets. As the KST design uses one periodic cell with linear optics, only the reference energy can be properly matched: as details for other energies are not given, presumably they are mismatched both geometrically and optically. Given these caveats, the KST lattice design still demonstrates the potential of a LEA BDS, establishing a good benchmark for further work.

Fenning and Machida \cite{Fenning2011} propose an alternate design methodology, using `scaling' FFA optics (where the phase advance as a function of energy is fixed using nonlinear magnetic fields) to create a dispersion suppressor that brings all trajectories together, to first order. This is then optimised to bring together all the closed-orbit positions at the end of the arc, with both zero dispersion and its derivative. In this work, several alternate gantry designs are presented with magnets assumed to be similar to those in the PAMELA study \cite{Peach2013}. It is found in all cases that sufficient dispersion suppression can be achieved, however the beta functions are severely mismatched leading to an energy-dependent spot size. It is suggested that this beam size variation can be countered as part of the scanning system, but detailed studies have not been performed.

Another design strategy is presented by Dascalu and Sheehy \cite{Dascalu2021}, using connected `adiabatic transitions' wherein the bending strengths of combined-function magnets are gradually increased/decreased over several magnets along the arc. This slow variation should ensure that the beta functions are not perturbed but the dispersion should smoothly come back to zero. Using adiabatic transitions leads to a large footprint -- approximately the same as the KST lattice -- and requires a very high packing factor, with 58 cells in total required to ensure good matching. In spite of the large number of magnets, the results indicate that the residual dispersion at the end of the arc is almost as much as in the main part of the arc itself, though the orbits converge well. With such a large number of magnets in close proximity, it is not clear that an adiabatic transition would be suitable for a CPT facility.

Trbojevic et al. \cite{Trbojevic2021} provides another technique to achieve geometric achromaticity while maintaining linear accelerator optics, making use of magnet edge angles to bring all trajectories back together. With this method, it is not possible to produce an optical achromat, as the small path length variations do not compensate for the differences in focusing from the quadrupoles. In addition, the lattice packing factor is just as high as for an adiabatic transition. The authors suggest that permanent magnet `Halbach' arrays can be used to provide the high field gradients necessary to ensure that beam excursions remain small, producing gradients in excess of 150~T/m, though this would preclude reductions to the facility footprint by adopting SC magnet technologies. Though this arc design technique shows some promise, detailed studies on the impacts of realistic magnet fringe fields and errors have not been performed, making it difficult to assess the feasibility of such a beamline under realistic conditions.

A 2019 design study from LBNL by Brouwer, Huggins, and Wan \cite{Brouwer2019} achieves a 70--220~MeV proton kinetic energy range using fixed-field Nb-Ti superconducting magnets up to 3.5~T, designed with racetrack coils. A fixed-field SC dipole bends the full energy beam into the gantry, which then passes through a gantry-mounted energy degrader. A geometrically achromatic bending section consists of two straight dipole magnets totalling a 155 degree bend, achieving achromaticity in a similar manner to the end regions of a racetrack microtron. For optical achromaticity, three fast-ramping resistive quadrupoles are placed symmetrically on either side of the dipoles to provide optical matching. The strength of each of these three magnets must be independently varied to ensure consistent beam spot shape and size, which may become a bottleneck for energy layer switching. The authors claim the design could bring the estimated ELST down to around 100~ms, and that the design is a few metres smaller than competing options. 

It is possible to devise other approaches for beam delivery, that do not resemble a conventional BDS. One particular design of note is the GaToroid study from CERN \cite{Bottura2020a}, which proposes to use a large SC toroid (similar to the ATLAS detector magnet), with the patient couch located at the centre of the toroid. The proton beam is directed into the field of the toroid, where its trajectory curves in the magnetic field toward isocentre, allowing for a wide range of irradiation field angles without needing to rotate the magnet structure or patient. The design requires at least one fast switching X/Y steering `vector magnet' dipole to direct each beam energy onto the necessary trajectory in the toroidal field. It is reasonable to expect from first principles that the tolerance requirements on the rapid switching dipole would be extremely challenging. The scale of the required vacuum and cryocooling systems to cover all beam angles around a patient may also be prohibitive.

The gantry design by Nesteruk et al. \cite{Nesteruk2019} comprises two combined function dipole-quadrupole-sextupole magnets and one combined function quadrupole-sextupole magnet in between, where each combined function magnet uses a pair of SC racetrack coils. The gantry has a momentum acceptance of $\pm15\%$, though the nonlinear fields result in some beam distortion. The gantry is combined with an ultra-fast degrader mounted in the gantry and a 2D lateral scanning system. The analysis presented in the paper demonstrates that a range of $\pm$30\% in energy acceptance is expected to enable 70\% of treatments to be delivered with a single magnet setting. The gantry design followed an iterative process using different codes for high-order beam transport calculations and particle tracking in the magnetic fields. Though this design reached the magnet fabrication-ready phase, further work towards this BDS has not taken place.

Liao et al. \cite{Liao2024} propose a similar solution, offering a $\pm$10\% momentum acceptance that incorporates a movable energy slit in the middle of the achromatic bending section. The slit, together with variable-size collimators in the degrader system, allows controlling the momentum spread and beam size; their proposed delivery method is discussed later in Section \ref{Section_BeamVariations}. For lateral scanning, the distortions at large scanning angles due to dispersion effects are of a bigger concern. This effect is relevant for low energies where the energy spread after passing the degrader is greater than $\pm$1-2\%. The main disadvantage of this method is a suboptimal use of the ultra-fast energy switching, as moving the slit and adjusting the collimator size increases the total beam delivery time. Alternative methods could be considered such as treatment planning strategies or adding an adjustable field-specific aperture at the end of the nozzle.

At the University of Melbourne, the Technology for Ultra Rapid Beam Operation (TURBO) project seeks to de-risk the techniques required to realise a LEA arc via a scaled-down technology demonstrator beamline. The initial design by Steinberg et al. \cite{Steinberg2024a} proceeds along similar lines to Fenning's earlier work, beginning with a scaling FFA lattice before using a multi-objective genetic algorithm to reduce the residual dispersion whilst also maintaining an approximately constant beam size. By employing a multi-objective routine, this methodology enables optimisation of ancillary variables, allowing a reduction in the maximum excursion (i.e. a smaller beam pipe and magnet aperture) without degrading the beam quality. However, this beamline requires highly nonlinear optics, and optimising the multipole components of several magnets up to decapole order, leading to some distortion of the delivered beam spot. Further work aims to improve this result by imposing boundary conditions that ensure geometric achromaticity, and future studies are expected to include a pathway to scaling up the TURBO design to clinical energies.
   
Given these LEA arc designs, it is apparent that there are two main routes to providing the additional degrees of freedom necessary for a large energy acceptance: having a large number of magnets; or increasing the nonlinearity of the allowed magnetic fields. No design concepts have yet been realised, however several proposals are being further developed, in addition to innovation in magnet technologies.

\subsubsection{A Prospective View of Magnet Technologies} 
\label{Section_FutureDev}
Future accelerator and beam delivery systems will rely on advanced magnet technologies: this is true for both proton and heavy ion systems (most prominently helium and carbon). For protons, one innovation already mentioned is the use of permanent magnet Halbach arrays. A number of recent accelerator test systems have demonstrated the potential for nonlinear fields and field correction techniques \cite{Brooks2024}, while further radiation resiliency tests are underway \cite{Bodenstein2024}.

While much of our focus has been on protons, heavy ions are even more challenging to reach a future scenario of smaller systems and faster treatments, or a vision of single-room facilities. What is clear at present is that the costs and physical requirements for CIBT gantry facilities are higher than for protons. This drives two potential directions: either gantry-less treatments, which is the standard choice at most of the existing CIBT facilities \cite{PTCOGfacilities}, or the use of innovative superconducting magnet gantries. Whether the future will see upright treatments for CIBT is uncertain \cite{Chinniah2023}, but partial patient rotation may allow robust angle selection \cite{Zhou2021} while easing gantry requirements.

To first order, the challenge for heavier ions lies in the much higher required magnetic rigidity (over 3 times that of clinical protons) to reach depths for treatment. This means either the magnetic field strength or facility size is naturally larger than for protons. For example, the first carbon gantry was the Heidelberg Ion Therapy gantry \cite{Haberer2004} which is normal conducting, 25~m long, with a radius around 6.5~m and weight around 600~t. It is clear that SC technology can help reduce the gantry size: a SC carbon ion gantry was made by NIRS/QST and Toshiba \cite{Iwata2017} and is now in clinical use, weighing around 300~t, 13~m long and is 5.5~m in radius. This lighter gantry utilises novel 2.7~T SC magnets which are combined function and (relatively) fast ramping (tested up to 0.3~T/second), contained in helium-free cryocoolers to overcome issues arising from the physical movement of SC cryostats. Ongoing work aims to achieve a design reaching a 1~T/m ramp rate \cite{Yang2022}.

\begin{figure}[hb!]
\centering
\includegraphics*[width=\textwidth]{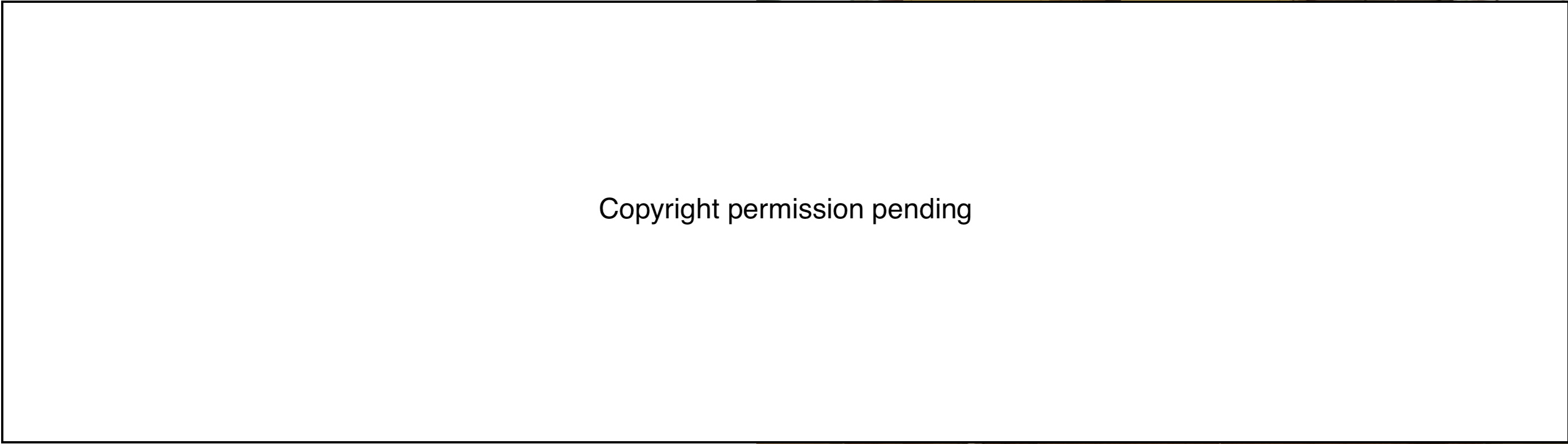}
\caption{Rendering of two curved CCT dipole layers (left) and photo of the fabricated mandrel assembly (right). Figures pending copyright permissions.} 
\label{F_CCT}
\end{figure}

In addition to direct-wound SC magnets (cosine-theta type), a possible configuration for SC magnets is Canted-Cosine-Theta (CCT) \cite{Brouwer2020}, which can produce pure dipole fields whilst cancelling unwanted solenoidal fields due to its winding geometry, over shorter lengths than conventional cosine-theta magnets \cite{Gupta2015,Meyer1970}. While CCTs have not yet been implemented in accelerators or gantries, other single-pass optics studies \cite{Wan2015,Pullia2024} have presented gantry designs based on curved, alternating gradient (AG)-CCT configurations, consisting of focusing-defocusing quadrupole layers inside CCT dipoles (Figure \ref{F_CCT}). As previously shown, Wan et al. report a momentum acceptance of $\pm$12.5\%, where an AG-CCT arrangement could achieve the strong focusing required to transport the wide range of rigidities without fast magnet ramping during treatment. The bore radius of the magnet must also have a good field region (min 10$\times$10~$\text{cm}^{2}$) to accommodate a suitable treatment field size at the patient isocentre \cite{Robin2011}. The exact design parameters for the CCT magnet will depend on the optical design of the BDS to determine what higher order field gradients are tolerable.  

With new technologies come new opportunities that are currently beyond current clinical realisation: a LEA BDS as an alternative to a gantry for carbon or multi-ions would decrease the ELST, potentially allow multi-ion transport, remove the need for fast-ramping magnets, and reduce the complexity of the cooling configurations required for rotation. The full energy ranges of He-4 and C-12 from 70~MeV/u to 430~MeV/u requires $\pm$46\% momentum acceptance, and to extend this to include protons\footnote{Note that proton CT which requires a maximum energy of 330~MeV \cite{Holder2014}, is easily covered by this acceptance.} becomes $\pm$69\% \cite{Yap2023}. By achieving the full range of carbon ions, this also offers the potential of using helium ions for in-vivo imaging \cite{Volz2020,Hardt2024} alongside C-12 as mixed beams for treatment and online imaging \cite{Mazzucconi2018}.

\section{Considerations for Clinical Implementation}
While each LEA proposal aims to increase the energy acceptance using various design techniques and technologies, it is also important to consider the different aspects which may need to either be addressed in conceptual stage, reactively in the clinic, or in reality be inconsequential and thus manageable with standard operation. Nevertheless, a LEA system will offer capabilities to deliver novel PBT treatments which have not yet been explored by currently existing systems. In practice, this may result in beam properties different to what is normally `standard' for a PBT facility where some flexibility may be expected with ideal beam parameters and standard machine operation \cite{Goma2024}. Although clinics experience day-to-day variations with accelerator operation and beam performance, it is important that this can still be accommodated with a LEA BDS whilst maintaining high performance and beam quality within the recommended clinical tolerances. 

An ideal LEA BDS design would be able to be retrofitted universally, without dependence on vendor specific components or complex adjustment to the existing system it joins. However, there may be unavoidable uncertainties and predicted variability with beam parameters, compared to conventional systems. We examine some aspects fundamental to clinical performance such as beam size and shape, discuss how these could be managed in practice, and the relevant considerations for implementation.

\subsection{Beam Quality} 
\label{Section_BeamVariations}
\begin{figure}[htb!]
\centering
\includegraphics*[width=\textwidth]{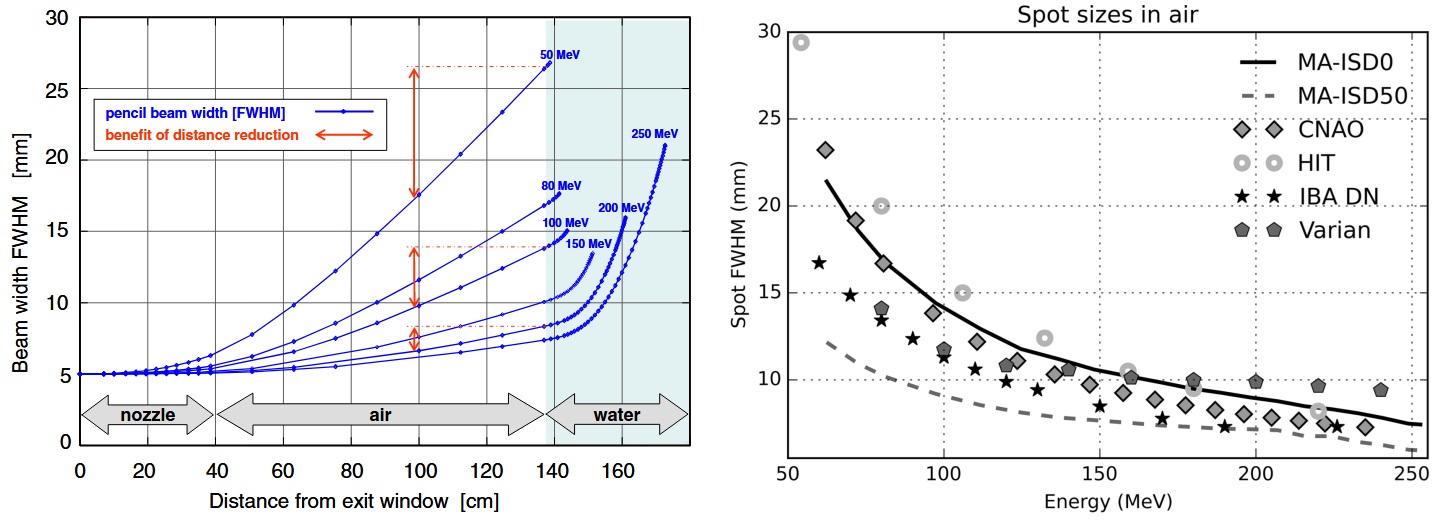} 
\caption{Calculated increases in beam size using the same nozzle geometry for various proton energies (left), also showing the effect of reducing the drift distance in air by 40~cm (red) \cite{Jelen2013}. Measured spot sizes (FWHM) at isocentre in air, at different PBT facilities (right), reproduced with permission \cite{Grevillot2020}.} 
\label{F_Spotvariation}
\end{figure}

A challenge for most LEA designs (Section \ref{Section_LEAproposals}) is the beam distortion caused by nonlinear and higher order effects due to the magnetic field requirements, resulting in non-circular, asymmetric beam spots. The distortion may be significant and have an energy dependence, as spot size and shape vary as a function of gantry angle \cite{Li2013,Lin2016}, and energy \cite{Grevillot2020}. This may also cause further complication due to amplification downstream at large transverse scanning angles \cite{Wang2023}. In addition to field effects from the various magnets through the transport line, the beam can also be altered during extraction (mechanism, septa or scrapers) \cite{Parodi2010}, by other components in the BDS such as beam instrumentation and vacuum chamber windows \cite{Titt2010}, the nozzle (configuration or type e.g. universal or dedicated) \cite{Chen2016}, the presence of patient specific devices, collimators \cite{Dowdell2012}, range shifters \cite{Romero-Exposito2023} or boluses \cite{Both2014}, and depending on other downstream factors such as the air gap and distance to isocentre \cite{Jelen2013,Grevillot2020,Kim2018} (Figure \ref{F_Spotvariation}). 

Although the beam properties are defined by the BDS optics, the contributions from external factors are likely to have a large impact on the beam spot. For example, exposure to air and scattering within the nozzle can `wash' out any shape irregularities toward a Gaussian distribution \cite{Pivi2024} where beam shape distortion may become insignificant when measured at isocentre. However, the extent will vary depending on characteristics such as the initial beam energy, size and shape. The BDS may also need further tuning to achieve ideal performance at isocentre: to produce a symmetric beam and spot size as small as possible at all energies \cite{Grevillot2020}.

\subsubsection{Spot Variation}
A small beam at every energy is ideal, given a small penumbra and therefore sharp lateral fall-off can reduce the integral dose \cite{Wang2014a}. This is challenging at lower energies as small pencil beams are particularly prone to multiple coulomb scattering (MCS) \cite{Newhauser2015} which contribute to the low dose `tail' regions, and can dominate the lateral dose distribution \cite{Lin2013a,Lin2013b}. This can cause beam broadening and penumbra growth \cite{Linz2012} where larger spot sizes can result in reduced conformity \cite{Jelen2013,Bues2005}, and lower treatment quality \cite{Moteabbed2016,Rana2021}. This increased scatter and beam size is particularly consequential for shallower targets (Figure \ref{F_Spotvariation}) as the increased lateral penumbra and dose fall-off may reduce tissue sparing at low energies \cite{Grewal2021a}. At higher energies, MCS is less prevalent and nuclear interactions dominate the spot size: an initial small beam with a sharp penumbra and fall-off is also ideal to minimise the spread which occurs within the patient itself \cite{Ko2024}. However, this can also result in dose inhomogeneities \cite{Guo2025} (hot or cold spots) and is more sensitive to motion \cite{Grassberger2013}. 

Penumbra growth within the patient may also be reduced with collimation \cite{Safai2008,Bues2005,Lomax2018} and patient specific apertures \cite{Dowdell2012,Yasui2015}. As multileaf collimators (MLCs) are common in XRT, collimators were previously explored for PBT \cite{Daartz2009a,Ko2024}; also with passive (wobbling) systems in the earlier days of PBS \cite{Torikoshi2007}. Improvements in conformity or tissue sparing were reported particularly for low energy beams \cite{Hyer2014}, fixed beamlines \cite{Baumer2021} and irradiation of cranial \cite{Winterhalter2018a,Kim2018,Moignier2016} and H\&N tumours \cite{Moignier2016a}. The use of any beam modification devices must also consider the consequences of additional contaminant scatter \cite{Smith2025a,Ueno2019}, secondary production, activation, neutrons \cite{VanGoethem2009,Grewal2021}, increased surface, physical, and out-of-field dose \cite{Hopfensperger2023}. Currently for PBT collimation, Mevion offers an adaptive aperture, a mini-MLC with a small air gap \cite{Kang2020a,Vilches-Freixas2020}, and in-development is a dynamic collimator system consisting of two pairs of orthogonal nickel blades \cite{Nelson2023,Hyer2021}. Both of these systems offer energy layer specific collimation however the sequencing and translation of these devices also introduces added time penalties \cite{Smith2019,Engwall2025}. A concept using a thin metal block with a cylindrical aperture attached to a robotic arm has also been proposed to trim individual proton beamlets \cite{Holmes2022}. The Mevion S250-FIT utilises range shifters for fast energy changes (Section \ref{Section_Minimising_ELST}) with AA but an additional 5--10~s for delivery of each field has been reported \cite{Maradia2025}. A BDS which implements any auxiliary devices requires that these supplementary components operate in synchronicity, sufficiently fast and with positional accuracy: presently, interventional methods of beam shaping by dynamic collimation is disadvantageous to the overall BDT.

Similarly with beam size, spot positions can also vary with energy. As described in Section \ref{Section_rescanning}, the control system uses a lookup table of parameters which correspond to accelerator, beamline and scanning magnet settings to ensure accurate delivery of each spot and maintain beam positioning at isocentre. A LEA BDS which does not need to rely successive magnetic field changes for each IES on may offer better stability in this respect\footnote{also for fixed-line systems where there are no gantry dependence errors such as beam deviations or offsets caused by mechanical strain, optics mismatches, or asymmetries at the gantry coupling point \cite{Gerbershagen2016b}.} however will need to manage any additional dispersive effects \cite{Tan2023}. A comprehensive QA program will be critical to ensure spot parameters remain within the recommended AAPM TG-224 tolerances ($\pm$10\% and $\pm$1mm, respectively) \cite{AAPM-TG224_2019}. 

\subsection{Delivery Strategies}
Although beam control and spot variability still poses as a challenge for LEA delivery, the possibility of rapid interchange between beams with variable parameters and different planning strategies may offer advantages for treatment. 

\begin{figure}[htb!]
\centering
\includegraphics*[width=0.8\textwidth]{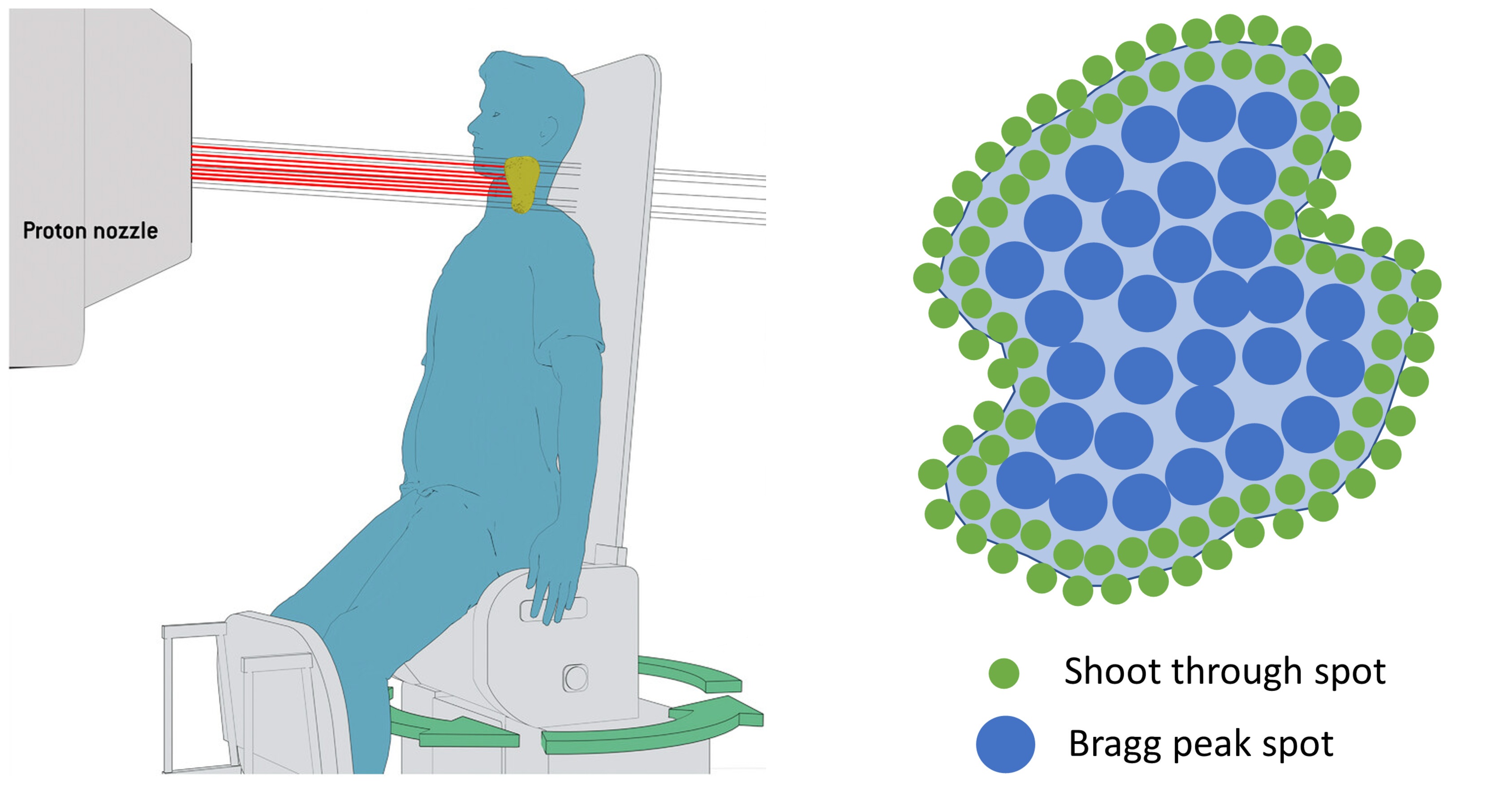}
\caption{Upright static arc delivery with TBs (left), adapted from \cite{Engwall2025}. Illustration showing the arrangement of transmission (`shoot through') beams to irradiate small spots at the target boundary with larger, conventional BP spots for the remainder of the volume (right) \cite{Penfold2024}.}
\label{F_mixedspotdelivery}
\end{figure}

To circumvent energy dependence and spot variation with their LEA BDS, Wang et al. \cite{Wang2024} proposed delivering mixed spot sizes for prostate cases by using adjustable collimators to produce fields with large spots centrally, surrounded by small spots at the periphery (Figure \ref{F_mixedspotdelivery}). A set of collimators and an energy slit upstream of the nozzle restricts the beam to $\pm$0.1-0.5\% to produce a small spot and are opened to return to the `natural' larger momentum spread, transporting larger spots. Small spots are used for the two outermost contours of the target boundary for each energy layer, and for the entire last energy layer, to obtain a sharper distal fall-off. The collimation addresses distortion effects due to dispersion at large scanning angles however adjusting the slit and collimator size compromises the delivery speed: this is a significant disadvantage of this mixed-size method. In practice, trimming the distal dose fall-off may also be counterproductive, as too sharp a dose gradient increases the sensitivity to machine range uncertainties and can result in hot spots or non-uniformities \cite{Rana2024}. Spot size has a different impact depending on disease site and there are several technical trade-offs with quality, speed and robustness \cite{Titt2010}. 

Alternatively, the use of transmission beams (TB) have been explored as an approach to instead capitalise on the steep distal fall-off, by improving the lateral penumbra and tissue sparing. Van Marlen et al. \cite{VanMarlen2021} initially demonstrated the applicability of multi-field TB plans (for FLASH delivery), where Kong et al. \cite{Kong2024} showed the potential for H\&N treatments by combining IMPT fields with 244~MeV TBs. Penfold et al. \cite{Penfold2024} examined the use of a TB (at max 226~MeV) in a single arc to shape the target boundary with IMPT fields for the interior volume. Engwall et al. \cite{Engwall2025} proposed delivering static arcs with TB for upright treatments (Figure \ref{F_mixedspotdelivery}), reporting high quality plans and improved efficiency over delivery slowed by collimators or ELSTs. Hytönen et al. \cite{Hytonen2025} demonstrated that hybrid IMPT and TB fields could significantly reduce BDT without compromising target coverage for breast and H\&N treatments. Similarly, Maradia et al. \cite{Maradia2025} proposed combined fields with an optimised scanning routine to further reduce BDT. 

Additional considerations with radiation protection and beam angle selection for TBs are needed, as the proton range may not exceed the patient. In general, the use of TBs for PBT may add potential improvements to organ at risk sparing and with a LEA BDS, could be delivered instantaneously alongside conventional BP layers. 

\subsection{Energy Modulation}
Given existing technologies, the most practical method to rapidly change beam energies is through the use of an energy degrader. The design of the degrader determines many beam properties, including: choice of material (lower Z generally results in less scattering and energy spread), geometry (i.e. wedge, wheel or blocks) \cite{Oponowicz2020}, and location in the beamline. It must be positioned upstream or configured such that there is sufficient shielding and minimal exposure to the patient, and also be robust enough to move quickly with low uncertainty of positioning error. As the degrader increases the angular divergence, energy spread, and emittance, typically an ESS or multiple downstream collimators are required to restrict the beam to a nominal, narrow distribution. However, this results in losses to particle intensity which is significant ($>$99\%) at low energies \cite{Schippers2009}. 

The absence of an ESS for a LEA BDS -- redundant in this case given restrictive magnet ramping speeds -- allows a much higher overall transmission however will also transport particles outside of conventional acceptance limits. This requires the design of the beamline optics and components configured for an optimal balance between transmission, emittance and energy spread\footnote{A note here to avoid confusion and reiterate: an advantage of a LEA BDS is the ability to deliver beams across a large range of energies rather than beams with a large energy spread.}. Slits or collimators may be necessary to ensure appropriate beam selection and shaping, but also introduce further practical and physics complexities. Another approach for energy modulation is to employ multiple wedges with different lengths and materials, grouping energy ranges into multiple bands to improve beam properties and optimise delivery efficiency \cite{Liao2024}. 

Given the range of possible errors due to the complex coordination of different beamline elements, rapid delivery requires that all system components to cooperate and communicate together quickly and reliably. The control system must be capable of maintaining effective beamline operation and fast synchronism with all ancillary components (degrader, slits, scanning magnets etc.). Ultra-fast energy switching may be on the order of 10~ms: the system and any interlocks must be able to respond instantaneously, interrupting the beam if there are any detected mismatches with beam parameters. A study performed at PSI investigated the feasibility of rapid and continuous energy modulation, significantly decreasing their ELST by 45\% to 27~ms (for small energy changes) \cite{Giovannelli2021}. The authors developed a strategy to exploit the full beamline momentum acceptance (typical few percent dp/p) by parametrising energy bands for range compensation, creating new beamline settings to allow continuous energy regulation. They were able to preserve clinical beam quality but suggest further development to translate experimental performance to clinical settings. For example, calibration of the scanning magnets with the new band settings and optimising the degrader motion and latency can further minimise dead time. 

Therefore, there are numerous sources of beam variation in a LEA BDS: additional to nonlinear field effects, the nozzle and auxiliary devices can have a non-trivial impact on the output beam characteristics. This could result in distinct and enduring beam characteristics, where inconsistencies in spot size or shape can lead to dose errors and deviation from planned dose calculations. This requires further study where beam irregularities need to be realistically implemented in the TPS, rather than standard beam models which typically assume a circular spot \cite{Chen2016}, or a Gaussian distribution in phase space \cite{Janson2024}. It is possible that some of these features will have a significant or an inconsequential effect on treatment: in practice this may mean any departure from standard operation could actually be negligible as long as variations are clinically tolerated. Nonetheless, the BDS must be able to deliver beams which are reproducible, accurate and effective for all clinical treatments.  

\section{Summary}
The role of PBT as a primary modality of cancer treatment continues to mature, as improvements in technology and costs drive the rapid expansion of facilities worldwide. The dosimetric and therapeutic benefits are well established, however even if parity with XRT is achieved in terms of accessibility and costs, several limitations still need to be addressed to fully exploit the advantages of PBT. As the future evolves toward emerging methodologies -- arc, upright, mixed beams or new particle types etc. -- improvements in the technological capabilities of existing BDS still need to be overcome: not being able to deliver a treatment sufficiently fast is a fundamental constraint, and a key obstacle to enabling superior and advanced treatments. 

Increasing the energy acceptance range of conventional beamlines enables rapid energy modulation, ultra-fast treatments, and a multitude of possibilities including volumetric rescanning, bidirectional delivery, PAT, and BP FLASH. A LEA BDS could significantly reduce treatment times and costs, whilst enhancing the treatment quality and utility of PBT. Ultra-fast delivery can immediately benefit current treatment by increasing delivery efficiency, throughput and enabling more effective use of modern motion management techniques. Further gains can also be achieved with novel delivery approaches and planning optimisation strategies, but the greatest advantages can only be realised with improvements to the BDS. 

Several PBT beamline designs with an increased rigidity acceptance range have been proposed and are under development, however none have yet been constructed. These will require further innovation in magnet technology, degrader and collimation systems, and the integration of these components in novel conditions. Many considerations also still remain with beam quality and clinical performance, warranting further study to model, and evaluate the potential impact on treatment quality. We have examined the existing literature to provide critical discussion in these areas, identifying opportunities and challenges for clinical implementation. A LEA BDS offers improved capabilities for PBT with ultra-fast beam delivery and beyond, enabling many exciting possibilities for current patient needs and future facilities. 

\section*{Conflict of Interest Statement}
The authors declare that the research was conducted in the absence of any commercial or financial relationships that could be construed as a potential conflict of interest.

\section*{Author Contributions}
JY: literature review, manuscript preparation, writing, editing and figures. AFS: literature review, writing, editing and figures. KN, HN, SS: writing, review and editing. All authors contributed to the manuscript and approved the submitted version.

\section*{Funding}
This work was supported by the Gross Foundation, Laby Foundation, Manchester Global Doctoral Research Network, EPSRC, and NIH NCI R37CA2883437.

\section*{Acknowledgments}
The authors thank and acknowledge support from ANSTO and the National Collaborative Research Infrastructure Strategy (NCRIS). 

\bibliographystyle{frontiersinHLTH_FPHY} 
\bibliography{MyLibrary}

\begin{thebibliography}{238}
\expandafter\ifx\csname natexlab\endcsname\relax\def\natexlab#1{#1}\fi
\expandafter\ifx\csname urlstyle\endcsname\relax
  \expandafter\ifx\csname doi\endcsname\relax
  \def\doi#1{doi:\discretionary{}{}{}#1}\fi \else
  \expandafter\ifx\csname doi\endcsname\relax
  \def\doi{doi:\discretionary{}{}{}\begingroup \urlstyle{rm}\Url}\fi \fi
\expandafter\ifx\csname selectlanguage\endcsname\relax
  \def\selectlanguage#1{}\fi

\bibitem[{Durante et~al.(2021)Durante, Debus, and Loeffler}]{Durante2021}
Durante M, Debus J, Loeffler JS.
\newblock Physics and biomedical challenges.
\newblock {\em Nature Reviews Physics\/} {\bf 0123456789} (2021).
\newblock \doi{10.1038/s42254-021-00368-5}.

\bibitem[{Delaney et~al.(2005)Delaney, Jacob, Featherstone, and Barton}]{Delaney2005}
Delaney G, Jacob S, Featherstone C, Barton M.
\newblock The role of radiotherapy in cancer treatment: {{Estimating}} optimal utilization from a review of evidence-based clinical guidelines.
\newblock {\em Cancer\/} {\bf 104} (2005) 1129--1137.
\newblock \doi{10.1002/cncr.21324}.

\bibitem[{Barton et~al.(2014)Barton, Jacob, Shafiq, Wong, Thompson, Hanna et~al.}]{Barton2014}
Barton MB, Jacob S, Shafiq J, Wong K, Thompson SR, Hanna TP, et~al.
\newblock Estimating the demand for radiotherapy from the evidence: {{A}} review of changes from 2003 to 2012.
\newblock {\em Radiotherapy and Oncology\/} {\bf 112} (2014) 140--144.
\newblock \doi{10.1016/j.radonc.2014.03.024}.

\bibitem[{Yan et~al.(2023)Yan, Ngoma, Ngwa, and Bortfeld}]{Yan2023}
Yan S, Ngoma TA, Ngwa W, Bortfeld TR.
\newblock Global democratisation of proton radiotherapy.
\newblock {\em The Lancet Oncology\/} {\bf 24} (2023) e245--e254.
\newblock \doi{10.1016/S1470-2045(23)00184-5}.

\bibitem[{{Particle Therapy Co-Operative Group (PTCOG)}(2025)}]{PTCOGfacilities}
[Dataset] {Particle Therapy Co-Operative Group (PTCOG)}.
\newblock Particle therapy facilities.
\newblock https://www.ptcog.site/index.php/facilities-in-operation-public (2025).

\bibitem[{{Abdel-Wahab} et~al.(2024){Abdel-Wahab}, Giammarile, Carrara, Paez, Hricak, Ayati et~al.}]{Abdel-Wahab2024}
{Abdel-Wahab} M, Giammarile F, Carrara M, Paez D, Hricak H, Ayati N, et~al.
\newblock Radiotherapy and theranostics: A {{Lancet Oncology Commission}}.
\newblock {\em The Lancet Oncology\/} {\bf 25} (2024) e545--e580.
\newblock \doi{10.1016/S1470-2045(24)00407-8}.

\bibitem[{Bortfeld and Loeffler(2017)}]{Bortfeld2017}
Bortfeld TR, Loeffler JS.
\newblock Three ways to make proton therapy affordable.
\newblock {\em Nature\/} {\bf 549} (2017) 451--453.
\newblock \doi{10.1038/549451a}.

\bibitem[{Pidikiti et~al.(2018)Pidikiti, Patel, Maynard, Dugas, Syh, Sahoo et~al.}]{Pidikiti2018}
Pidikiti R, Patel BC, Maynard MR, Dugas JP, Syh J, Sahoo N, et~al.
\newblock Commissioning of the world's first compact pencil-beam scanning proton therapy system.
\newblock {\em Journal of Applied Clinical Medical Physics\/} {\bf 19} (2018) 94--105.
\newblock \doi{10.1002/acm2.12225}.

\bibitem[{Rossi(2022)}]{Rossi2022}
Rossi S.
\newblock Hadron {{Therapy Achievements}} and {{Challenges}}: {{The CNAO Experience}}.
\newblock {\em Physics\/} {\bf 4} (2022) 229--257.
\newblock \doi{10.3390/physics4010017}.

\bibitem[{Feldman et~al.(2024)Feldman, Pryanichnikov, Achkienasi, Polyansky, Hillman, Raskin et~al.}]{Feldman2024}
Feldman J, Pryanichnikov A, Achkienasi A, Polyansky I, Hillman Y, Raskin S, et~al.
\newblock Commissioning of a novel gantry-less proton therapy system.
\newblock {\em Frontiers in Oncology\/}  (2024).

\bibitem[{Balakin et~al.(2021)Balakin, Pryanichnikov, Alexandrov, Bazhan, Belikhin, Chashurin et~al.}]{Balakin2021}
Balakin V, Pryanichnikov A, Alexandrov V, Bazhan A, Belikhin M, Chashurin V, et~al.
\newblock Updated {{Status}} of {{Protom Synchrotrons}} for {{Radiation Therapy}}.
\newblock {\em Proceedings of the 27th {{Russian Particle Accelerator Conference}}\/} (2021), vol. RuPAC2021, 4 pages, 0.468 MB.
\newblock \doi{10.18429/JACOW-RUPAC2021-FRB05}.

\bibitem[{Kerstiens et~al.(2018)Kerstiens, Johnstone, and Johnstone}]{Kerstiens2018a}
Kerstiens J, Johnstone GP, Johnstone PA.
\newblock Proton {{Facility Economics}}: {{Single-Room Centers}}.
\newblock {\em Journal of the American College of Radiology\/} {\bf 15} (2018) 1704--1708.
\newblock \doi{10.1016/j.jacr.2018.07.020}.

\bibitem[{{Canadian Agency for Drugs and Technologies in Health (CADTH)}(2017)}]{CADTH_145}
{Canadian Agency for Drugs and Technologies in Health (CADTH)}.
\newblock Proton {{Beam Therapy}} for the {{Treatment}} of {{Cancer}} in {{Children}} and {{Adults}}: {{A Health Technology Assessment}}.
\newblock {{CADTH HEALTH TECHNOLOGY ASSESSMENT}} no.145, {Canadian Agency for Drugs and Technologies in Health} (2017).

\bibitem[{Xia et~al.(2022)Xia, Wang, Xia, Wang, and Cheng}]{Xia2022}
Xia Z, Wang J, Xia J, Wang M, Cheng Z.
\newblock Inequality in {{Accessibility}} of {{Proton Therapy}} for {{Cancers}} and {{Its Economic Determinants}}: {{A Cross-Sectional Study}}.
\newblock {\em Frontiers in Oncology\/} {\bf 12} (2022).
\newblock \doi{10.3389/fonc.2022.876368}.

\bibitem[{Bortfeld et~al.(2020)Bortfeld, de~Viana, and Yan}]{Bortfeld2020}
Bortfeld TR, de~Viana MF, Yan S.
\newblock The societal impact of ion beam therapy.
\newblock {\em Zeitschrift fur Medizinische Physik\/}  (2020) 6--8.
\newblock \doi{10.1016/j.zemedi.2020.06.007}.

\bibitem[{Clasie et~al.(2022)Clasie, Bortfeld, Kooy, Seybolt, Sharp, and Winey}]{Clasie2022}
Clasie BM, Bortfeld TR, Kooy HM, Seybolt K, Sharp GC, Winey B.
\newblock Retrofitting {{LINAC Vaults}} for {{Compact Proton Systems}}---{{Experiences Learned}}.
\newblock {\em IEEE Transactions on Radiation and Plasma Medical Sciences\/} {\bf 6} (2022) 282--287.
\newblock \doi{10.1109/TRPMS.2021.3083982}.

\bibitem[{Yap et~al.(2021)Yap, De~Franco, and Sheehy}]{Yap2021}
Yap J, De~Franco A, Sheehy S.
\newblock Future {{Developments}} in {{Charged Particle Therapy}}: {{Improving Beam Delivery}} for {{Efficiency}} and {{Efficacy}}.
\newblock {\em Frontiers in Oncology\/} {\bf 11} (2021) 1--25.
\newblock \doi{10.3389/fonc.2021.780025}.

\bibitem[{Paganetti et~al.(2021)Paganetti, Beltran, Both, Dong, Flanz, Furutani et~al.}]{Paganetti2021}
Paganetti H, Beltran C, Both S, Dong L, Flanz J, Furutani K, et~al.
\newblock Roadmap: Proton therapy physics and biology.
\newblock {\em Physics in Medicine \& Biology\/} {\bf 66} (2021) 05RM01.
\newblock \doi{10.1088/1361-6560/abcd16}.

\bibitem[{Schreuder and Shamblin(2020)}]{Schreuder2020}
Schreuder AN, Shamblin J.
\newblock Proton therapy delivery : What is needed in the next ten years ?
\newblock {\em British Journal of Radiology\/} {\bf 2019} (2020).

\bibitem[{Myers et~al.(2019)Myers, Degiovanni, and Farr}]{Myers2019}
Myers S, Degiovanni A, Farr JB.
\newblock Future {{Prospects}} for {{Particle Therapy Accelerators}}.
\newblock {\em Reviews of Accelerator Science and Technology\/} {\bf 10} (2019) 49--92.
\newblock \doi{10.1142/s1793626819300056}.

\bibitem[{Vidal et~al.(2021)Vidal, Moignier, Patriarca, Sotiropoulos, Schneider, and Marzi}]{Vidal2021}
Vidal M, Moignier C, Patriarca A, Sotiropoulos M, Schneider T, Marzi LD.
\newblock Future technological developments in proton therapy - {{A}} predicted technological breakthrough.
\newblock {\em Cancer / Radiotherapie\/}  (2021).
\newblock \doi{10.1016/j.canrad.2021.06.017}.

\bibitem[{Flanz(2017)}]{Flanz2017}
Flanz J.
\newblock ({{The}}) future (of) synchrotrons for particle therapy.
\newblock {\em CERN Yellow Reports: School Proceedings\/} {\bf 1} (2017) 293--307.
\newblock \doi{10.23730/CYRSP-2017-001.293}.

\bibitem[{Mohan and Grosshans(2017)}]{Mohan2017}
Mohan R, Grosshans D.
\newblock Proton therapy -- {{Present}} and future.
\newblock {\em Advanced Drug Delivery Reviews\/} {\bf 109} (2017) 26--44.
\newblock \doi{10.1016/j.addr.2016.11.006}.

\bibitem[{Farr et~al.(2018)Farr, Flanz, Gerbershagen, and Moyers}]{Farr2018}
Farr JB, Flanz JB, Gerbershagen A, Moyers MF.
\newblock New horizons in particle therapy systems.
\newblock {\em Medical Physics\/} {\bf 45} (2018) e953--e983.
\newblock \doi{10.1002/mp.13193}.

\bibitem[{Collings et~al.(2022)Collings, Lu, Gupta, and Sumption}]{Collings2022}
Collings EW, Lu L, Gupta N, Sumption MD.
\newblock Accelerators, {{Gantries}}, {{Magnets}} and {{Imaging Systems}} for {{Particle Beam Therapy}}: {{Recent Status}} and {{Prospects}} for {{Improvement}}.
\newblock {\em Frontiers in Oncology\/} {\bf 11} (2022) 1--20.
\newblock \doi{10.3389/fonc.2021.737837}.

\bibitem[{Graeff et~al.(2023)Graeff, Volz, and Durante}]{Graeff2023}
Graeff C, Volz L, Durante M.
\newblock Emerging technologies for cancer therapy using accelerated particles.
\newblock {\em Progress in Particle and Nuclear Physics\/} {\bf 131} (2023) 104046.
\newblock \doi{10.1016/j.ppnp.2023.104046}.

\bibitem[{Mohan(2022)}]{Mohan2022}
Mohan R.
\newblock A review of proton therapy -- {{Current}} status and future directions.
\newblock {\em Precision Radiation Oncology\/} {\bf 6} (2022) 164--176.
\newblock \doi{10.1002/pro6.1149}.

\bibitem[{Zenklusen et~al.(2010)Zenklusen, Pedroni, and Meer}]{Zenklusen2010}
Zenklusen SM, Pedroni E, Meer D.
\newblock A study on repainting strategies for treating moderately moving targets with proton pencil beam scanning at the new gantry 2 at {{PSI}}.
\newblock {\em Physics in Medicine and Biology\/} {\bf 55} (2010) 5103--5121.
\newblock \doi{10.1088/0031-9155/55/17/014}.

\bibitem[{Giordanengo et~al.(2017)Giordanengo, Manganaro, and Vignati}]{Giordanengo2017}
Giordanengo S, Manganaro L, Vignati A.
\newblock Review of technologies and procedures of clinical dosimetry for scanned ion beam radiotherapy.
\newblock {\em Physica Medica\/} {\bf 43} (2017) 79--99.
\newblock \doi{10.1016/j.ejmp.2017.10.013}.

\bibitem[{{van de Water} et~al.(2012){van de Water}, Lomax, Bijl, Schilstra, Hug, and Langendijk}]{VanDeWater2012}
{van de Water} TA, Lomax AJ, Bijl HP, Schilstra C, Hug EB, Langendijk JA.
\newblock Using a {{Reduced Spot Size}} for {{Intensity-Modulated Proton Therapy Potentially Improves Salivary Gland-Sparing}} in {{Oropharyngeal Cancer}}.
\newblock {\em International Journal of Radiation Oncology*Biology*Physics\/} {\bf 82} (2012) e313--e319.
\newblock \doi{10.1016/j.ijrobp.2011.05.005}.

\bibitem[{Kraan et~al.(2018)Kraan, Depauw, Clasie, Giunta, Madden, and Kooy}]{Kraan2018}
Kraan AC, Depauw N, Clasie B, Giunta M, Madden T, Kooy HM.
\newblock Effects of spot parameters in pencil beam scanning treatment planning.
\newblock {\em Medical Physics\/} {\bf 45} (2018) 60--73.
\newblock \doi{10.1002/mp.12675}.

\bibitem[{Liu et~al.(2024)Liu, Kolano, Gray, Stephans, Videtic, Farr et~al.}]{Liu2024}
Liu CW, Kolano AM, Gray T, Stephans KL, Videtic GMM, Farr JB, et~al.
\newblock Cyclotron and linear accelerator generated scanning proton beams for lung cancer {{SBRT}}: {{Interplay}} effects and mitigations.
\newblock {\em Medical Physics\/}  (2024) mp.17082.
\newblock \doi{10.1002/mp.17082}.

\bibitem[{Alshaikhi et~al.(2019)Alshaikhi, Doolan, D'Souza, Holloway, Amos, and Royle}]{Alshaikhi2019}
Alshaikhi J, Doolan PJ, D'Souza D, Holloway SM, Amos RA, Royle G.
\newblock Impact of varying planning parameters on proton pencil beam scanning dose distributions in four commercial treatment planning systems.
\newblock {\em Medical Physics\/} {\bf 46} (2019) 1150--1162.
\newblock \doi{10.1002/mp.13382}.

\bibitem[{Hsi et~al.(2009)Hsi, Moyers, Nichiporov, Anferov, Wolanski, Allgower et~al.}]{Hsi2009}
Hsi WC, Moyers MF, Nichiporov D, Anferov V, Wolanski M, Allgower CE, et~al.
\newblock Energy spectrum control for modulated proton beams.
\newblock {\em Medical Physics\/} {\bf 36} (2009) 2297--2308.
\newblock \doi{10.1118/1.3132422}.

\bibitem[{Van~Goethem et~al.(2009)Van~Goethem, Van Der~Meer, Reist, and Schippers}]{VanGoethem2009}
Van~Goethem MJ, Van Der~Meer R, Reist HW, Schippers JM.
\newblock Geant4 simulations of proton beam transport through a carbon or beryllium degrader and following a beam line.
\newblock {\em Physics in Medicine and Biology\/} {\bf 54} (2009) 5831--5846.
\newblock \doi{10.1088/0031-9155/54/19/011}.

\bibitem[{Pakela et~al.(2022)Pakela, Knopf, Dong, Rucinski, and Zou}]{Pakela2022}
Pakela JM, Knopf A, Dong L, Rucinski A, Zou W.
\newblock Management of {{Motion}} and {{Anatomical Variations}} in {{Charged Particle Therapy}}: {{Past}}, {{Present}}, and {{Into}} the {{Future}}.
\newblock {\em Frontiers in Oncology\/} {\bf 12} (2022) 1--16.
\newblock \doi{10.3389/fonc.2022.806153}.

\bibitem[{Schaub et~al.(2020)Schaub, Harrabi, and Debus}]{Schaub2020}
Schaub L, Harrabi SB, Debus J.
\newblock Particle therapy in the future of precision therapy.
\newblock {\em The British journal of radiology\/} {\bf 93} (2020) 20200183.
\newblock \doi{10.1259/bjr.20200183}.

\bibitem[{Suzuki et~al.(2016)Suzuki, Palmer, Sahoo, Zhang, Poenisch, MacKin et~al.}]{Suzuki2016}
Suzuki K, Palmer MB, Sahoo N, Zhang X, Poenisch F, MacKin DS, et~al.
\newblock Quantitative analysis of treatment process time and throughput capacity for spot scanning proton therapy.
\newblock {\em Medical Physics\/} {\bf 43} (2016) 3975--3986.
\newblock \doi{10.1118/1.4952731}.

\bibitem[{Shen et~al.(2017)Shen, Tryggestad, Younkin, Keole, Furutani, Kang et~al.}]{Shen2017}
Shen J, Tryggestad E, Younkin JE, Keole SR, Furutani KM, Kang Y, et~al.
\newblock Technical {{Note}}: {{Using}} experimentally determined proton spot scanning timing parameters to accurately model beam delivery time.
\newblock {\em Medical Physics\/} {\bf 44} (2017) 5081--5088.
\newblock \doi{10.1002/mp.12504}.

\bibitem[{Mah et~al.(2020)Mah, Chen, Nawaz, Galbreath, Shmulenson, Lee et~al.}]{Mah2020}
Mah D, Chen CC, Nawaz AO, Galbreath G, Shmulenson R, Lee N, et~al.
\newblock Retrospective analysis of reduced energy switching and room switching times on throughput efficiency of a multi-room proton therapy center.
\newblock {\em British Journal of Radiology\/} {\bf 93} (2020).
\newblock \doi{10.1259/bjr.20190820}.

\bibitem[{Pedroni et~al.(2011)Pedroni, Meer, Bula, Safai, and Zenklusen}]{Pedroni2011}
Pedroni E, Meer D, Bula C, Safai S, Zenklusen S.
\newblock Pencil beam characteristics of the next-generation proton scanning gantry of {{PSI}}: {{Design}} issues and initial commissioning results.
\newblock {\em European Physical Journal Plus\/} {\bf 126} (2011) 1--27.
\newblock \doi{10.1140/epjp/i2011-11066-0}.

\bibitem[{Zhu et~al.(2023)Zhu, Flampouri, Stanforth, Slopsema, Diamond, LePain et~al.}]{Zhu2023}
Zhu M, Flampouri S, Stanforth A, Slopsema R, Diamond Z, LePain W, et~al.
\newblock Effect of the initial energy layer and spot placement parameters on {{IMPT}} delivery efficiency and plan quality.
\newblock {\em Journal of Applied Clinical Medical Physics\/}  (2023) e13997.
\newblock \doi{10.1002/acm2.13997}.

\bibitem[{Mizushima et~al.(2017)Mizushima, Furukawa, Iwata, Hara, Saotome, Saraya et~al.}]{Mizushima2017}
Mizushima K, Furukawa T, Iwata Y, Hara Y, Saotome N, Saraya Y, et~al.
\newblock Performance of the {{HIMAC}} beam control system using multiple-energy synchrotron operation.
\newblock {\em Nuclear Instruments and Methods in Physics Research, Section B: Beam Interactions with Materials and Atoms\/} {\bf 406} (2017) 347--351.
\newblock \doi{10.1016/j.nimb.2017.03.051}.

\bibitem[{Younkin et~al.(2018)Younkin, Bues, Sio, Liu, Ding, Keole et~al.}]{Younkin2018}
Younkin JE, Bues M, Sio TT, Liu W, Ding X, Keole SR, et~al.
\newblock Multiple energy extraction reduces beam delivery time for a synchrotron-based proton spot-scanning system.
\newblock {\em Advances in Radiation Oncology\/} {\bf 3} (2018) 412--420.
\newblock \doi{10.1016/j.adro.2018.02.006}.

\bibitem[{Lebbink et~al.(2022)Lebbink, Stock, Georg, and Kn{\"a}usl}]{Lebbink2022}
Lebbink F, Stock M, Georg D, Kn{\"a}usl B.
\newblock The {{Influence}} of {{Motion}} on the {{Delivery Accuracy When Comparing Actively Scanned Carbon Ions}} versus {{Protons}} at a {{Synchrotron-Based Radiotherapy Facility}}.
\newblock {\em Cancers\/} {\bf 14} (2022) 1--14.

\bibitem[{Zhao et~al.(2022)Zhao, Liu, Chen, Shen, Zheng, Qin et~al.}]{Zhao2022}
Zhao L, Liu G, Chen S, Shen J, Zheng W, Qin A, et~al.
\newblock Developing an accurate model of spot-scanning treatment delivery time and sequence for a compact superconducting synchrocyclotron proton therapy system.
\newblock {\em Radiation Oncology\/} {\bf 17} (2022) 87.
\newblock \doi{10.1186/s13014-022-02055-w}.

\bibitem[{{Vilches-Freixas} et~al.(2020){Vilches-Freixas}, Unipan, Rinaldi, Martens, Roijen, Almeida et~al.}]{Vilches-Freixas2020}
{Vilches-Freixas} G, Unipan M, Rinaldi I, Martens J, Roijen E, Almeida IP, et~al.
\newblock Beam commissioning of the first compact proton therapy system with spot scanning and dynamic field collimation.
\newblock {\em The British journal of radiology\/} {\bf 93} (2020) 20190598.
\newblock \doi{10.1259/bjr.20190598}.

\bibitem[{De~Franco et~al.(2021)De~Franco, Schmitzer, Gambino, Glatzl, Myalski, and Pivi}]{DeFranco2021}
De~Franco A, Schmitzer C, Gambino N, Glatzl T, Myalski S, Pivi M.
\newblock Optimization of synchrotron based ion beam therapy facilities for treatment time reduction, options and the {{MedAustron}} development roadmap.
\newblock {\em Physica Medica\/} {\bf 81} (2021) 264--272.
\newblock \doi{10.1016/j.ejmp.2020.11.029}.

\bibitem[{Iwata et~al.(2010)Iwata, Kadowaki, Uchiyama, Fujimoto, Takada, Shirai et~al.}]{Iwata2010}
Iwata Y, Kadowaki T, Uchiyama H, Fujimoto T, Takada E, Shirai T, et~al.
\newblock Multiple-energy operation with extended flattops at {{HIMAC}}.
\newblock {\em Nuclear Instruments and Methods in Physics Research, Section A: Accelerators, Spectrometers, Detectors and Associated Equipment\/} {\bf 624} (2010) 33--38.
\newblock \doi{10.1016/j.nima.2010.09.016}.

\bibitem[{Noda et~al.(2017)Noda, Furukawa, Fujimoto, Hara, Inaniwa, Iwata et~al.}]{Noda2017}
Noda K, Furukawa T, Fujimoto T, Hara Y, Inaniwa T, Iwata Y, et~al.
\newblock Recent progress and future plans of heavy-ion cancer radiotherapy with {{HIMAC}}.
\newblock {\em Nuclear Instruments and Methods in Physics Research, Section B: Beam Interactions with Materials and Atoms\/} {\bf 406} (2017) 374--378.
\newblock \doi{10.1016/j.nimb.2017.04.021}.

\bibitem[{Simeonov et~al.(2017)Simeonov, Weber, Penchev, Ringb{\ae}k, Schuy, Brons et~al.}]{Simeonov2017}
Simeonov Y, Weber U, Penchev P, Ringb{\ae}k TP, Schuy C, Brons S, et~al.
\newblock {{3D}} range-modulator for scanned particle therapy: {{Development}}, {{Monte Carlo}} simulations and experimental evaluation.
\newblock {\em Physics in Medicine and Biology\/} {\bf 62} (2017) 7075--7096.
\newblock \doi{10.1088/1361-6560/aa81f4}.

\bibitem[{Weber and Kraft(1999)}]{Weber1999}
Weber U, Kraft G.
\newblock Design and construction of a ripple filter for a smoothed depth dose distribution in conformal particle therapy.
\newblock {\em Physics in Medicine and Biology\/} {\bf 44} (1999) 2765--2775.
\newblock \doi{10.1088/0031-9155/44/11/306}.

\bibitem[{Gerbershagen et~al.(2016{\natexlab{a}})Gerbershagen, Calzolaio, Meer, Sanfilippo, and Schippers}]{Gerbershagen2016}
Gerbershagen A, Calzolaio C, Meer D, Sanfilippo S, Schippers M.
\newblock The advantages and challenges of superconducting magnets in particle therapy.
\newblock {\em Superconductor Science and Technology\/} {\bf 29} (2016{\natexlab{a}}).
\newblock \doi{10.1088/0953-2048/29/8/083001}.

\bibitem[{Psoroulas et~al.(2018)Psoroulas, Bula, Actis, Weber, and Meer}]{Psoroulas2018}
Psoroulas S, Bula C, Actis O, Weber DC, Meer D.
\newblock A predictive algorithm for spot position corrections after fast energy switching in proton pencil beam scanning.
\newblock {\em Medical Physics\/} {\bf 45} (2018) 4806--4815.
\newblock \doi{10.1002/mp.13217}.

\bibitem[{Chaudhri et~al.(2010)Chaudhri, Saito, Bert, Franczak, Steidl, Durante et~al.}]{Chaudhri2010}
Chaudhri N, Saito N, Bert C, Franczak B, Steidl P, Durante M, et~al.
\newblock Ion-optical studies for a range adaptation method in ion beam therapy using a static wedge degrader combined with magnetic beam deflection.
\newblock {\em Physics in Medicine and Biology\/} {\bf 55} (2010) 3499--3513.
\newblock \doi{10.1088/0031-9155/55/12/015}.

\bibitem[{Fattori et~al.(2020)Fattori, Zhang, Meer, Weber, Lomax, and Safai}]{Fattori2020}
Fattori G, Zhang Y, Meer D, Weber DC, Lomax AJ, Safai S.
\newblock The potential of {{Gantry}} beamline large momentum acceptance for real time tumour tracking in pencil beam scanning proton therapy.
\newblock {\em Scientific Reports\/} {\bf 10} (2020) 1--13.
\newblock \doi{10.1038/s41598-020-71821-1}.

\bibitem[{Nesteruk et~al.(2019{\natexlab{a}})Nesteruk, Calzolaio, Meer, Rizzoglio, Seidel, and Schippers}]{Nesteruk2019}
Nesteruk KP, Calzolaio C, Meer D, Rizzoglio V, Seidel M, Schippers JM.
\newblock Large energy acceptance gantry for proton therapy utilizing superconducting technology.
\newblock {\em Physics in Medicine and Biology\/} {\bf 64} (2019{\natexlab{a}}).
\newblock \doi{10.1088/1361-6560/ab2f5f}.

\bibitem[{Kang and Pang(2020)}]{Kang2020a}
Kang M, Pang D.
\newblock Commissioning and beam characterization of the first gantry-mounted accelerator pencil beam scanning proton system.
\newblock {\em Medical Physics\/} {\bf 47} (2020) 3496--3510.
\newblock \doi{10.1002/mp.13972}.

\bibitem[{Silvus et~al.(2024)Silvus, Haefner, Altman, Zhao, Perkins, and Zhang}]{Silvus2024}
Silvus A, Haefner J, Altman MB, Zhao T, Perkins S, Zhang T.
\newblock Dosimetric evaluation of dose shaping by adaptive aperture and its impact on plan quality.
\newblock {\em Medical Dosimetry\/} {\bf 49} (2024) 30--36.
\newblock \doi{10.1016/j.meddos.2023.10.011}.

\bibitem[{Grewal et~al.(2021{\natexlab{a}})Grewal, Ahmad, and Jin}]{Grewal2021}
Grewal HS, Ahmad S, Jin H.
\newblock Performance evaluation of adaptive aperture's static and dynamic collimation in a compact pencil beam scanning proton therapy system: {{A}} dosimetric comparison study for multiple disease sites.
\newblock {\em Medical Dosimetry\/} {\bf 46} (2021{\natexlab{a}}) 179--187.
\newblock \doi{10.1016/j.meddos.2020.11.001}.

\bibitem[{B{\"a}umer et~al.(2021)B{\"a}umer, Plaude, Khalil, Geismar, Kramer, Kr{\"o}ninger et~al.}]{Baumer2021}
B{\"a}umer C, Plaude S, Khalil DA, Geismar D, Kramer PH, Kr{\"o}ninger K, et~al.
\newblock Clinical {{Implementation}} of {{Proton Therapy Using Pencil-Beam Scanning Delivery Combined With Static Apertures}}.
\newblock {\em Frontiers in Oncology\/} {\bf 11} (2021) 1--11.
\newblock \doi{10.3389/fonc.2021.599018}.

\bibitem[{Grewal et~al.(2021{\natexlab{b}})Grewal, Ahmad, and Jin}]{Grewal2021a}
Grewal HS, Ahmad S, Jin H.
\newblock Characterization of penumbra sharpening and scattering by adaptive aperture for a compact pencil beam scanning proton therapy system.
\newblock {\em Medical Physics\/} {\bf 48} (2021{\natexlab{b}}) 1508--1519.
\newblock \doi{10.1002/mp.14771}.

\bibitem[{Schippers and Seidel(2015)}]{Schippers2015}
Schippers JM, Seidel M.
\newblock Operational and design aspects of accelerators for medical applications.
\newblock {\em Physical Review Special Topics - Accelerators and Beams\/} {\bf 18} (2015) 1--7.
\newblock \doi{10.1103/PhysRevSTAB.18.034801}.

\bibitem[{Gerbershagen et~al.(2016{\natexlab{b}})Gerbershagen, Meer, Schippers, and Seidel}]{Gerbershagen2016b}
Gerbershagen A, Meer D, Schippers JM, Seidel M.
\newblock A novel beam optics concept in a particle therapy gantry utilizing the advantages of superconducting magnets.
\newblock {\em Zeitschrift fur Medizinische Physik\/} {\bf 26} (2016{\natexlab{b}}) 224--237.
\newblock \doi{10.1016/j.zemedi.2016.03.006}.

\bibitem[{Giovannelli et~al.(2021)Giovannelli, Maradia, Meer, Safai, Psoroulas, Togno et~al.}]{Giovannelli2021}
Giovannelli AC, Maradia V, Meer D, Safai S, Psoroulas S, Togno M, et~al.
\newblock Beam properties within the momentum acceptance of a clinical gantry beamline for proton therapy.
\newblock {\em Medical Physics\/}  (2021) 1--15.
\newblock \doi{10.1002/mp.15449}.

\bibitem[{Actis et~al.(2018)Actis, Mayor, Meer, and Weber}]{Actis2018}
Actis O, Mayor A, Meer D, Weber D.
\newblock Precise beam delivery for proton therapy with dynamic energy modulation.
\newblock {\em Journal of Physics: Conference Series\/} {\bf 1067} (2018).
\newblock \doi{10.1088/1742-6596/1067/9/092002}.

\bibitem[{Actis et~al.(2023)Actis, Mayor, Meer, Rechsteiner, Bolsi, Lomax et~al.}]{Actis2023}
Actis O, Mayor A, Meer D, Rechsteiner U, Bolsi A, Lomax AJ, et~al.
\newblock A bi-directional beam-line energy ramping for efficient patient treatment with scanned proton therapy.
\newblock {\em Physics in Medicine \& Biology\/} {\bf 68} (2023) 175001.
\newblock \doi{10.1088/1361-6560/acebb2}.

\bibitem[{Artz et~al.(2025)Artz, Grewal, Zhang, Johnson, Saki, Schreuder et~al.}]{Artz2025}
Artz ME, Grewal HS, Zhang Y, Johnson PB, Saki M, Schreuder N, et~al.
\newblock Operational {{Improvement}} of a {{Proton Therapy System From Reduced Energy Layer Switching Time}}.
\newblock {\em International Journal of Particle Therapy\/} {\bf 16} (2025) 100742.
\newblock \doi{10.1016/j.ijpt.2025.100742}.

\bibitem[{Fu et~al.(2024)Fu, Taasti, and Zarepisheh}]{Fu2024}
Fu A, Taasti VT, Zarepisheh M.
\newblock Simultaneous reduction of number of spots and energy layers in intensity modulated proton therapy for rapid spot scanning delivery.
\newblock {\em Medical Physics\/}  (2024) mp.17070.
\newblock \doi{10.1002/mp.17070}.

\bibitem[{O'Grady et~al.(2023)O'Grady, Janson, Rao, Chawla, Stephenson, Fan et~al.}]{OGrady2023}
O'Grady F, Janson M, Rao AD, Chawla AK, Stephenson L, Fan J, et~al.
\newblock The use of a mini-ridge filter with cyclotron-based pencil beam scanning proton therapy.
\newblock {\em Medical Physics\/}  (2023) mp.16254.
\newblock \doi{10.1002/mp.16254}.

\bibitem[{M{\"u}ller and Wilkens(2016)}]{Muller2016}
M{\"u}ller BS, Wilkens JJ.
\newblock Prioritized efficiency optimization for intensity modulated proton therapy.
\newblock {\em Physics in Medicine and Biology\/} {\bf 61} (2016) 8249--8265.
\newblock \doi{10.1088/0031-9155/61/23/8249}.

\bibitem[{Taasti et~al.(2025)Taasti, Kneepkens, {van der Stoep}, Velders, Cobben, Vullings et~al.}]{Taasti2025}
Taasti VT, Kneepkens E, {van der Stoep} J, Velders M, Cobben M, Vullings A, et~al.
\newblock Proton therapy of lung cancer patients -- {{Treatment}} strategies and clinical experience from a medical physicist's perspective.
\newblock {\em Physica Medica\/} {\bf 130} (2025) 104890.
\newblock \doi{10.1016/j.ejmp.2024.104890}.

\bibitem[{Mori et~al.(2018)Mori, Knopf, and Umegaki}]{Mori2018}
Mori S, Knopf AC, Umegaki K.
\newblock Motion management in particle therapy.
\newblock {\em Medical Physics\/} {\bf 45} (2018) e994--e1010.
\newblock \doi{10.1002/mp.12679}.

\bibitem[{Graeff(2014)}]{Graeff2014}
Graeff C.
\newblock Motion mitigation in scanned ion beam therapy through {{4D-optimization}}.
\newblock {\em Physica Medica\/} {\bf 30} (2014) 570--577.
\newblock \doi{10.1016/j.ejmp.2014.03.011}.

\bibitem[{Kn{\"a}usl et~al.(2024)Kn{\"a}usl, Belotti, Bertholet, Daartz, Flampouri, Hoogeman et~al.}]{Knausl2024}
Kn{\"a}usl B, Belotti G, Bertholet J, Daartz J, Flampouri S, Hoogeman M, et~al.
\newblock A review of the clinical introduction of {{4D}} particle therapy research concepts.
\newblock {\em Physics and Imaging in Radiation Oncology\/}  (2024) 100535.
\newblock \doi{10.1016/j.phro.2024.100535}.

\bibitem[{Engelsman et~al.(2013)Engelsman, Schwarz, and Dong}]{Engelsman2013}
Engelsman M, Schwarz M, Dong L.
\newblock Physics {{Controversies}} in {{Proton Therapy}}.
\newblock {\em Seminars in Radiation Oncology\/} {\bf 23} (2013) 88--96.
\newblock \doi{10.1016/j.semradonc.2012.11.003}.

\bibitem[{Phillips et~al.(1992)Phillips, Pedroni, Blattmann, Boehringer, Coray, and Scheib}]{Phillips1992}
Phillips MH, Pedroni E, Blattmann H, Boehringer T, Coray A, Scheib S.
\newblock Effects of respiratory motion on dose uniformity with a charged particle scanning method.
\newblock {\em Physics in Medicine and Biology\/} {\bf 37} (1992) 223--234.
\newblock \doi{10.1088/0031-9155/37/1/016}.

\bibitem[{Bert et~al.(2008)Bert, Gr{\"o}zinger, and Rietzel}]{Bert2008}
Bert C, Gr{\"o}zinger SO, Rietzel E.
\newblock Quantification of interplay effects of scanned particle beams and moving targets.
\newblock {\em Physics in Medicine and Biology\/} {\bf 53} (2008) 2253--2265.
\newblock \doi{10.1088/0031-9155/53/9/003}.

\bibitem[{Shan et~al.(2020)Shan, Yang, Schild, Daniels, Wong, Fatyga et~al.}]{Shan2020}
Shan J, Yang Y, Schild SE, Daniels TB, Wong WW, Fatyga M, et~al.
\newblock Intensity-modulated proton therapy ({{IMPT}}) interplay effect evaluation of asymmetric breathing with simultaneous uncertainty considerations in patients with non-small cell lung cancer.
\newblock {\em Medical Physics\/} {\bf 47} (2020) 5428--5440.
\newblock \doi{10.1002/mp.14491}.

\bibitem[{Kraus et~al.(2011)Kraus, Heath, and Oelfke}]{Kraus2011}
Kraus KM, Heath E, Oelfke U.
\newblock Dosimetric consequences of tumour motion due to respiration for a scanned proton beam.
\newblock {\em Physics in Medicine and Biology\/} {\bf 56} (2011) 6563--6581.
\newblock \doi{10.1088/0031-9155/56/20/003}.

\bibitem[{Li et~al.(2015)Li, Zhu, and Zhang}]{Li2015}
Li H, Zhu XR, Zhang X.
\newblock Reducing {{Dose Uncertainty}} for {{Spot-Scanning Proton Beam Therapy}} of {{Moving Tumors}} by {{Optimizing}} the {{Spot Delivery Sequence}}.
\newblock {\em International Journal of Radiation Oncology Biology Physics\/} {\bf 93} (2015) 547--556.
\newblock \doi{10.1016/j.ijrobp.2015.06.019}.

\bibitem[{Kang et~al.(2017)Kang, Huang, Solberg, Mayer, Thomas, Teo et~al.}]{Kang2017}
Kang M, Huang S, Solberg TD, Mayer R, Thomas A, Teo BKK, et~al.
\newblock A study of the beam-specific interplay effect in proton pencil beam scanning delivery in lung cancer.
\newblock {\em Acta Oncologica\/} {\bf 56} (2017) 531--540.
\newblock \doi{10.1080/0284186X.2017.1293287}.

\bibitem[{Liu and Chang(2011)}]{Liu2011}
Liu H, Chang JY.
\newblock Proton therapy in clinical practice.
\newblock {\em Chinese Journal of Cancer\/} {\bf 30} (2011) 315--326.
\newblock \doi{10.5732/cjc.010.10529}.

\bibitem[{K{\"o}the et~al.(2022)K{\"o}the, Lomax, Giovannelli, Safai, Bizzocchi, Roelofs et~al.}]{Kothe2022}
K{\"o}the A, Lomax AJ, Giovannelli AC, Safai S, Bizzocchi N, Roelofs E, et~al.
\newblock The impact of organ motion and the appliance of mitigation strategies on the effectiveness of hypoxia-guided proton therapy for non-small cell lung cancer.
\newblock {\em Radiotherapy and Oncology\/} {\bf 176} (2022) 208--214.
\newblock \doi{10.1016/j.radonc.2022.09.021}.

\bibitem[{Keall(2004)}]{Keall2004}
Keall P.
\newblock 4-{{Dimensional Computed Tomography Imaging}} and {{Treatment Planning}}.
\newblock {\em Seminars in Radiation Oncology\/} {\bf 14} (2004) 81--90.
\newblock \doi{10.1053/j.semradonc.2003.10.006}.

\bibitem[{Hoekstra et~al.(2021)Hoekstra, Habraken, {Swaak-Kragten}, Hoogeman, and Pignol}]{Hoekstra2021}
Hoekstra N, Habraken S, {Swaak-Kragten} A, Hoogeman M, Pignol JP.
\newblock Intrafraction motion during partial breast irradiation depends on treatment time.
\newblock {\em Radiotherapy and Oncology\/} {\bf 159} (2021) 176--182.
\newblock \doi{10.1016/j.radonc.2021.03.029}.

\bibitem[{Hoogeman et~al.(2008)Hoogeman, Nuyttens, Levendag, and Heijmen}]{Hoogeman2008}
Hoogeman MS, Nuyttens JJ, Levendag PC, Heijmen BJ.
\newblock Time {{Dependence}} of {{Intrafraction Patient Motion Assessed}} by {{Repeat Stereoscopic Imaging}}.
\newblock {\em International Journal of Radiation Oncology*Biology*Physics\/} {\bf 70} (2008) 609--618.
\newblock \doi{10.1016/j.ijrobp.2007.08.066}.

\bibitem[{Bert and Durante(2011)}]{Bert2011}
Bert C, Durante M.
\newblock Motion in radiotherapy: Particle therapy.
\newblock {\em Physics in Medicine and Biology\/} {\bf 56} (2011) R113--R144.
\newblock \doi{10.1088/0031-9155/56/16/R01}.

\bibitem[{Kraan(2015)}]{Kraan2015}
Kraan AC.
\newblock Range {{Verification Methods}} in {{Particle Therapy}}: {{Underlying Physics}} and {{Monte Carlo Modeling}}.
\newblock {\em Frontiers in Oncology\/} {\bf 5} (2015) 150.
\newblock \doi{10.3389/fonc.2015.00150}.

\bibitem[{Knopf and Lomax(2013)}]{Knopf2013}
Knopf AC, Lomax A.
\newblock In vivo proton range verification: A review.
\newblock {\em Physics in Medicine and Biology\/} {\bf 58} (2013) R131--R160.
\newblock \doi{10.1088/0031-9155/58/15/R131}.

\bibitem[{Zhang et~al.(2023)Zhang, Trnkova, Toshito, Heijmen, Richter, Aznar et~al.}]{Zhang2023}
Zhang Y, Trnkova P, Toshito T, Heijmen B, Richter C, Aznar M, et~al.
\newblock A survey of practice patterns for real-time intrafractional motion-management in particle therapy.
\newblock {\em Physics and Imaging in Radiation Oncology\/} {\bf 26} (2023) 100439.
\newblock \doi{10.1016/j.phro.2023.100439}.

\bibitem[{Rietzel and Bert(2010)}]{Rietzel2010}
Rietzel E, Bert C.
\newblock Respiratory motion management in particle therapy.
\newblock {\em Medical Physics\/} {\bf 37} (2010) 449--460.
\newblock \doi{10.1118/1.3250856}.

\bibitem[{Poulsen et~al.(2018)Poulsen, Eley, Langner, Simone, and Langen}]{Poulsen2018}
Poulsen PR, Eley J, Langner U, Simone CB, Langen K.
\newblock Efficient {{Interplay Effect Mitigation}} for {{Proton Pencil Beam Scanning}} by {{Spot-Adapted Layered Repainting Evenly Spread}} out {{Over}} the {{Full Breathing Cycle}}.
\newblock {\em International Journal of Radiation Oncology Biology Physics\/} {\bf 100} (2018) 226--234.
\newblock \doi{10.1016/j.ijrobp.2017.09.043}.

\bibitem[{Younkin et~al.(2021)Younkin, Morales, Shen, Ding, Stoker, Yu et~al.}]{Younkin2021}
Younkin JE, Morales DH, Shen J, Ding X, Stoker JB, Yu NY, et~al.
\newblock Technical {{Note}}: {{Multiple}} energy extraction techniques for synchrotron-based proton delivery systems may exacerbate motion interplay effects in lung cancer treatments.
\newblock {\em Medical Physics\/} {\bf 48} (2021) 4812--4823.
\newblock \doi{10.1002/mp.15056}.

\bibitem[{Sengbusch et~al.(2009)Sengbusch, {P{\'e}rez-And{\'u}jar}, DeLuca, and Mackie}]{Sengbusch2009}
Sengbusch E, {P{\'e}rez-And{\'u}jar} A, DeLuca PM, Mackie TR.
\newblock Maximum proton kinetic energy and patient-generated neutron fluence considerations in proton beam arc delivery radiation therapy.
\newblock {\em Medical Physics\/} {\bf 36} (2009) 364--372.
\newblock \doi{10.1118/1.3049787}.

\bibitem[{Beltran et~al.(2024)Beltran, Perales, and Furutani}]{Beltran2024}
Beltran CJ, Perales A, Furutani KM.
\newblock Does the {{Maximum Initial Beam Energy}} for {{Proton Therapy Have}} to {{Be}} 230 {{MeV}}?
\newblock {\em Quantum Beam Science\/} {\bf 8} (2024) 23.
\newblock \doi{10.3390/qubs8030023}.

\bibitem[{Wang et~al.(2023)Wang, Liu, Yang, Liao, Li, Zhao et~al.}]{Wang2023}
Wang W, Liu X, Yang Z, Liao Y, Li P, Zhao R, et~al.
\newblock Improving delivery efficiency using spots and energy layers reduction algorithms based on a large momentum acceptance beamline.
\newblock {\em Medical Physics\/}  (2023) mp.16420.
\newblock \doi{10.1002/mp.16420}.

\bibitem[{Giovannelli et~al.(2023)Giovannelli, K{\"o}the, Safai, Meer, Zhang, Weber et~al.}]{Giovannelli2023}
Giovannelli AC, K{\"o}the A, Safai S, Meer D, Zhang Y, Weber DC, et~al.
\newblock Exploring beamline momentum acceptance for tracking respiratory variability in lung cancer proton therapy: A simulation study.
\newblock {\em Physics in Medicine \& Biology\/}  (2023).
\newblock \doi{10.1088/1361-6560/acf5c4}.

\bibitem[{Jensen et~al.(2018)Jensen, Hoffmann, Petersen, M{\o}ller, and Alber}]{Jensen2018}
Jensen MF, Hoffmann L, Petersen JBB, M{\o}ller DS, Alber M.
\newblock Energy layer optimization strategies for intensity-modulated proton therapy of lung cancer patients.
\newblock {\em Medical Physics\/} {\bf 45} (2018) 4355--4363.
\newblock \doi{10.1002/mp.13139}.

\bibitem[{Cao et~al.(2014)Cao, Lim, Liao, Li, Jiang, Li et~al.}]{Cao2014}
Cao W, Lim G, Liao L, Li Y, Jiang S, Li X, et~al.
\newblock Proton energy optimization and reduction for intensity-modulated proton therapy.
\newblock {\em Physics in Medicine and Biology\/} {\bf 59} (2014) 6341--6354.
\newblock \doi{10.1088/0031-9155/59/21/6341}.

\bibitem[{Lin et~al.(2019)Lin, Clasie, Liu, McDonald, Langen, and Gao}]{Lin2019}
Lin Y, Clasie B, Liu T, McDonald M, Langen KM, Gao H.
\newblock Minimum-{{MU}} and sparse-energy-layer ({{MMSEL}}) constrained inverse optimization method for efficiently deliverable {{PBS}} plans.
\newblock {\em Physics in Medicine \& Biology\/} {\bf 64} (2019) 205001.
\newblock \doi{10.1088/1361-6560/ab4529}.

\bibitem[{Gao et~al.(2020)Gao, Clasie, McDonald, Langen, Liu, and Lin}]{Gao2020}
Gao H, Clasie B, McDonald M, Langen KM, Liu T, Lin Y.
\newblock Technical {{Note}}: {{Plan}}-delivery-time constrained inverse optimization method with minimum-{{MU}}-per-energy-layer ({{MMPEL}}) for efficient pencil beam scanning proton therapy.
\newblock {\em Medical Physics\/} {\bf 47} (2020) 3892--3897.
\newblock \doi{10.1002/mp.14363}.

\bibitem[{Van De~Water et~al.(2015)Van De~Water, Kooy, Heijmen, and Hoogeman}]{VanDeWater2015}
Van De~Water S, Kooy HM, Heijmen BJ, Hoogeman MS.
\newblock Shortening delivery times of intensity modulated proton therapy by reducing proton energy layers during treatment plan optimization.
\newblock {\em International Journal of Radiation Oncology Biology Physics\/} {\bf 92} (2015) 460--468.
\newblock \doi{10.1016/j.ijrobp.2015.01.031}.

\bibitem[{Kang et~al.(2008)Kang, Wilkens, and Oelfke}]{Kang2008}
Kang JH, Wilkens JJ, Oelfke U.
\newblock Non-uniform depth scanning for proton therapy systems employing active energy variation.
\newblock {\em Physics in Medicine and Biology\/} {\bf 53} (2008) 149--155.
\newblock \doi{10.1088/0031-9155/53/9/N01}.

\bibitem[{Lin et~al.(2024)Lin, Li, Liu, Fu, Lin, and Gao}]{Lin2024}
Lin B, Li Y, Liu B, Fu S, Lin Y, Gao H.
\newblock Cardinality-constrained plan-quality and delivery-time optimization method for proton therapy.
\newblock {\em Medical Physics\/} {\bf 51} (2024) 4567--4580.
\newblock \doi{10.1002/mp.17249}.

\bibitem[{Kang et~al.(2022)Kang, Wei, Choi, Lin, and Simone}]{Kang2022}
Kang M, Wei S, Choi JI, Lin H, Simone CB.
\newblock A {{Universal Range Shifter}} and {{Range Compensator Can Enable Proton Pencil Beam Scanning Single-Energy Bragg Peak FLASH-RT Treatment Using Current Commercially Available Proton Systems}}.
\newblock {\em International Journal of Radiation Oncology*Biology*Physics\/} {\bf 113} (2022) 203--213.
\newblock \doi{10.1016/j.ijrobp.2022.01.009}.

\bibitem[{Schwarz(2025)}]{Schwarz2025}
Schwarz M.
\newblock Bridging research and clinical practice in ultra-high dose rate proton therapy.
\newblock {\em The European Physical Journal Plus\/}  (2025).

\bibitem[{Zhang et~al.(2022)Zhang, Gao, and Peng}]{Zhang2022}
Zhang G, Gao W, Peng H.
\newblock Design of static and dynamic ridge filters for {{FLASH}}--{{IMPT}}: {{A}} simulation study.
\newblock {\em Medical Physics\/} {\bf 49} (2022) 5387--5399.
\newblock \doi{10.1002/mp.15717}.

\bibitem[{Hagmann et~al.(2025)Hagmann, Colizzi, Ambruosi, Lomax, Meer, Psoroulas et~al.}]{Hagmann2025}
Hagmann V, Colizzi I, Ambruosi A, Lomax AJ, Meer D, Psoroulas S, et~al.
\newblock A {{Novel Energy Modulator Design Concept}} for {{FLASH Proton Therapy}}.
\newblock {\em Current Directions in Biomedical Engineering\/} {\bf 11} (2025) 1--4.
\newblock \doi{10.1515/cdbme-2025-0101}.

\bibitem[{Hotoiu et~al.(2025)Hotoiu, Stappen, Pin, Nilsson, Ivoc, Kim et~al.}]{Hotoiu2025}
Hotoiu L, Stappen FV, Pin A, Nilsson R, Ivoc J, Kim M, et~al.
\newblock Experimental validation of coarse ridge filters for {{FLASH}} proton therapy.
\newblock {\em Medical Physics\/} {\bf 52} (2025) e18044.
\newblock \doi{10.1002/mp.18044}.

\bibitem[{Nesteruk et~al.(2021)Nesteruk, Togno, Grossmann, Lomax, Weber, Schippers et~al.}]{Nesteruk2021b}
Nesteruk KP, Togno M, Grossmann M, Lomax AJ, Weber DC, Schippers JM, et~al.
\newblock Commissioning of a clinical pencil beam scanning proton therapy unit for ultra-high dose rates ({{FLASH}}).
\newblock {\em Medical Physics\/} {\bf 48} (2021) 4017--4026.
\newblock \doi{10.1002/mp.14933}.

\bibitem[{Zeng et~al.(2024)Zeng, Li, Wang, Liu, Qin, Dai et~al.}]{Zeng2024}
Zeng Y, Li H, Wang W, Liu X, Qin B, Dai S, et~al.
\newblock Feasibility study of multiple-energy {{Bragg}} peak proton {{FLASH}} on a superconducting gantry with large momentum acceptance.
\newblock {\em Medical Physics\/}  (2024) mp.16932.
\newblock \doi{10.1002/mp.16932}.

\bibitem[{Zeng et~al.(2025)Zeng, Quan, Zhang, Wang, Liu, Qin et~al.}]{Zeng2025}
Zeng Y, Quan H, Zhang Q, Wang W, Liu X, Qin B, et~al.
\newblock Treatment parameters consideration for universal range shifter-based multi-energy proton {{FLASH-RT}}.
\newblock {\em Medical Physics\/} {\bf 52} (2025) e70039.
\newblock \doi{10.1002/mp.70039}.

\bibitem[{Farr et~al.(2022)Farr, Grilj, Malka, Sudharsan, and Schippers}]{Farr2022}
Farr J, Grilj V, Malka V, Sudharsan S, Schippers M.
\newblock Ultra-{{High}} dose rate radiation production and delivery systems intended for {{FLASH}}.
\newblock {\em Medical Physics\/}  (2022) 1--37.
\newblock \doi{10.1002/mp.15659}.

\bibitem[{Mazal et~al.(2021)Mazal, Vera~Sanchez, {Sanchez-Parcerisa}, Udias, Espa{\~n}a, {Sanchez-Tembleque} et~al.}]{Mazal2021}
Mazal A, Vera~Sanchez JA, {Sanchez-Parcerisa} D, Udias JM, Espa{\~n}a S, {Sanchez-Tembleque} V, et~al.
\newblock Biological and {{Mechanical Synergies}} to {{Deal With Proton Therapy Pitfalls}}: {{Minibeams}}, {{FLASH}}, {{Arcs}}, and {{Gantryless Rooms}}.
\newblock {\em Frontiers in Oncology\/} {\bf 10} (2021) 1--14.
\newblock \doi{10.3389/fonc.2020.613669}.

\bibitem[{Jolly et~al.(2020)Jolly, Owen, Schippers, and Welsch}]{Jolly2020}
Jolly S, Owen H, Schippers M, Welsch C.
\newblock Technical challenges for {{FLASH}} proton therapy.
\newblock {\em Physica Medica\/} {\bf 78} (2020) 71--82.
\newblock \doi{10.1016/j.ejmp.2020.08.005}.

\bibitem[{Esplen et~al.(2020)Esplen, Mendonca, and {Bazalova-Carter}}]{Esplen2020}
Esplen N, Mendonca MS, {Bazalova-Carter} M.
\newblock Physics and biology of ultrahigh dose-rate ({{FLASH}}) radiotherapy: {{A}} topical review.
\newblock {\em Physics in Medicine and Biology\/} {\bf 65} (2020).
\newblock \doi{10.1088/1361-6560/abaa28}.

\bibitem[{Fenwick et~al.(2024)Fenwick, Mayhew, Jolly, Amos, and Hawkins}]{Fenwick2024}
Fenwick JD, Mayhew C, Jolly S, Amos RA, Hawkins MA.
\newblock Navigating the straits: Realizing the potential of proton {{FLASH}} through physics advances and further pre-clinical characterization.
\newblock {\em Frontiers in Oncology\/} {\bf 14} (2024) 1420337.
\newblock \doi{10.3389/fonc.2024.1420337}.

\bibitem[{Zhao et~al.(2023)Zhao, You, Liu, Wuyckens, Lu, and Ding}]{Zhao2023}
Zhao L, You J, Liu G, Wuyckens S, Lu X, Ding X.
\newblock The first direct method of spot sparsity optimization for proton arc therapy.
\newblock {\em Acta Oncologica\/} {\bf 62} (2023) 48--52.
\newblock \doi{10.1080/0284186X.2023.2172689}.

\bibitem[{Engwall et~al.(2022)Engwall, Battinelli, Wase, Marthin, Glimelius, Bokrantz et~al.}]{Engwall2022}
Engwall E, Battinelli C, Wase V, Marthin O, Glimelius L, Bokrantz R, et~al.
\newblock Fast robust optimization of proton {{PBS}} arc therapy plans using early energy layer selection and spot assignment.
\newblock {\em Physics in Medicine \& Biology\/} {\bf 67} (2022) 065010.
\newblock \doi{10.1088/1361-6560/ac55a6}.

\bibitem[{Liu et~al.(2020)Liu, Li, Zhao, Zheng, Qin, Zhang et~al.}]{Liu2020a}
Liu G, Li X, Zhao L, Zheng W, Qin A, Zhang S, et~al.
\newblock A novel energy sequence optimization algorithm for efficient spot-scanning proton arc ({{SPArc}}) treatment delivery.
\newblock {\em Acta Oncologica\/} {\bf 59} (2020) 1178--1185.
\newblock \doi{10.1080/0284186X.2020.1765415}.

\bibitem[{Gu et~al.(2020)Gu, Ruan, Lyu, Zou, Dong, and Sheng}]{Gu2020}
Gu W, Ruan D, Lyu Q, Zou W, Dong L, Sheng K.
\newblock A novel energy layer optimization framework for spot-scanning proton arc therapy.
\newblock {\em Medical Physics\/} {\bf 47} (2020) 2072--2084.
\newblock \doi{10.1002/mp.14083}.

\bibitem[{Janson et~al.(2024)Janson, Glimelius, Fredriksson, Traneus, and Engwall}]{Janson2024}
Janson M, Glimelius L, Fredriksson A, Traneus E, Engwall E.
\newblock Treatment planning of scanned proton beams in {{RayStation}}.
\newblock {\em Medical Dosimetry\/} {\bf 49} (2024) 2--12.
\newblock \doi{10.1016/j.meddos.2023.10.009}.

\bibitem[{Fracchiolla et~al.(2025)Fracchiolla, Engwall, Mikhalev, Cianchetti, Giacomelli, Siniscalchi et~al.}]{Fracchiolla2025}
Fracchiolla F, Engwall E, Mikhalev V, Cianchetti M, Giacomelli I, Siniscalchi B, et~al.
\newblock Static proton arc therapy: {{Comprehensive}} plan quality evaluation and first clinical treatments in patients with complex head and neck targets.
\newblock {\em Medical Physics\/}  (2025) mp.17669.
\newblock \doi{10.1002/mp.17669}.

\bibitem[{Cong et~al.(2024)Cong, Liu, Liu, Zhao, Chen, Li et~al.}]{Cong2024}
Cong X, Liu G, Liu P, Zhao L, Chen S, Li X, et~al.
\newblock Explore the feasibility of using spot-scanning proton arc therapy for a synchrotron accelerator-based proton therapy system -- {{A}} simulation study.
\newblock {\em Journal of Applied Clinical Medical Physics\/}  (2024) e14526.
\newblock \doi{10.1002/acm2.14526}.

\bibitem[{Ding et~al.(2016)Ding, Li, Zhang, Kabolizadeh, Stevens, and Yan}]{Ding2016}
Ding X, Li X, Zhang JM, Kabolizadeh P, Stevens C, Yan D.
\newblock Spot-{{Scanning Proton Arc}} ({{SPArc}}) {{Therapy}}: {{The First Robust}} and {{Delivery-Efficient Spot-Scanning Proton Arc Therapy}}.
\newblock {\em International Journal of Radiation Oncology Biology Physics\/} {\bf 96} (2016) 1107--1116.
\newblock \doi{10.1016/j.ijrobp.2016.08.049}.

\bibitem[{Schreuder et~al.(2022)Schreuder, Ding, and Li}]{Schreuder2022}
Schreuder N, Ding X, Li Z.
\newblock Fixed beamlines can replace gantries for particle therapy.
\newblock {\em Medical Physics\/}  (2022) 1--4.
\newblock \doi{10.1002/mp.15531}.

\bibitem[{Volz et~al.(2024)Volz, Korte, Martire, Zhang, Hardcastle, Durante et~al.}]{Volz2024}
Volz L, Korte JC, Martire MC, Zhang Y, Hardcastle N, Durante M, et~al.
\newblock Opportunities and challenges of upright patient positioning in radiotherapy.
\newblock {\em Physics in Medicine \& Biology\/}  (2024).
\newblock \doi{10.1088/1361-6560/ad70ee}.

\bibitem[{Underwood et~al.(2025)Underwood, Yan, Bortfeld, Gr{\'e}goire, and Lomax}]{Underwood2025}
Underwood T, Yan S, Bortfeld T, Gr{\'e}goire V, Lomax T.
\newblock A {{Strengths}}, {{Weaknesses}}, {{Opportunities}}, and {{Threats}} ({{SWOT}}) {{Analysis}} for {{Gantry-Less Upright Radiation Therapy}}.
\newblock {\em International Journal of Radiation Oncology*Biology*Physics\/} {\bf 122} (2025) 1337--1343.
\newblock \doi{10.1016/j.ijrobp.2025.02.039}.

\bibitem[{Volz(2022)}]{Volz2022}
Volz L.
\newblock Considerations for {{Upright Particle Therapy Patient Positioning}} and {{Associated Image Guidance}}.
\newblock {\em Frontiers in Oncology\/} {\bf 12} (2022) 13.

\bibitem[{Mein et~al.(2024)Mein, Wuyckens, Li, Both, Carabe, Vera et~al.}]{Mein2024}
Mein S, Wuyckens S, Li X, Both S, Carabe A, Vera MC, et~al.
\newblock Particle arc therapy: {{Status}} and potential.
\newblock {\em Radiotherapy and Oncology\/}  (2024) 110434.
\newblock \doi{10.1016/j.radonc.2024.110434}.

\bibitem[{Zhang et~al.(2016)Zhang, Huth, Wegner, Weber, and Lomax}]{Zhang2016}
Zhang Y, Huth I, Wegner M, Weber DC, Lomax AJ.
\newblock An evaluation of rescanning technique for liver tumour treatments using a commercial {{PBS}} proton therapy system.
\newblock {\em Radiotherapy and Oncology\/} {\bf 121} (2016) 281--287.
\newblock \doi{10.1016/j.radonc.2016.09.011}.

\bibitem[{Sch{\"a}tti et~al.(2013)Sch{\"a}tti, Zakova, Meer, and Lomax}]{Schatti2013a}
Sch{\"a}tti A, Zakova M, Meer D, Lomax AJ.
\newblock Experimental verification of motion mitigation of discrete proton spot scanning by re-scanning.
\newblock {\em Physics in Medicine and Biology\/} {\bf 58} (2013) 8555--8572.
\newblock \doi{10.1088/0031-9155/58/23/8555}.

\bibitem[{Lambert et~al.(2005)Lambert, Suchowerska, McKenzie, and Jackson}]{Lambert2005}
Lambert J, Suchowerska N, McKenzie DR, Jackson M.
\newblock Intrafractional motion during proton beam scanning.
\newblock {\em Physics in Medicine and Biology\/} {\bf 50} (2005) 4853--4862.
\newblock \doi{10.1088/0031-9155/50/20/008}.

\bibitem[{Han(2019)}]{Han2019a}
Han Y.
\newblock Current status of proton therapy techniques for lung cancer.
\newblock {\em Radiation Oncology Journal\/} {\bf 37} (2019) 232--248.
\newblock \doi{10.3857/roj.2019.00633}.

\bibitem[{Engwall et~al.(2018)Engwall, Glimelius, and Hynning}]{Engwall2018}
Engwall E, Glimelius L, Hynning E.
\newblock Effectiveness of different rescanning techniques for scanned proton radiotherapy in lung cancer patients.
\newblock {\em Physics in Medicine and Biology\/} {\bf 63} (2018) aabb7b.
\newblock \doi{10.1088/1361-6560/aabb7b}.

\bibitem[{Bernatowicz et~al.(2013)Bernatowicz, Lomax, and Knopf}]{Bernatowicz2013}
Bernatowicz K, Lomax AJ, Knopf A.
\newblock Comparative study of layered and volumetric rescanning for different scanning speeds of proton beam in liver patients.
\newblock {\em Physics in Medicine and Biology\/} {\bf 58} (2013) 7905--7920.
\newblock \doi{10.1088/0031-9155/58/22/7905}.

\bibitem[{Sch{\"a}tti et~al.(2014)Sch{\"a}tti, Zakova, Meer, and Lomax}]{Schatti2014b}
Sch{\"a}tti A, Zakova M, Meer D, Lomax AJ.
\newblock The effectiveness of combined gating and re-scanning for treating mobile targets with proton spot scanning. {{An}} experimental and simulation-based investigation.
\newblock {\em Physics in Medicine and Biology\/} {\bf 59} (2014) 3813--3828.
\newblock \doi{10.1088/0031-9155/59/14/3813}.

\bibitem[{Rana et~al.(2020{\natexlab{a}})Rana, Bennouna, Gutierrez, and Rosenfeld}]{Rana2020a}
Rana S, Bennouna J, Gutierrez A, Rosenfeld A.
\newblock Evaluation of spot size using volumetric repainting technique on a {{ProteusPLUS PBS Proton Therapy System}}.
\newblock {\em Journal of Physics: Conference Series\/} {\bf 1662} (2020{\natexlab{a}}) 012027.
\newblock \doi{10.1088/1742-6596/1662/1/012027}.

\bibitem[{Rana et~al.(2020{\natexlab{b}})Rana, Bennouna, Gutierrez, and Rosenfeld}]{Rana2020}
Rana S, Bennouna J, Gutierrez AN, Rosenfeld AB.
\newblock Impact of magnetic field regulation in conjunction with the volumetric repainting technique on the spot positions and beam range in pencil beam scanning proton therapy.
\newblock {\em Journal of Applied Clinical Medical Physics\/} {\bf 21} (2020{\natexlab{b}}) 124--131.
\newblock \doi{10.1002/acm2.13045}.

\bibitem[{Giordanengo and Palmans(2018)}]{Giordanengo2018}
Giordanengo S, Palmans H.
\newblock Dose detectors, sensors, and their applications.
\newblock {\em Medical Physics\/} {\bf 45} (2018) e1051--e1072.

\bibitem[{Schoemers et~al.(2015)Schoemers, Feldmeier, Naumann, Panse, Peters, and Haberer}]{Schoemers2015}
Schoemers C, Feldmeier E, Naumann J, Panse R, Peters A, Haberer T.
\newblock The intensity feedback system at {{Heidelberg Ion-Beam Therapy Centre}}.
\newblock {\em Nuclear Instruments and Methods in Physics Research, Section A: Accelerators, Spectrometers, Detectors and Associated Equipment\/} {\bf 795} (2015) 92--99.
\newblock \doi{10.1016/j.nima.2015.05.054}.

\bibitem[{Degiovanni et~al.(2024)Degiovanni, De~Michele, Bonomi, Cabaleiro~Magallanes, Caldara, Dimov et~al.}]{Degiovanni2024}
Degiovanni A, De~Michele G, Bonomi R, Cabaleiro~Magallanes F, Caldara M, Dimov V, et~al.
\newblock Design, integration, and commissioning of the first linac for image guided hadron therapy prototype.
\newblock {\em Physical Review Accelerators and Beams\/} {\bf 27} (2024) 054701.
\newblock \doi{10.1103/PhysRevAccelBeams.27.054701}.

\bibitem[{Peach et~al.(2013)Peach, Aslaninejad, Barlow, Beard, Bliss, Cobb et~al.}]{Peach2013}
Peach KJ, Aslaninejad M, Barlow RJ, Beard CD, Bliss N, Cobb JH, et~al.
\newblock Conceptual design of a nonscaling fixed field alternating gradient accelerator for protons and carbon ions for charged particle therapy.
\newblock {\em Physical Review Special Topics - Accelerators and Beams\/} {\bf 16} (2013) 1--34.
\newblock \doi{10.1103/PhysRevSTAB.16.030101}.

\bibitem[{Amaldi(2019)}]{Amaldi2019}
Amaldi U.
\newblock Oblique raster scanning: {{An}} ion dose delivery procedure with variable energy layers.
\newblock {\em Physics in Medicine and Biology\/} {\bf 64} (2019).
\newblock \doi{10.1088/1361-6560/ab0920}.

\bibitem[{Tsubouchi et~al.(2023)Tsubouchi, Beltran, Yagi, Hamatani, Takashina, Shimizu et~al.}]{Tsubouchi2023}
Tsubouchi T, Beltran CJ, Yagi M, Hamatani N, Takashina M, Shimizu S, et~al.
\newblock Beam delivery characteristics of the {{Hitachi}} carbon ion scanning system at {{Osaka Heavy Ion Medical Accelerator}} in {{Kansai}} ({{HIMAK}}).
\newblock {\em Medical Physics\/}  (2023) mp.16791.
\newblock \doi{10.1002/mp.16791}.

\bibitem[{Tsubouchi et~al.(2024)Tsubouchi, Furutani, Yagi, Nomura, Shimizu, Kanai et~al.}]{Tsubouchi2024}
Tsubouchi T, Furutani KM, Yagi M, Nomura T, Shimizu S, Kanai T, et~al.
\newblock Dosimetric consequences of flap dose due to rapid beam off control for a high intensity carbon ion radiation therapy synchrotron.
\newblock {\em Medical Physics\/}  (2024) mp.17309.
\newblock \doi{10.1002/mp.17309}.

\bibitem[{Kang et~al.(2023)Kang, Ding, and Rong}]{Kang2023}
Kang M, Ding X, Rong Y.
\newblock {{FLASH}} instead of proton arc therapy is a more promising advancement for the next generation proton radiotherapy.
\newblock {\em Journal of Applied Clinical Medical Physics\/} {\bf 24} (2023) e14091.
\newblock \doi{10.1002/acm2.14091}.

\bibitem[{Fukumoto(1995)}]{Fukumoto1995}
Fukumoto S.
\newblock Cyclotron {{Versus Synchrotron}} for {{Proton Beam Therapy}}.
\newblock {\em Proceedings of the 14th International Conference on Cyclotrons and their Applications\/}  (1995) 533--536.

\bibitem[{Schippers and Lomax(2011)}]{Schippers2011a}
Schippers JM, Lomax AJ.
\newblock Emerging technologies in proton therapy.
\newblock {\em Acta Oncologica\/} {\bf 50} (2011) 838--850.
\newblock \doi{10.3109/0284186X.2011.582513}.

\bibitem[{Schippers(2009)}]{Schippers2009}
Schippers M.
\newblock Beam {{Delivery Systems}} for {{Particle Therapy}}: {{Current Status}} and {{Recent Developments}}.
\newblock {\em Reviews of Accelerator Science and Technology\/} {\bf 2} (2009) 179--200.

\bibitem[{Anferov et~al.(2007)Anferov, Ball, Collins, and Derenchuk}]{Anferov2007}
Anferov VA, Ball MS, Collins JC, Derenchuk VP.
\newblock Indiana {{University Cyclotron Operation For Proton Therapy Facility}}.
\newblock {\em Cyclotrons and Their Applications\/}  (2007).

\bibitem[{Gerbershagen et~al.(2017)Gerbershagen, Adelmann, D{\"o}lling, Meer, Rizzoglio, and Schippers}]{Gerbershagen2017}
Gerbershagen A, Adelmann A, D{\"o}lling R, Meer D, Rizzoglio V, Schippers JM.
\newblock Simulations and measurements of proton beam energy spectrum after energy degradation.
\newblock {\em Journal of Physics: Conference Series\/} {\bf 874} (2017) 012108.
\newblock \doi{10.1088/1742-6596/874/1/012108}.

\bibitem[{Schippers(2017)}]{Schippers2017}
Schippers JM.
\newblock Beam-{{Transport Systems}} for {{Particle Therapy}}.
\newblock {\em Proceedings of the {{CAS-CERN Accelerator School}}: {{Accelerators}} for {{Medical Applications}}, {{V\"osendorf}}, {{Austria}}, 26 {{May}}--5 {{June}} 2015\/} (2017), vol.~1, 241--252.
\newblock \doi{10.23730/CYRSP-2017-001.241}.

\bibitem[{Coutrakon et~al.(1994)Coutrakon, Hubbard, Johanning, Maudsley, Slaton, and Morton}]{Coutrakon1994}
Coutrakon G, Hubbard J, Johanning J, Maudsley G, Slaton T, Morton P.
\newblock A performance study of the {{Loma Linda}} proton medical accelerator.
\newblock {\em Medical Physics\/} {\bf 21} (1994) 1691--1701.
\newblock \doi{10.1118/1.597270}.

\bibitem[{Pavlovi{\v c} et~al.(2024)Pavlovi{\v c}, Pivi, Stra{\v s}{\'i}k, Rizzoglio, Pullia, Adler et~al.}]{Pavlovic2024}
Pavlovi{\v c} M, Pivi MTF, Stra{\v s}{\'i}k I, Rizzoglio V, Pullia MG, Adler L, et~al.
\newblock Rotatorlike gantry optics.
\newblock {\em Physical Review Accelerators and Beams\/} {\bf 27} (2024) 073502.
\newblock \doi{10.1103/PhysRevAccelBeams.27.073502}.

\bibitem[{Keil et~al.(2007)Keil, Sessler, and Trbojevic}]{Keil2007}
Keil E, Sessler AM, Trbojevic D.
\newblock Hadron cancer therapy complex using nonscaling fixed field alternating gradient accelerator and gantry design.
\newblock {\em Physical Review Special Topics - Accelerators and Beams\/} {\bf 10} (2007) 4--11.
\newblock \doi{10.1103/PhysRevSTAB.10.054701}.

\bibitem[{Garland et~al.(2015)Garland, Appleby, Owen, and Tygier}]{Garland2015}
Garland JM, Appleby RB, Owen H, Tygier S.
\newblock Normal-conducting scaling fixed field alternating gradient accelerator for proton therapy.
\newblock {\em Physical Review Special Topics - Accelerators and Beams\/} {\bf 18} (2015) 094701.
\newblock \doi{10.1103/PhysRevSTAB.18.094701}.

\bibitem[{Taylor et~al.(2018)Taylor, Edgecock, and Johnstone}]{Taylor2018}
Taylor J, Edgecock TR, Johnstone C.
\newblock {{HEATHER}} -- {{HElium}} ion {{Accelerator}} for {{radioTHERapy}}.
\newblock {\em 13th International Topical Meeting on Nuclear Applications of Accelerators 2017, AccApp 2017: The Expanding Universe of Accelerator Applications\/}  (2018) 368--376.

\bibitem[{Meot(2019)}]{Meot2019}
Meot F.
\newblock {{RACCAM}}: {{An}} example of spiral sector scaling {{FFA}} technology.
\newblock Tech. Rep. BNL--211536-2019-NEWS, 1507116, Brookhaven National Laboratory (2019).

\bibitem[{Aymar et~al.(2020)Aymar, Becker, Boogert, Borghesi, Bingham, Brenner et~al.}]{Aymar2020}
Aymar G, Becker T, Boogert S, Borghesi M, Bingham R, Brenner C, et~al.
\newblock {{LhARA}}: {{The Laser-hybrid Accelerator}} for {{Radiobiological Applications}}.
\newblock {\em Frontiers in Physics\/} {\bf 8} (2020) 1--21.
\newblock \doi{10.3389/fphy.2020.567738}.

\bibitem[{Fenning(2011)}]{Fenning2011}
Fenning R.
\newblock {\em Novel {{FFAG Gantry}} and {{Transport Line Designs}} for {{Charged Particle Therapy}}\/}.
\newblock Ph.D. thesis, Brunel University (2011).

\bibitem[{Wan et~al.(2015)Wan, Brouwer, Caspi, Prestemon, Gerbershagen, Schippers et~al.}]{Wan2015}
Wan W, Brouwer L, Caspi S, Prestemon S, Gerbershagen A, Schippers JM, et~al.
\newblock Alternating-gradient canted cosine theta superconducting magnets for future compact proton gantries.
\newblock {\em Physical Review Special Topics - Accelerators and Beams\/} {\bf 18} (2015) 103501.
\newblock \doi{10.1103/PhysRevSTAB.18.103501}.

\bibitem[{Brouwer et~al.(2019)Brouwer, Huggins, and Wan}]{Brouwer2019}
Brouwer L, Huggins A, Wan W.
\newblock An achromatic gantry for proton therapy with fixed-field superconducting magnets.
\newblock {\em International Journal of Modern Physics A\/} {\bf 34} (2019).
\newblock \doi{10.1142/S0217751X19420235}.

\bibitem[{Nesteruk et~al.(2019{\natexlab{b}})Nesteruk, Calzolaio, Seidel, and Schippers}]{Nesteruk2019a}
Nesteruk KP, Calzolaio C, Seidel M, Schippers JM.
\newblock Beam optics of a superconducting proton-therapy gantry with a large momentum acceptance.
\newblock {\em International Journal of Modern Physics A\/} {\bf 34} (2019{\natexlab{b}}) 1--8.
\newblock \doi{10.1142/S0217751X19420247}.

\bibitem[{Bottura et~al.(2020)Bottura, Felcini, De~Rijk, and Dutoit}]{Bottura2020a}
Bottura L, Felcini E, De~Rijk G, Dutoit B.
\newblock {{GaToroid}}: {{A}} novel toroidal gantry for hadron therapy.
\newblock {\em Nuclear Instruments and Methods in Physics Research, Section A: Accelerators, Spectrometers, Detectors and Associated Equipment\/} {\bf 983} (2020) 164588.
\newblock \doi{10.1016/j.nima.2020.164588}.

\bibitem[{Trbojevic et~al.(2021)Trbojevic, Brooks, Roser, and Tsoupas}]{Trbojevic2021}
Trbojevic D, Brooks S, Roser T, Tsoupas N.
\newblock Superb {{Fixed Field Permanent Magnet Proton Therapy Gantry}}.
\newblock {\em Proceedings of {{IPAC2021}}\/} (Campinas, Brazil: JACoW Publishing) (2021), 1405--1408.
\newblock \doi{10.18429/JACoW-IPAC2021-TUPAB030}.

\bibitem[{Dascalu and Sheehy(2021)}]{Dascalu2021}
[Dataset] Dascalu TS, Sheehy SL.
\newblock Beam delivery systems for linac-based proton therapy (2021).

\bibitem[{Liao et~al.(2024)Liao, Liu, Wang, Yang, Chen, and Qin}]{Liao2024}
Liao YC, Liu X, Wang W, Yang ZY, Chen QS, Qin B.
\newblock Design of a large momentum acceptance proton therapy gantry utilizing {{AG-CCT}} magnets.
\newblock {\em Nuclear Science and Techniques\/} {\bf 35} (2024) 172.
\newblock \doi{10.1007/s41365-024-01522-1}.

\bibitem[{Steinberg et~al.(2024)Steinberg, Appleby, Yap, and Sheehy}]{Steinberg2024a}
Steinberg AF, Appleby RB, Yap JSL, Sheehy SL.
\newblock Design of a large energy acceptance beamline using fixed field accelerator optics.
\newblock {\em Physical Review Accelerators and Beams\/} {\bf 27} (2024) 071601.
\newblock \doi{10.1103/PhysRevAccelBeams.27.071601}.

\bibitem[{Brooks(2024)}]{Brooks2024}
Brooks SJ.
\newblock {{OPTIMISATION OF A PERMANENT MAGNET MULTI-ENERGY FFA ARC FOR THE CEBAF ENERGY UPGRADE}}.
\newblock {\em Proceedings of {{IPAC2024}}\/} (Nashville, TN: JACoW Publishing) (2024).

\bibitem[{Bodenstein et~al.(2024)Bodenstein, Deitrick, Gamage, Hamlette, Meyers, Nissen et~al.}]{Bodenstein2024}
Bodenstein R, Deitrick K, Gamage B, Hamlette D, Meyers J, Nissen E, et~al.
\newblock Permanent magnet resiliency in {{CEBAF}}'s radiation environment: {{LDRD}} grant status and plans.
\newblock {\em Proceedings of {{IPAC2024}}\/} (Nashville, TN) (2024).
\newblock \doi{10.18429/JACoW-IPAC2024-THPS58}.

\bibitem[{Chinniah et~al.(2023)Chinniah, Deisher, Herman, Johnson, Mahajan, and Foote}]{Chinniah2023}
Chinniah S, Deisher AJ, Herman MG, Johnson JE, Mahajan A, Foote RL.
\newblock Rotating {{Gantries Provide Individualized Beam Arrangements}} for {{Charged Particle Therapy}}.
\newblock {\em Cancers\/} {\bf 15} (2023) 2044.
\newblock \doi{10.3390/cancers15072044}.

\bibitem[{Zhou et~al.(2021)Zhou, Li, Kubota, Sakai, and Ohno}]{Zhou2021}
Zhou Y, Li Y, Kubota Y, Sakai M, Ohno T.
\newblock Robust {{Angle Selection}} in {{Particle Therapy}}.
\newblock {\em Frontiers in Oncology\/} {\bf 11} (2021) 715025.
\newblock \doi{10.3389/fonc.2021.715025}.

\bibitem[{Haberer et~al.(2004)Haberer, Debus, Eickhoff, J{\"a}kel, {Schulz-Ertner}, and Weber}]{Haberer2004}
Haberer T, Debus J, Eickhoff H, J{\"a}kel O, {Schulz-Ertner} D, Weber U.
\newblock The {{Heidelberg}} ion therapy center.
\newblock {\em Radiotherapy and Oncology\/} {\bf 73} (2004) 186--190.
\newblock \doi{10.1016/S0167-8140(04)80046-X}.

\bibitem[{Iwata et~al.(2017)Iwata, Fujimoto, Matsuba, Fujita, Sato, Furukawa et~al.}]{Iwata2017}
Iwata Y, Fujimoto T, Matsuba S, Fujita T, Sato S, Furukawa T, et~al.
\newblock Recent progress of a superconducting rotating-gantry for carbon-ion radiotherapy.
\newblock {\em Nuclear Instruments and Methods in Physics Research, Section B: Beam Interactions with Materials and Atoms\/} {\bf 406} (2017) 338--342.
\newblock \doi{10.1016/j.nimb.2016.10.040}.

\bibitem[{Yang et~al.(2022)Yang, Mizushima, Matsuba, Fujimoto, Noda, Urata et~al.}]{Yang2022}
Yang Y, Mizushima K, Matsuba S, Fujimoto T, Noda E, Urata M, et~al.
\newblock Thermal characteristics of a helium-free superconducting magnet system for a fast-ramping heavy-ion synchrotron.
\newblock {\em Cryogenics\/} {\bf 126} (2022) 103524.
\newblock \doi{10.1016/j.cryogenics.2022.103524}.

\bibitem[{Brouwer et~al.(2020)Brouwer, Caspi, Edwards, Godeke, Hafalia, Hodgkinson et~al.}]{Brouwer2020}
Brouwer L, Caspi S, Edwards K, Godeke A, Hafalia R, Hodgkinson A, et~al.
\newblock Design and test of a curved superconducting dipole magnet for proton therapy.
\newblock {\em Nuclear Instruments and Methods in Physics Research Section A: Accelerators, Spectrometers, Detectors and Associated Equipment\/} {\bf 957} (2020) 163414.
\newblock \doi{10.1016/j.nima.2020.163414}.

\bibitem[{Gupta et~al.(2015)Gupta, Anerella, Ghosh, Kolonko, Larson, Rey et~al.}]{Gupta2015}
Gupta R, Anerella M, Ghosh A, Kolonko J, Larson D, Rey C, et~al.
\newblock {{HTS}}/{{lTS Hybrid High Field Superconducting Magnet Designs}} for the {{Proposed}} 100 {{TeV Proton Colliders}}.
\newblock {\em Proceedings of the 6th Int. Particle Accelerator Conf.\/} {\bf IPAC2015} (2015) 3 pages, 1.201 MB.
\newblock \doi{10.18429/JACOW-IPAC2015-WEPWI053}.

\bibitem[{Meyer and Flasck(1970)}]{Meyer1970}
Meyer DI, Flasck R.
\newblock A new configuration for a dipole magnet for use in high energy physics applications.
\newblock {\em Nuclear Instruments and Methods\/} {\bf 80} (1970) 339--341.
\newblock \doi{10.1016/0029-554X(70)90784-6}.

\bibitem[{Pullia et~al.(2024)Pullia, Felcini, Benedetto, Dassa, De~Matteis, Donetti et~al.}]{Pullia2024}
Pullia MG, Felcini E, Benedetto E, Dassa L, De~Matteis E, Donetti M, et~al.
\newblock Gantries for carbon ions.
\newblock {\em Health and Technology\/}  (2024).
\newblock \doi{10.1007/s12553-024-00870-7}.

\bibitem[{Robin et~al.(2011)Robin, Arbelaez, Caspi, Sun, Sessler, Wan et~al.}]{Robin2011}
Robin DS, Arbelaez D, Caspi S, Sun C, Sessler A, Wan W, et~al.
\newblock Superconducting toroidal combined-function magnet for a compact ion beam cancer therapy gantry.
\newblock {\em Nuclear Instruments and Methods in Physics Research, Section A: Accelerators, Spectrometers, Detectors and Associated Equipment\/} {\bf 659} (2011) 484--493.
\newblock \doi{10.1016/j.nima.2011.08.049}.

\bibitem[{Holder(2014)}]{Holder2014}
Holder DJ.
\newblock A {{Compact Superconducting}} 330 {{MeV Proton Gantry}} for {{Radiotherapy}} and {{Computed Tomography A COMPACT SUPERCONDUCTING}} 330 {{MeV PROTON GANTRY FOR RADIOTHERAPY AND COMPUTED TOMOGRAPHY}}*.
\newblock {\em Proceedings of {{IPAC2014}}\/} (2014).

\bibitem[{Yap et~al.(2023)Yap, Sheehy, Steinberg, Norman, and Appleby}]{Yap2023}
Yap JS, Sheehy SL, Steinberg AF, Norman HX, Appleby RB.
\newblock {{TURBO}}: {{A}} novel beam delivery system enabling rapid depth scanning for charged particle therapy.
\newblock {\em Journal of Physics: Conference Series\/} {\bf 2420} (2023) 012094.
\newblock \doi{10.1088/1742-6596/2420/1/012094}.

\bibitem[{Volz et~al.(2020)Volz, Kelleter, Brons, Burigo, Graeff, Niebuhr et~al.}]{Volz2020}
Volz L, Kelleter L, Brons S, Burigo L, Graeff C, Niebuhr NI, et~al.
\newblock Experimental exploration of a mixed helium/carbon beam for online treatment monitoring in carbon ion beam therapy.
\newblock {\em Physics in Medicine and Biology\/} {\bf 65} (2020) 0--14.
\newblock \doi{10.1088/1361-6560/ab6e52}.

\bibitem[{Hardt et~al.(2024)Hardt, Pryanichnikov, Homolka, DeJongh, DeJongh, Cristoforetti et~al.}]{Hardt2024}
Hardt JJ, Pryanichnikov AA, Homolka N, DeJongh EA, DeJongh DF, Cristoforetti R, et~al.
\newblock The potential of mixed carbon-helium beams for online treatment verification: A simulation and treatment planning study.
\newblock {\em Physics in Medicine \& Biology\/}  (2024).
\newblock \doi{10.1088/1361-6560/ad46db}.

\bibitem[{Mazzucconi et~al.(2018)Mazzucconi, Agosteo, Ferrarini, Fontana, Lante, Pullia et~al.}]{Mazzucconi2018}
Mazzucconi D, Agosteo S, Ferrarini M, Fontana L, Lante V, Pullia M, et~al.
\newblock Mixed particle beam for simultaneous treatment and online range verification in carbon ion therapy: {{Proof}}-of-concept study.
\newblock {\em Medical Physics\/} {\bf 45} (2018) 5234--5243.
\newblock \doi{10.1002/mp.13219}.

\bibitem[{Gom{\`a} et~al.(2024)Gom{\`a}, Henkner, J{\"a}kel, Lorentini, Magro, Mirandola et~al.}]{Goma2024}
Gom{\`a} C, Henkner K, J{\"a}kel O, Lorentini S, Magro G, Mirandola A, et~al.
\newblock {{ESTRO-EPTN}} radiation dosimetry guidelines for the acquisition of proton pencil beam modelling data.
\newblock {\em Physics and Imaging in Radiation Oncology\/} {\bf 31} (2024) 100621.
\newblock \doi{10.1016/j.phro.2024.100621}.

\bibitem[{Li et~al.(2013)Li, Sahoo, Poenisch, Suzuki, Li, Li et~al.}]{Li2013}
Li H, Sahoo N, Poenisch F, Suzuki K, Li Y, Li X, et~al.
\newblock Use of treatment log files in spot scanning proton therapy as part of patient-specific quality assurance.
\newblock {\em Medical Physics\/} {\bf 40} (2013) 021703.
\newblock \doi{10.1118/1.4773312}.

\bibitem[{Lin et~al.(2016)Lin, Clasie, Lu, Flanz, Shen, and Jee}]{Lin2016}
Lin Y, Clasie B, Lu HM, Flanz J, Shen T, Jee KW.
\newblock Impacts of gantry angle dependent scanning beam properties on proton {{PBS}} treatment.
\newblock {\em Physics in Medicine \& Biology\/} {\bf 62} (2016) 344.
\newblock \doi{10.1088/1361-6560/aa5084}.

\bibitem[{Grevillot et~al.(2020)Grevillot, Osorio~Moreno, Letellier, Dreindl, Elia, Fuchs et~al.}]{Grevillot2020}
Grevillot L, Osorio~Moreno J, Letellier V, Dreindl R, Elia A, Fuchs H, et~al.
\newblock Clinical implementation and commissioning of the {{MedAustron Particle Therapy Accelerator}} for non-isocentric scanned proton beam treatments.
\newblock {\em Medical Physics\/} {\bf 47} (2020) 380--392.
\newblock \doi{10.1002/mp.13928}.

\bibitem[{Parodi et~al.(2010)Parodi, Mairani, Brons, Naumann, Kr{\"a}mer, Sommerer et~al.}]{Parodi2010}
Parodi K, Mairani A, Brons S, Naumann J, Kr{\"a}mer M, Sommerer F, et~al.
\newblock The influence of lateral beam profile modifications in scanned proton and carbon ion therapy: A {{Monte Carlo}} study.
\newblock {\em Physics in Medicine and Biology\/} {\bf 55} (2010) 5169--87.
\newblock \doi{10.1088/0031-9155/55/17/018}.

\bibitem[{Titt et~al.(2010)Titt, Mirkovic, Sawakuchi, Perles, Newhauser, Taddei et~al.}]{Titt2010}
Titt U, Mirkovic D, Sawakuchi GO, Perles LA, Newhauser WD, Taddei PJ, et~al.
\newblock Adjustment of the lateral and longitudinal size of scanned proton beam spots using a pre-absorber to optimize penumbrae and delivery efficiency.
\newblock {\em Phys Med Biol\/} {\bf 55} (2010) 7097--7106.
\newblock \doi{10.1088/0031-9155/55/23/S10.Adjustment}.

\bibitem[{Chen et~al.(2016)Chen, Chang, Moyers, Gao, and Mah}]{Chen2016}
Chen CC, Chang C, Moyers MF, Gao M, Mah D.
\newblock Technical {{Note}}: {{Spot}} characteristic stability for proton pencil beam scanning a).
\newblock {\em Medical Physics\/} {\bf 43} (2016) 777--782.
\newblock \doi{10.1118/1.4939663}.

\bibitem[{Dowdell et~al.(2012)Dowdell, Clasie, Depauw, Metcalfe, Rosenfeld, Kooy et~al.}]{Dowdell2012}
Dowdell SJ, Clasie B, Depauw N, Metcalfe P, Rosenfeld AB, Kooy HM, et~al.
\newblock Monte {{Carlo}} study of the potential reduction in out-of-field dose using a patient-specific aperture in pencil beam scanning proton therapy.
\newblock {\em Physics in Medicine and Biology\/} {\bf 57} (2012) 2829--2842.
\newblock \doi{10.1088/0031-9155/57/10/2829}.

\bibitem[{Romero-Exp{\'o}sito et~al.(2023)Romero-Exp{\'o}sito, Liszka, Christou, Toma-Dasu, and Dasu}]{Romero-Exposito2023}
Romero-Exp{\'o}sito M, Liszka M, Christou A, Toma-Dasu I, Dasu A.
\newblock Range shifter contribution to neutron exposure of patients undergoing proton pencil beam scanning.
\newblock {\em Medical Physics\/}  (2023) mp.16897.
\newblock \doi{10.1002/mp.16897}.

\bibitem[{Both et~al.(2014)Both, Shen, Kirk, Lin, Tang, {Alonso-Basanta} et~al.}]{Both2014}
Both S, Shen J, Kirk M, Lin L, Tang S, {Alonso-Basanta} M, et~al.
\newblock Development and {{Clinical Implementation}} of a {{Universal Bolus}} to {{Maintain Spot Size During Delivery}} of {{Base}} of {{Skull Pencil Beam Scanning Proton Therapy}}.
\newblock {\em International Journal of Radiation Oncology*Biology*Physics\/} {\bf 90} (2014) 79--84.
\newblock \doi{10.1016/j.ijrobp.2014.05.005}.

\bibitem[{Jelen et~al.(2013)Jelen, Bubula, Ammazzalorso, {Engenhart-Cabillic}, Weber, and Wittig}]{Jelen2013}
Jelen U, Bubula ME, Ammazzalorso F, {Engenhart-Cabillic} R, Weber U, Wittig A.
\newblock Dosimetric impact of reduced nozzle-to-isocenter distance in intensity-modulated proton therapy of intracranial tumors in combined proton-carbon fixed-nozzle treatment facilities.
\newblock {\em Radiation Oncology\/} {\bf 8} (2013) 218.
\newblock \doi{10.1186/1748-717X-8-218}.

\bibitem[{Kim et~al.(2018)Kim, Park, Jo, Cho, Shin, Lim et~al.}]{Kim2018}
Kim DH, Park S, Jo K, Cho S, Shin E, Lim DH, et~al.
\newblock Investigations of line scanning proton therapy with dynamic multi-leaf collimator.
\newblock {\em Physica Medica\/} {\bf 55} (2018) 47--55.
\newblock \doi{10.1016/j.ejmp.2018.10.009}.

\bibitem[{Pivi et~al.(2024)Pivi, Adler, Guidoboni, Kowarik, Kurf{\"u}rst, Maderb{\"o}ck et~al.}]{Pivi2024}
Pivi MTF, Adler L, Guidoboni G, Kowarik G, Kurf{\"u}rst C, Maderb{\"o}ck C, et~al.
\newblock Commissioning of a gantry beamline with rotator at a synchrotron-based ion therapy center.
\newblock {\em Physical Review Accelerators and Beams\/} {\bf 27} (2024) 023503.
\newblock \doi{10.1103/PhysRevAccelBeams.27.023503}.

\bibitem[{Wang et~al.(2014)Wang, Dirksen, Hyer, Buatti, Sheybani, Dinges et~al.}]{Wang2014a}
Wang D, Dirksen B, Hyer DE, Buatti JM, Sheybani A, Dinges E, et~al.
\newblock Impact of spot size on plan quality of spot scanning proton radiosurgery for peripheral brain lesions.
\newblock {\em Medical Physics\/} {\bf 41} (2014) 121705.
\newblock \doi{10.1118/1.4901260}.

\bibitem[{Newhauser and Zhang(2015)}]{Newhauser2015}
Newhauser WD, Zhang R.
\newblock The physics of proton therapy.
\newblock {\em Physics in Medicine \& Biology\/} {\bf 60} (2015).
\newblock \doi{10.1088/0031-9155/60/8/R155}.

\bibitem[{Lin et~al.(2013{\natexlab{a}})Lin, Ainsley, and McDonough}]{Lin2013a}
Lin L, Ainsley CG, McDonough JE.
\newblock Experimental characterization of two-dimensional pencil beam scanning proton spot profiles.
\newblock {\em Physics in Medicine and Biology\/} {\bf 58} (2013{\natexlab{a}}) 6193--6204.
\newblock \doi{10.1088/0031-9155/58/17/6193}.

\bibitem[{Lin et~al.(2013{\natexlab{b}})Lin, Ainsley, Solberg, and McDonough}]{Lin2013b}
Lin L, Ainsley CG, Solberg TD, McDonough JE.
\newblock Experimental characterization of two-dimensional spot profiles for two proton pencil beam scanning nozzles.
\newblock {\em Physics in Medicine and Biology\/} {\bf 59} (2013{\natexlab{b}}) 493--504.
\newblock \doi{10.1088/0031-9155/59/2/493}.

\bibitem[{Linz(2012)}]{Linz2012}
Linz U, editor.
\newblock {\em Ion {{Beam Therapy}}: {{Fundamentals}}, {{Technology}}, {{Clinical Applications}}\/} (Springer) (2012).

\bibitem[{Bues et~al.(2005)Bues, Newhauser, Titt, and Smith}]{Bues2005}
Bues M, Newhauser WD, Titt U, Smith AR.
\newblock Therapeutic step and shoot proton beam spot-scanning with a multi-leaf collimator: {{A Monte Carlo}} study.
\newblock {\em Radiation Protection Dosimetry\/} {\bf 115} (2005) 164--169.
\newblock \doi{10.1093/rpd/nci259}.

\bibitem[{Moteabbed et~al.(2016)Moteabbed, Yock, Depauw, Madden, Kooy, and Paganetti}]{Moteabbed2016}
Moteabbed M, Yock TI, Depauw N, Madden TM, Kooy HM, Paganetti H.
\newblock Impact of {{Spot Size}} and {{Beam-Shaping Devices}} on the {{Treatment Plan Quality}} for {{Pencil Beam Scanning Proton Therapy}}.
\newblock {\em International Journal of Radiation Oncology Biology Physics\/} {\bf 95} (2016) 190--198.
\newblock \doi{10.1016/j.ijrobp.2015.12.368}.

\bibitem[{Rana and Rosenfeld(2021)}]{Rana2021}
Rana S, Rosenfeld AB.
\newblock Impact of errors in spot size and spot position in robustly optimized pencil beam scanning proton-based stereotactic body radiation therapy ({{SBRT}}) lung plans.
\newblock {\em Journal of Applied Clinical Medical Physics\/} {\bf 22} (2021) 147--154.
\newblock \doi{10.1002/acm2.13293}.

\bibitem[{Ko et~al.(2024)Ko, Jeon, Ahn, Han, Chung, Cho et~al.}]{Ko2024}
Ko M, Jeon C, Ahn SH, Han Y, Chung K, Cho S, et~al.
\newblock Utility of a patient-specific aperture collimator and multi-leaf collimator in line scanning proton therapy.
\newblock {\em Journal of the Korean Physical Society\/}  (2024).
\newblock \doi{10.1007/s40042-024-01166-9}.

\bibitem[{Guo et~al.(2025)Guo, Zhang, Cheng, Zhou, and Sheng}]{Guo2025}
Guo J, Zhang F, Cheng J, Zhou R, Sheng Y.
\newblock Dosimetric impact of spot size variations in proton pencil beam scanning --- a {{Monte Carlo}} simulation study.
\newblock {\em Medical Engineering \& Physics\/} {\bf 140} (2025) 104346.
\newblock \doi{10.1016/j.medengphy.2025.104346}.

\bibitem[{Grassberger et~al.(2013)Grassberger, Dowdell, Lomax, Sharp, Shackleford, Choi et~al.}]{Grassberger2013}
Grassberger C, Dowdell S, Lomax A, Sharp G, Shackleford J, Choi N, et~al.
\newblock Motion interplay as a function of patient parameters and spot size in spot scanning proton therapy for lung cancer.
\newblock {\em International Journal of Radiation Oncology Biology Physics\/} {\bf 86} (2013) 380--386.
\newblock \doi{10.1016/j.ijrobp.2013.01.024}.

\bibitem[{Safai et~al.(2008)Safai, Bortfeld, and Engelsman}]{Safai2008}
Safai S, Bortfeld T, Engelsman M.
\newblock Comparison between the lateral penumbra of a collimated double-scattered beam and uncollimated scanning beam in proton radiotherapy.
\newblock {\em Physics in Medicine and Biology\/} {\bf 53} (2008) 1729--1750.
\newblock \doi{10.1088/0031-9155/53/6/016}.

\bibitem[{Lomax(2018)}]{Lomax2018}
Lomax A.
\newblock What will the medical physics of proton therapy look like 10~yr from now? {{A}} personal view.
\newblock {\em Medical Physics\/} {\bf 45} (2018) e984--e993.
\newblock \doi{10.1002/mp.13206}.

\bibitem[{Yasui et~al.(2015)Yasui, Toshito, Omachi, Kibe, Hayashi, Shibata et~al.}]{Yasui2015}
Yasui K, Toshito T, Omachi C, Kibe Y, Hayashi K, Shibata H, et~al.
\newblock A patient-specific aperture system with an energy absorber for spot scanning proton beams: {{Verification}} for clinical application.
\newblock {\em Medical Physics\/} {\bf 42} (2015) 6999--7010.
\newblock \doi{10.1118/1.4935528}.

\bibitem[{Daartz et~al.(2009)Daartz, Bangert, Bussi{\`e}re, Engelsman, and Kooy}]{Daartz2009a}
Daartz J, Bangert M, Bussi{\`e}re MR, Engelsman M, Kooy HM.
\newblock Characterization of a mini-multileaf collimator in a proton beamline.
\newblock {\em Medical Physics\/} {\bf 36} (2009) 1886--1894.
\newblock \doi{10.1118/1.3116382}.

\bibitem[{Torikoshi et~al.(2007)Torikoshi, Minohara, Kanematsu, Komori, Kanazawa, Noda et~al.}]{Torikoshi2007}
Torikoshi M, Minohara S, Kanematsu N, Komori M, Kanazawa M, Noda K, et~al.
\newblock Irradiation {{System}} for {{HIMAC}}.
\newblock {\em Journal of Radiation Research\/} {\bf 48} (2007) A15--A25.
\newblock \doi{10.1269/jrr.48.A15}.

\bibitem[{Hyer et~al.(2014)Hyer, Hill, Wang, Smith, and Flynn}]{Hyer2014}
Hyer DE, Hill PM, Wang D, Smith BR, Flynn RT.
\newblock A dynamic collimation system for penumbra reduction in spot-scanning proton therapy: {{Proof}} of concept.
\newblock {\em Medical Physics\/} {\bf 41} (2014) 1--9.
\newblock \doi{10.1118/1.4837155}.

\bibitem[{Winterhalter et~al.(2018)Winterhalter, Meier, Oxley, Weber, Lomax, and Safai}]{Winterhalter2018a}
Winterhalter C, Meier G, Oxley D, Weber DC, Lomax AJ, Safai S.
\newblock Contour scanning, multi-leaf collimation and the combination thereof for proton pencil beam scanning.
\newblock {\em Physics in Medicine \& Biology\/} {\bf 64} (2018) 015002.
\newblock \doi{10.1088/1361-6560/aaf2e8}.

\bibitem[{Moignier et~al.(2016{\natexlab{a}})Moignier, Gelover, Wang, Smith, Flynn, Kirk et~al.}]{Moignier2016}
Moignier A, Gelover E, Wang D, Smith B, Flynn R, Kirk M, et~al.
\newblock Theoretical {{Benefits}} of {{Dynamic Collimation}} in {{Pencil Beam Scanning Proton Therapy}} for {{Brain Tumors}}: {{Dosimetric}} and {{Radiobiological Metrics}}.
\newblock {\em International Journal of Radiation Oncology*Biology*Physics\/} {\bf 95} (2016{\natexlab{a}}) 171--180.
\newblock \doi{10.1016/j.ijrobp.2015.08.030}.

\bibitem[{Moignier et~al.(2016{\natexlab{b}})Moignier, Gelover, Wang, Smith, Flynn, Kirk et~al.}]{Moignier2016a}
Moignier A, Gelover E, Wang D, Smith B, Flynn R, Kirk M, et~al.
\newblock Improving {{Head}} and {{Neck Cancer Treatments Using Dynamic Collimation}} in {{Spot Scanning Proton Therapy}}.
\newblock {\em International Journal of Particle Therapy\/} {\bf 2} (2016{\natexlab{b}}) 544--554.
\newblock \doi{10.14338/IJPT-15-00026.1}.

\bibitem[{Smith and Hyer(2025)}]{Smith2025a}
Smith BR, Hyer DE.
\newblock The {{LET}} enhancement of energy-specific collimation in pencil beam scanning proton therapy.
\newblock {\em Journal of Applied Clinical Medical Physics\/} {\bf 26} (2025) e14477.
\newblock \doi{10.1002/acm2.14477}.

\bibitem[{Ueno et~al.(2019)Ueno, Matsuura, Hirayama, Takao, Ueda, Matsuo et~al.}]{Ueno2019}
Ueno K, Matsuura T, Hirayama S, Takao S, Ueda H, Matsuo Y, et~al.
\newblock Physical and biological impacts of collimator-scattered protons in spot-scanning proton therapy.
\newblock {\em Journal of Applied Clinical Medical Physics\/} {\bf 20} (2019) 48--57.
\newblock \doi{10.1002/acm2.12653}.

\bibitem[{Hopfensperger et~al.(2023)Hopfensperger, Li, Paxton, Sarkar, Su, Price et~al.}]{Hopfensperger2023}
Hopfensperger KM, Li X, Paxton A, Sarkar V, Su FCF, Price RG, et~al.
\newblock Measurements of fetal dose with {{Mevion S250i}} proton therapy system with {{HYPERSCAN}}.
\newblock {\em Journal of Applied Clinical Medical Physics\/} {\bf 24} (2023) e13957.
\newblock \doi{10.1002/acm2.13957}.

\bibitem[{Nelson et~al.(2023)Nelson, Culberson, Hyer, Geoghegan, Patwardhan, Smith et~al.}]{Nelson2023}
Nelson NP, Culberson WS, Hyer DE, Geoghegan TJ, Patwardhan KA, Smith BR, et~al.
\newblock Dosimetric delivery validation of dynamically collimated pencil beam scanning proton therapy.
\newblock {\em Physics in Medicine \& Biology\/} {\bf 68} (2023) 055003.
\newblock \doi{10.1088/1361-6560/acb6cd}.

\bibitem[{Hyer et~al.(2021)Hyer, Bennett, Geoghegan, Bues, and Smith}]{Hyer2021}
Hyer DE, Bennett LC, Geoghegan TJ, Bues M, Smith BR.
\newblock Innovations and the use of collimators in the delivery of pencil beam scanning proton therapy.
\newblock {\em International Journal of Particle Therapy\/} {\bf 8} (2021) 73--83.
\newblock \doi{10.14338/IJPT-20-00039.1}.

\bibitem[{Smith et~al.(2019)Smith, Hyer, Flynn, Hill, and Culberson}]{Smith2019}
Smith BR, Hyer DE, Flynn RT, Hill PM, Culberson WS.
\newblock Trimmer sequencing time minimization during dynamically collimated proton therapy using a colony of cooperating agents.
\newblock {\em Physics in Medicine \& Biology\/} {\bf 64} (2019) 205025.
\newblock \doi{10.1088/1361-6560/ab416d}.

\bibitem[{Engwall et~al.(2025)Engwall, Mikhalev, Sundstr{\"o}m, Marthin, and Wase}]{Engwall2025}
Engwall E, Mikhalev V, Sundstr{\"o}m J, Marthin O, Wase V.
\newblock Shoot-through layers in upright proton arcs unlock advantages in plan quality and range verification.
\newblock {\em Medical Physics\/} {\bf 52} (2025) e18051.
\newblock \doi{10.1002/mp.18051}.

\bibitem[{Holmes et~al.(2022)Holmes, Shen, Patel, Wong, Foote, Bues et~al.}]{Holmes2022}
Holmes J, Shen J, Patel SH, Wong WW, Foote RL, Bues M, et~al.
\newblock Collimating individual beamlets in pencil beam scanning proton therapy, a dosimetric investigation.
\newblock {\em Frontiers in Oncology\/} {\bf 12} (2022).
\newblock \doi{10.3389/fonc.2022.1031340}.

\bibitem[{Maradia et~al.(2025)Maradia, Yue, Molzahn, Wang, Pankuch, Charyyev et~al.}]{Maradia2025}
Maradia V, Yue N, Molzahn A, Wang J, Pankuch M, Charyyev S, et~al.
\newblock High-speed proton therapy within a short breath-hold  (2025).
\newblock ArXiv [preprint]. arXiv:2510.06766. Available at: \url{https://arxiv.org/pdf/2510.06766} (Accessed Jan 3, 2026).

\bibitem[{Tan et~al.(2023)Tan, Lew, Koh, Wibawa, Yeap, Master et~al.}]{Tan2023}
Tan HQ, Lew KS, Koh CWY, Wibawa A, Yeap PL, Master Z, et~al.
\newblock Implementing dispersion measurement as part of scanning proton therapy commissioning and quality assurance.
\newblock {\em Physics in Medicine \& Biology\/}  (2023).
\newblock \doi{10.1088/1361-6560/ad0536}.

\bibitem[{Arjomandy et~al.(2019)Arjomandy, Taylor, Ainsley, Safai, Sahoo, Pankuch et~al.}]{AAPM-TG224_2019}
Arjomandy B, Taylor P, Ainsley C, Safai S, Sahoo N, Pankuch M, et~al.
\newblock {{AAPM}} task group 224: {{Comprehensive}} proton therapy machine quality assurance.
\newblock {\em Medical Physics\/} {\bf 46} (2019) e678--e705.
\newblock \doi{10.1002/mp.13622}.

\bibitem[{Wang et~al.(2024)Wang, Liu, Liao, Zeng, Chen, Yu et~al.}]{Wang2024}
Wang W, Liu X, Liao Y, Zeng Y, Chen Y, Yu B, et~al.
\newblock Mixed-size spot scanning with a compact large momentum acceptance superconducting ({{LMA-SC}}) gantry beamline for proton therapy.
\newblock {\em Physics in Medicine \& Biology\/} {\bf 69} (2024) 115011.
\newblock \doi{10.1088/1361-6560/ad45a6}.

\bibitem[{Rana and Rosenfeld(2024)}]{Rana2024}
Rana S, Rosenfeld AB.
\newblock Effects of spot size errors in {{DynamicARC}} pencil beam scanning proton therapy planning.
\newblock {\em Physics in Medicine \& Biology\/} {\bf 69} (2024) 235008.
\newblock \doi{10.1088/1361-6560/ad8feb}.

\bibitem[{{van Marlen} et~al.(2021){van Marlen}, Dahele, Folkerts, Abel, Slotman, and Verbakel}]{VanMarlen2021}
{van Marlen} P, Dahele M, Folkerts M, Abel E, Slotman BJ, Verbakel W.
\newblock Ultra-high dose rate transmission beam proton therapy for conventionally fractionated head and neck cancer: {{Treatment}} planning and dose rate distributions.
\newblock {\em Cancers\/} {\bf 13} (2021) 1--12.
\newblock \doi{10.3390/cancers13081859}.

\bibitem[{Kong et~al.(2024)Kong, Huiskes, Habraken, Astreinidou, Rasch, Heijmen et~al.}]{Kong2024}
Kong W, Huiskes M, Habraken S, Astreinidou E, Rasch C, Heijmen B, et~al.
\newblock Reducing the lateral dose penumbra in {{IMPT}} by incorporating transmission pencil beams.
\newblock {\em Radiotherapy and Oncology\/} {\bf 198} (2024) 110388.
\newblock \doi{10.1016/j.radonc.2024.110388}.

\bibitem[{Penfold et~al.(2024)Penfold, Santos, Penfold, Shierlaw, and Crain}]{Penfold2024}
Penfold SN, Santos AMC, Penfold M, Shierlaw E, Crain R.
\newblock Single high-energy arc proton therapy with {{Bragg}} peak boost ({{SHARP}}).
\newblock {\em Journal of Medical Radiation Sciences\/}  (2024) jmrs.769.
\newblock \doi{10.1002/jmrs.769}.

\bibitem[{Hyt{\"o}nen et~al.(2025)Hyt{\"o}nen, Vanderstraeten, and Verbakel}]{Hytonen2025}
Hyt{\"o}nen R, Vanderstraeten R, Verbakel WFAR.
\newblock Hybrid proton planning combining spread-out {{Bragg}} peak beams with transmission beams to shorten field delivery times while maintaining plan quality.
\newblock {\em Physics and Imaging in Radiation Oncology\/} {\bf 35} (2025) 100809.
\newblock \doi{10.1016/j.phro.2025.100809}.

\bibitem[{Oponowicz et~al.(2020)Oponowicz, Owen, Psoroulas, and Meer}]{Oponowicz2020}
Oponowicz E, Owen H, Psoroulas S, Meer D.
\newblock Geometry optimisation of graphite energy degrader for proton therapy.
\newblock {\em Physica Medica\/} {\bf 76} (2020) 227--235.
\newblock \doi{10.1016/j.ejmp.2020.06.023}.

\end{thebibliography}

\end{document}